\documentclass[journal]{IEEEtran}
\usepackage{amsmath, amssymb, amsfonts}
\usepackage{algorithmic}
\usepackage{algorithm}
\usepackage{array}
\usepackage{subfigure}
\usepackage{textcomp}
\usepackage{stfloats}
\usepackage{url}
\usepackage{verbatim}
\usepackage{graphicx}
\usepackage{cite}
\usepackage{textcomp}
\usepackage{makecell}
\usepackage{flushend}
\usepackage{bm} 
\usepackage{booktabs}
\usepackage{multirow}
\usepackage{balance}
\usepackage{hyperref}
\usepackage{enumitem}
\usepackage{threeparttable}
\usepackage[numbers, sort&compress]{natbib}

\hypersetup{colorlinks = true,
  linkcolor = black, 
  urlcolor = black,
  citecolor = black
} 

\begin{document}

\title{Unifying Tree-Reweighted Belief Propagation and Mean Field for Tracking Extended Targets}

\author{\IEEEauthorblockN{Weizhen Ma, Zhongliang Jing, \IEEEmembership{Senior Member, IEEE}, Peng Dong, Henry Leung, \IEEEmembership{Fellow, IEEE}}
\thanks{This work is partially supported by the National Natural Science Foundation of China under Grant 61673262, the National GF Basic Research Program under Grant JCKY2021110B134, the Fundamental Research Funds for the Central Universities, and the Engineering Research Center of Aerospace Science and Technology, Ministry of Education. \emph{(Corresponding authors: Zhongliang Jing.)}} 
\thanks{Weizhen Ma, Zhongliang Jing and Peng Dong are with the School of Aeronautics and Astronautics, Shanghai Jiao Tong University, Shanghai 200240, China (e-mail: \href{mailto:weizhenma@sjtu.edu.cn}{weizhenma@sjtu.edu.cn}, \href{mailto:zljing@sjtu.edu.cn}{zljing@sjtu.edu.cn}, \href{mailto:dongpengkty@sjtu.edu.cn}{dongpengkty@sjtu.edu.cn}).}
\thanks{Henry Leung is with the Department of Electrical and Computer Engineering, University of Calgary, Calgary AB T2N 1N4, Canada (e-mail: \href{mailto:leungh@ucalgary.ca}{leungh@ucalgary.ca}).}}

\markboth{Journal of \LaTeX\ Class Files,~Vol.~14, No.~8, August~2021}%
{Shell \MakeLowercase{\textit{et al.}}: A Sample Article Using IEEEtran.cls for IEEE Journals}


\maketitle

\begin{abstract}
	This paper proposes a unified tree-reweighted belief propagation (BP) and mean field (MF) approach for scalable detection and tracking of extended targets within the framework of factor graph. The factor graph is partitioned into a BP region and an MF region so that the messages in each region are updated according to the corresponding region rules. The BP region exploits the tree-reweighted BP, which offers improved convergence than the standard BP for graphs with massive cycles, to resolve data association. The MF region approximates the posterior densities of the measurement rate, kinematic state and extent. For linear Gaussian target models and gamma Gaussian inverse Wishart distributed state density, the unified approach provides a closed-form recursion for the state density. Hence, the proposed algorithm is more efficient than particle-based BP algorithms for extended target tracking. This method also avoids measurement clustering and gating since it solves the data association problem in a probabilistic fashion. We compare the proposed approach with algorithms such as the Poisson multi-Bernoulli mixture filter and the BP-based Poisson multi-Bernoulli filter. Simulation results demonstrate that the proposed algorithm achieves enhanced tracking performance.
\end{abstract}

\vspace{-0.3cm}
\section{Introduction} \label{section_one}
\IEEEPARstart{I}{n} multi-target tracking, a point target generates at most one measurement per scan, whereas an extended target can produce multiple measurements per scan. Extended targets are commonly involved in tracking applications that utilize high-resolution or near-field sensors, such as lidar. A widely used measurement model for extended targets is the inhomogeneous Poisson point process \cite{GS2005}, which assumes that the number of measurements generated by a target follows a Poisson distribution and that measurements are statistically independent. 

Tracking multiple extended targets is more challenging than tracking multiple point targets. The Poisson rate of the measurement model and the target extent, which determines the spatial distribution of measurements, are typically unknown and must be estimated together with the kinematic state. Additionally, data association, which correlates measurements with targets, becomes computationally prohibitive because of the vast number of possible association events. This paper proposes a scalable algorithm for tracking extended targets, offering an efficient solution to data association and state estimation.  
\vspace{-0.4cm}
\subsection{Related work}
The extent of an extended target is typically modeled as an ellipse \cite{K2008, FFK2011, YB2019, TO2021}, a rectangular \cite{GLO2011}, or a star-convex shape \cite{BH2014, WO2015}. The Poisson measurement rate admits a closed-form Bayesian recursion for the gamma conjugate prior \cite{GO2012C}. Under ellipsoidal target shape assumption, the gamma Gaussian inverse Wishart (GGIW) distribution models the joint density of measurement rate, kinematic state and extent, which facilitates the development of many multiple extended target tracking (ETT) algorithms \cite{GO2012, LGO2013, BRGVVS2016, GFS2020, XGSFFW2022}.

One heuristic strategy to reduce the computational complexity of data association involves partitioning measurements into disjoint clusters and associating clusters with targets, assuming that all measurements within a cluster originate from the same target. This two-stage approach has been widely adopted in many algorithms, yielding satisfactory results for well-spaced targets \cite{GO2012, LGO2013, BRGVVS2016, VB2016, CC2018, GFS2020, XGSFFW2022}. However, the performance of clustering-based algorithms deteriorates significantly for scenarios with closely-spaced targets \cite{GSRXF2018, MW2021}. 

Sampling-based and marginal-based algorithms that avoid measurement partition have been proposed. In \cite{GSRXF2018}, the Poisson multi-Bernoulli mixture (PMBM) filter employs various sampling methods to draw association events. This filter truncates most of the association events to reduce the computational complexity of sampling, such that it may fail for scenarios with high association uncertainty. A modified scalable Gibbs sampling-based tracker for a known number of targets is proposed in \cite{LGLG2023}. Marginal probabilities of association events or variables are computed using combinatorics in \cite{YWB2020} and variational Bayes in \cite{GLG2024}. However, the algorithm in \cite{YWB2020} updates the state sequentially such that the results rely on the order of measurements, and the algorithm in \cite{GLG2024} is designed for scenarios with a known number of targets. 

An emerging scalable tracking framework visualizes the tracking problem using factor graph (FG) \cite{KFL2001} and computes the state density via belief propagation (BP) \cite{P1988}. It has been extensively applied to point targets \cite{WL2014, MBWH2017, MKWLHBW2018}. BP-based algorithms for tracking known and unknown numbers of extended targets are proposed in \cite{MW2020} and \cite{MW2021}. A Poisson multi-Bernoulli (PMB) filter using BP is proposed in \cite{XGMWGS2023}. Although these methods outperform clustering-based algorithms, they have no analytical solutions and resort to particle implementations, hence their performance relies on the number of particles. A closed-form BP-based tracker is proposed in \cite{MJDL2024}, which heuristically combines the sum-product and max-sum algorithms \cite{BMN2006}. 

One feasible strategy for acquiring analytical solutions to FG-based target tracking problems is to calculate the single-target density using mean filed (MF) approximation \cite{BMN2006} and calculate the marginal distributions of association variables using BP. This concept stems from the merged BP and MF (BP-MF) approach \cite{EKMBF2013}, which integrates BP and MF into a unified message-passing algorithm based on the region-based free energy approximation \cite{YFW2005}. The merged BP-MF approach leverages the advantages of both methods: BP accommodates hard constraints (e.g., one-to-one association assumption) and provides exact marginals for acyclic graphs, while MF guarantees convergence and supports conjugate priors for exponential family distributions. This unified approach has been investigated for tracking point targets in \cite{LW2016, LMWPX2020, LWBPL2021, BLWPHL2024}, showing significant potential for developing closed-form ETT algorithms.

\subsection{Contributions and Notations}
In this paper, we propose a unified message-passing algorithm for scalable detection and tracking of extended targets. The proposed algorithm, referred to as TRWBP-MF, integrates tree-reweighted BP (TRWBP) \cite{WJW2005} and MF using the region-based free energy approximation \cite{YFW2005}, extending the merged BP-MF framework presented in \cite{EKMBF2013}. The TRWBP is a generalized BP that achieves a unique fixed-point or global optimum for any graph with parameters appropriately tuned, and offers improved convergence than BP for cyclic graphs \cite{WJW2005, RWS2008}. Hence, it is more suitable to apply TRWBP to the FGs of ETT problems, which often contain numerous loops such that ordinary BP may fail to converge \cite{MW2020}. 

We extend the system models presented in \cite{MW2020} to the tracking of an unknown number of targets and introduce a new FG representation for ETT. Note that the algorithm in \cite{MW2020} deals with known number of targets and only estimates the kinematic state. Using the unified TRWBP-MF approach, the FG is divided into a BP region and an MF region. The BP region computes the marginals of association variables while the MF region approximates the densities of state variables. Our method jointly estimates the target existence probability, measurement rate, kinematic state and extent. The single-target density is modeled by the GGIW distribution for target existence and a ``dummy'' probability density function (PDF) for target nonexistence. The proposed approach provides an analytical solution for liner Gaussian target models, significantly reducing the computational load compared to the particle-based BP algorithms \cite{MW2020, MW2021, XGMWGS2023}. In summary, the contributions are as follows. 
\begin{enumerate} 
    \item We propose a unified TRWBP-MF message-passing algorithm using the region-based free energy approximation.  
    \item A new FG representation for tracking an unknown number of extended targets is introduced. By applying the unified TRWBP-MF approach to this FG, we derive a closed-form tracker under linear Gaussian assumptions. 
    \item The performance advantages of the proposed algorithm are validated over comparative methods through two challenging simulated scenarios.
\end{enumerate}

The remaining sections are organized as follows. Section \ref{section_two} describes the target models, data association uncertainty and message-passing algorithms. Section \ref{section_three} introduces the FG representation for ETT and presents the unified TRWBP-MF massage-passing algorithm. Section \ref{section_four} provides the details of the proposed algorithm for ETT. Section \ref{section_five} demonstrates the simulation results and performance comparisons. Finally, Section \ref{section_six} concludes the paper.  

\textit{Notation:} Lowercase letters denote scalars, e.g., $l$,, bolded lowercase letters denote vectors, e.g., $\boldsymbol{x}$, bolded uppercase letters denote matrices, e.g., $\boldsymbol{F}$, uppercase calligraphic letters denote finite sets, e.g., $\mathcal{I}$. The cardinality of a set $\mathcal{I}$ is denoted by $|\mathcal{I}|$. For $i\in\mathcal{I}$, we write $\mathcal{I}\backslash i$ for $\mathcal{I}\backslash \{i\}$. The set-membership indicator function $\mathbb{I}_{\mathcal{I}}(i) = 1$ if $i\in\mathcal{I}$, otherwise $\mathbb{I}_{\mathcal{I}}(i) = 0$. Let $\|\boldsymbol{x}\|_0$ represent the zero norm of a vector $\boldsymbol{x}$, $d_{\boldsymbol{x}}$ and $d_{\boldsymbol{X}}$ denote the dimension of vector $\boldsymbol{x}$ and square matrix $\boldsymbol{X}$, respectively, and $\operatorname{tr}(\cdot)$ denote the trace of a matrix. For $x\in\mathbb{R}$, $\delta(x)=1$ if $x=0$, otherwise $\delta(x)=0$. Moreover, let $\mathrm{N}(\boldsymbol{x}; \boldsymbol{\mu}, \boldsymbol{\Sigma})$ denote the multivariate Gaussian distribution with mean $\boldsymbol{\mu}$ and covariance $\boldsymbol{\Sigma}$, $\mathrm{P}(x; \lambda)$ the Poisson distribution with rate $\lambda$, $\mathrm{G}(x; \alpha, \beta)$ the gamma distribution with shape $\alpha$ and rate $\beta$, $\mathrm{W}(\boldsymbol{X}; \nu, \boldsymbol{V})$ the Wishart distribution with degrees of freedom $\nu$ and scale matrix $\boldsymbol{V}$, $\mathrm{IW}(\boldsymbol{X}; \nu, \boldsymbol{V})$ the inverse Wishart (IW) distribution with degrees of freedom $\nu$ and scale matrix $\boldsymbol{V}$.   
\vspace{-0.2cm}
\section{Background} \label{section_two}
In this section, we describe the dynamic and measurement models and the data association uncertainty for extended target, and briefly review TRWBP and MF approximation.
\vspace{-0.2cm}
\subsection{Extended Target Models} \label{subsection_21}
We use potential targets (PTs) to represent both newly detected targets and surviving targets \cite{MKWLHBW2018, MW2021}. Let $N_k$ and $M_k$ denote the numbers of PTs and measurements at time $k$. For a PT  $n\in\{1,\dots, N_k\}$, we introduce the following state variables: measurement rate $\gamma^k_n$, kinematic state $\boldsymbol{\xi}^k_n$, ellipsoidal extent $\boldsymbol{E}^k_n$ and a binary existence variable $r^k_n$ that indicates target existence and nonexistence with values 1 and 0. We model the extent $\boldsymbol{E}^k_n$ by a symmetric semi-positive definite matrix using random matrix \cite{K2008}. Alternative to $\boldsymbol{E}^k_n$, we also consider a vector representation $\boldsymbol{e}^k_n$ of extent by concatenating the diagonal and unique off-diagonal elements of $\boldsymbol{E}^k_n$ \cite{MW2021}. We will interchangeably use the vector and matrix representations of extent in what follows. Let $\boldsymbol{y}_n^k=[(\boldsymbol{x}_n^k)^T~r^k_n]^T$ denote the augmented state of PT $n$, where $\boldsymbol{x}_n^k=[\gamma^k_n~(\boldsymbol{\xi}^k_n)^T~(\boldsymbol{e}^k_n)^T]^T$. The single target density $p(\boldsymbol{y}_n^k)=p(\boldsymbol{x}_n^k,r^k_n)$ is defined as
\begin{equation} \label{eqn_singleTargetDensity}
	p(\boldsymbol{x}_n^k, 0) = p^{k,0}_n p_\mathrm{d}(\boldsymbol{x}_n^k), \quad p(\boldsymbol{x}_n^k, 1) = p^{k,1}_n p(\boldsymbol{x}_n^k)
\end{equation}
where $p^{k,1}_n$ and $p^{k,0}_n$ denote the probabilities of target existence and nonexistence with normalization constraint $p^{k,0}_n + p^{k,1}_n \!=\! 1$, $p_\mathrm{d}(\boldsymbol{x}_n^k)$ is an arbitrary ``dummy" PDF integrated to one. Under the assumption that the state components $\gamma^k_n$, $\boldsymbol{\xi}^k_n$ and $\boldsymbol{e}^k_n$ are independent, $p(\boldsymbol{x}^k_n) = \mathrm{GGIW}(\boldsymbol{x}^k_n;\boldsymbol{\theta}^k_n)$ is a GGIW distribution, where $\boldsymbol{\theta}^k_n=(\alpha^k_n,\beta^k_n,\boldsymbol{\mu}^k_n,\boldsymbol{\Sigma}^k_n,\nu^k_n,\boldsymbol{V}^k_n)$ encapsulates all its parameters.  

In order to distinguish between surviving targets and newly detected targets, we represent surviving targets from last step as legacy PTs with states $\underline{\boldsymbol{y}}_n^k$, $n\in\{1,\dots, N_{k-1}\}$, and newly detected targets at current step as new PTs with states $\overline{\boldsymbol{y}}_m^k$, $m\in\{1,\dots, M_k\}$. Note that each new PT corresponds to a measurement \cite{MW2021}. We denote the joint PT state as $\boldsymbol{y}^k=[(\underline{\boldsymbol{y}}^k)^T~(\overline{\boldsymbol{y}}^k)^T]^T$ where $\underline{\boldsymbol{y}}^k=[(\underline{\boldsymbol{y}}_1^k)^T \cdots (\underline{\boldsymbol{y}}_{N_{k-1}}^k)^T]^T$ and $\overline{\boldsymbol{y}}^k=[(\overline{\boldsymbol{y}}_1^k)^T \cdots (\overline{\boldsymbol{y}}_{M_{k}}^k)^T]^T$. Hence the number of PTs at step $k$ is $N_k = N_{k-1}+M_k$.  

The transition density $p(\underline{\boldsymbol{y}}_n^k\vert\boldsymbol{y}_n^{k-1}) = p(\underline{\boldsymbol{x}}_n^k, \underline{r}_n^k\vert\boldsymbol{x}_n^{k-1}, \underline{r}_n^{k-1})$ for legacy PT $n$ is given by \cite{MW2021}
\begin{equation} \label{eqn_singleTarget_stateTransition}
	\begin{aligned}
		p(\underline{\boldsymbol{x}}_n^k, \underline{r}_n^k\vert\boldsymbol{x}_n^{k-1},0) &=
		\begin{cases}
			p_\mathrm{d}(\boldsymbol{x}_n^k), & \underline{r}^k_n=0 \\
			0, & \underline{r}^k_n=1
		\end{cases} \\
		p(\underline{\boldsymbol{x}}_n^k, \underline{r}_n^k\vert\boldsymbol{x}_n^{k-1},1) &=
		\begin{cases}
			(1-p_S)p_\mathrm{d}(\boldsymbol{x}_n^k), & \underline{r}^k_n \! = \! 0 \\
			p_S p(\underline{\boldsymbol{x}}_n^k\vert\boldsymbol{x}_n^{k-1}), & \underline{r}^k_n\! = \! 1
		\end{cases} 
	\end{aligned}
\end{equation}  
where $p_S$ and $p(\underline{\boldsymbol{x}}_n^k\vert\boldsymbol{x}_n^{k-1})$ are the survival probability and state transition density. Since the state components are independent, the density $p(\underline{\boldsymbol{x}}_n^k\vert\boldsymbol{x}_n^{k-1})$ can be decomposed as
\begin{equation} \label{eqn_transitionDensity}
  	p(\underline{\boldsymbol{x}}_n^k\vert\boldsymbol{x}_n^{k-1}) = p_\gamma(\underline{\gamma}_n^k\vert\gamma_n^{k-1}) p_{\boldsymbol{\xi}}(\underline{\boldsymbol{\xi}}_n^k\vert\boldsymbol{\xi}_n^{k-1}) p_{\boldsymbol{e}}(\underline{\boldsymbol{e}}_n^k\vert\boldsymbol{e}_n^{k-1}).
\end{equation}
The evolution density $p_\gamma(\underline{\gamma}_n^k\vert\gamma_n^{k-1})$ is generally unknown, and the prediction of measurement rate usually relies on heuristic method \cite{GO2012C}. We assume a linear Gaussian kinematic model $p_{\boldsymbol{\xi}}(\underline{\boldsymbol{\xi}}_n^k\vert\boldsymbol{\xi}_n^{k-1})=\mathrm{N}(\underline{\boldsymbol{\xi}}_n^k;\boldsymbol{F}\boldsymbol{\xi}_n^{k-1},\boldsymbol{Q})$, where $\boldsymbol{F}$ and $\boldsymbol{Q}$ are the transition matrix and process noise covariance matrix, respectively. The evolution of target extent follows a Wishart distribution $p_{\boldsymbol{e}}(\underline{\boldsymbol{e}}^k_n\vert\boldsymbol{e}_n^{k-1})=\mathrm{W}(\underline{\boldsymbol{E}}^k_n;\nu,\boldsymbol{E}^{k-1}_n/\nu)$, where a small value of $\nu$ indicates a large process noise. 

Let $\boldsymbol{z}^k = [(\boldsymbol{z}_1^k)^T \cdots (\boldsymbol{z}_{M_k}^k)^T]^T$ denote the joint measurement vector at time $k$. The sensor detects a target with detection probability $p_D$ and generates target-oriented measurements according to an inhomogeneous Poisson point process with Poisson rate $\gamma^k_n$. The measurement likelihood is given by
\begin{equation} \label{eqn_singleMeasLikelihood}
 	p(\boldsymbol{z}^k_m \vert \boldsymbol{\xi}^k_n, \boldsymbol{e}^k_n) = \mathrm{N}(\boldsymbol{z}^k_m;\boldsymbol{H}\boldsymbol{\xi}^k_n, s\boldsymbol{E}^k_n+\boldsymbol{R})
\end{equation}
where $\boldsymbol{H}$ and $\boldsymbol{R}$ represent the observation matrix and measurement noise covariance matrix, respectively, and $s$ is a scaling factor that controls the distribution of measurement sources on target extent. Clutter measurements are uniformly distributed over the surveillance region (SR) with density $p_\mathrm{c} = p_\mathrm{c}(\boldsymbol{z}^k_m)$ and intensity $\kappa_\mathrm{c} = \lambda_\mathrm{c} p_\mathrm{c}$, where $\lambda_\mathrm{c}$ is rate parameter of the Poisson distributed number of false alarms. 
\vspace{-0.3cm}
\subsection{Data Association Uncertainty}
We have introduced new PTs, which correspond to each measurement at current step, to represent newly detected targets. In this way, a new target can be represented by multiple new PTs if it generates multiple measurements. To properly associate each new target with a single new PT, we adopt the mapping rule between measurements and new PTs from \cite{MW2021}. For measurements $\boldsymbol{z}_{m_1},\dots,\boldsymbol{z}_{m_L}$ generated by the same new target, let $\overline{r}_{m_{\text{min}}}=1$ for $m_{\text{min}}=\min(m_1,\dots,m_L)$ and $\overline{r}_m=0$ for all $m\in\{m_1,\dots,m_L\}\backslash\{m_{\text{min}}\}$. This mapping ensures that a measurement with index $m$ is not associated with a new PT with index $m'>m$, and its realization relies on measurement clustering and reordering \cite{MW2021}.

Following the notations from \cite{MW2020}, we model the number of measurements $l$ generated by legacy PT $n$ using a truncated Poisson distribution $\mathrm{P}_{\mathrm{t}}(l|\underline{\gamma}_n) = \underline{c}^{\gamma}_n\underline{\gamma}_n^l\exp(-\underline{\gamma}_n)/l!$, where $\underline{c}^{\gamma}_n$ is a normalizing constant. This truncated PMF implies that theres exists an upper bound $\underline{\ell}_n$ such that $\mathrm{P}_{\mathrm{t}}(l|\underline{\gamma}_n)=0$ for $l>\underline{\ell}_n$. Similarly, a truncated Poisson distribution $\mathrm{P}_{\mathrm{t}}(l|\overline{\gamma}_m) = \overline{c}^{\gamma}_m\overline{\gamma}_m^l\exp(-\overline{\gamma}_m)/l!$ with upper bound $\overline{\ell}_m$ is defined for a new PT $m$. 

To describe the relationship between measurements and PTs, we define the target-oriented association vectors $\underline{\boldsymbol{a}}_n = (\underline{a}_{nl}| l\in\underline{\mathcal{L}}_n)^T$ and $\overline{\boldsymbol{a}}_m = (\overline{a}_{m l}| l \in \overline{\mathcal{L}}_{m} )^T$ for the legacy PT $n$ and the new PT $m$, respectively, where $\underline{\mathcal{L}}_n = \{1,\dots, \underline{\ell}_n\}$ and $\overline{\mathcal{L}}_{m} = \{1,\dots, \overline{\ell}_m\}$. The element $\underline{a}_{nl}$ takes a value $m\!\in\!\{1,\dots,M\}$ or 0, indicating that the $l$th measurement generated by the legacy PT $n$ is $\boldsymbol{z}_m$ or a dummy measurement \cite{MW2020}. And the element $\overline{a}_{m l}$  takes a value $o \in\{m,\dots,M\}$ or 0. We encapsulate all target-oriented association vectors into a joint association vector $\boldsymbol{a} = [\underline{\boldsymbol{a}}^T~\overline{\boldsymbol{a}}^T]^T$, where $\underline{\boldsymbol{a}} = [\underline{\boldsymbol{a}}_1^T \cdots \underline{\boldsymbol{a}}_{\underline{N}}^T]^T$, $\underline{N} = N_{k-1}$ and $\overline{\boldsymbol{a}}=[\overline{\boldsymbol{a}}_1^T \cdots \overline{\boldsymbol{a}}_M^T]^T$. 

Since redundant description of data association improves the scalability of BP-based multi-target tracking algorithms \cite{WL2014, MKWLHBW2018, MW2020, MW2021}, we consider the measurement-oriented association vector $\boldsymbol{b} = [\boldsymbol{b}_1^T \cdots \boldsymbol{b}_M^T]$, as defined in \cite{MW2020}. The element $\boldsymbol{b}_m = [b_{m1}~b_{m2}]^T$ takes the value $[n~l]^T$ or ${[0~0]^T}$, implying that $\boldsymbol{z}_m$ is the $l$th measurement generated by PT $n$ or a false alarm. Note that if $b_{m1} = n \in\{1,\dots,\underline{N}\}$ then $b_{m2} = l\in \underline{\mathcal{L}}_n$, and if $b_{m1} = n \in\{\underline{N}+1,\dots,\underline{N}+M\}$ then $b_{m2} = l\in \overline{\mathcal{L}}_{o}$ where $o=n-\underline{N}$.  

To ensure consistency between the target-oriented and measurement-oriented association variables, we introduce the following indicator functions
\begin{equation} \label{eqn_consistentConstraints}
	\begin{aligned}
		\underline{\varphi}_{nlm}(\underline{a}_{nl}, \boldsymbol{b}_m) & =
		\begin{cases}
			0, & \underline{a}_{nl} = m, \boldsymbol{b}_m\neq[n~l]^T \text{ or } \\
			&\underline{a}_{nl} \neq m, \boldsymbol{b}_m = [n~l]^T \\
			1, & \text{otherwise},
		\end{cases} \\
		\overline{\varphi}_{mlo}(\overline{a}_{ml}, \boldsymbol{b}_o) &=
		\begin{cases}
			0,& \overline{a}_{ml} = o, \boldsymbol{b}_o\neq[\underline{N} + m~l]^T \text{ or} \\
			& \underline{a}_{ml} \neq o , \boldsymbol{b}_o = [\underline{N} + m~l]^T \\
			1, & \text{otherwise}.
		\end{cases} \\
	\end{aligned}
\end{equation}
If all indicators $\underline{\varphi}_{nlm}(\underline{a}_{nl} \boldsymbol{b}_m)$, for $n\in\{1,\dots,\underline{N}\}$, $l\in\underline{\mathcal{L}}_n$, $m\in\{1,\dots,M\}$, and $\overline{\varphi}_{mlo}(\overline{a}_{ml}, \boldsymbol{b}_o)$, for $m\in\{1,\dots,M\}$, $l\in\overline{\mathcal{L}}_{m}$, $o\in\{m, \dots,M\}$, evaluate to one, the association vectors $\boldsymbol{a}$ and $\boldsymbol{b}$ describe a consistent or valid association event, i.e., no measurement is associated with multiple targets.

\subsection{TRWBP}
A FG is a graphical representation for the factorization of a joint probability mass function (PMF) or PDF $p(\boldsymbol{v})$. It consists of a set of variables $v_i$ with $i\in\mathcal{I}$, a set of factors $f_j(\cdot)$ with $j\in\mathcal{J}$, and a set of edges that illustrate the dependencies of factors on variables. The factors show how the global function $p(\boldsymbol{v})$ factorizes into local factors, i.e., $p(\boldsymbol{v})=\prod_{j\in\mathcal{J}}f_j(\boldsymbol{v}_j)$, where $\boldsymbol{v}_j = (v_i|i\in\mathcal{N}(j))^T$ with $\mathcal{N}(\cdot)$ being the index set of neighboring nodes of a given variable or factor node. 

TRWBP is a generalized BP developed for cyclic graphs. It aims to compute an upper bound of the partition function or the normalizing constant of $p(\boldsymbol{v})$ \cite{WJW2005}. It can also be viewed as an inference algorithm that approximates $p(\boldsymbol{v})$ by minimizing $\alpha$-divergence, or as a special case of fractional BP \cite{M2005}. The key motivation behind TRWBP is to represent the cyclic graph as a convex combination of spanning trees \cite{C1989}, which contain no cycles and can be solved exactly using ordinary BP. Due to the convexity, TRWBP has a unique global optimum or fixed-points and provides more stable message-passing updates \cite{RWS2008}.

The main difference between TRWBP and the standard BP is that TRWBP adjusts the messages using the strictly positive factor appearance probabilities (FAPs). The FAP $\rho_j$ represents the probability that factor $f_j(\cdot)$ appears in a spanning tree. These FAPs must be optimized for computing the marginals. The original optimization algorithm in \cite{WJW2005} is designed for graphs with pairwise interactions. However, extending this approach to graphs with higher-order interactions is not straightforward. Alternatively, setting $\rho_j$ as a constant can still improve performance in comparison to BP \cite{D2011, WPS2011, WPS2012}. Note that when $\rho_j=1$ for all $j\in\mathcal{J}$, TRWBP is simplified to the standard BP.

\subsection{MF Approximation}
With MF approximation, the function $p(\boldsymbol{v})$ is approximated by a fully factorized distribution $q(\boldsymbol{v}) = \prod_{i\in\mathcal{I}} q_i(v_i)$. The optimal approximation is obtained by minimizing the Kullback-Leibler divergence (KLD) between $q(\boldsymbol{v})$ and $p(\boldsymbol{v})$:
\begin{equation} \label{eqn_mf_KLD}
	q(\boldsymbol{v}) = \arg\min\mathrm{KLD}(q(\boldsymbol{v}) || p(\boldsymbol{v}))
\end{equation}
where $\mathrm{KLD}(q(\boldsymbol{v}) || p(\boldsymbol{v})) = \sum_{\boldsymbol{v}} q(\boldsymbol{v}) \ln\frac{q(\boldsymbol{v})}{p(\boldsymbol{v})}$. Using variational calculus, the solution for $q_i(v_i)$ fulfills \cite{BMN2006}
\begin{equation} \label{eqn_mf_posteriorUpdate}
	\ln q_i(v_i) = \mathbb{E}_{q_{\sim i}} [\ln p(\boldsymbol{v})] + c_i
\end{equation}
where $\mathbb{E}_{q_{\sim i}}[\cdot]$ denotes the expectation with respect to $q$ over $\boldsymbol{v}$ excluding $v_i$, and $c_i$ is a constant regarding $v_i$. Since the variables are usually coupled in $p(\boldsymbol{v})$, fixed-point iterations are used to resolve \eqref{eqn_mf_posteriorUpdate}. A local optimum of \eqref{eqn_mf_posteriorUpdate} is reached when the iteration converges. The MF approximation can also be interpreted as a message-passing algorithm \cite{WB2005, D2007}. 

\section{FG and Unified TRWBP-MF} \label{section_three}
In this section, we present the problem formulation and the joint PDF with the corresponding FG for ETT, as well as the unified TRWBP-MF message-passing algorithm for inference in the FG. Since we only consider a single time step of tracking recursion, the time index $k$ is omitted for notational brevity.

\subsection{Problem Formulation}
The considered ETT problem involves detecting and tracking PTs, i.e., estimating the existence variables $r_n$ and the state variables $\boldsymbol{x}_n$ based on the measurements up to the current step. In the Bayesian tracking framework, target detection and state estimation are equivalent to computing the existence probability $p^1_n$ and the marginal state density $p(\boldsymbol{x}_n)$ defined in \eqref{eqn_singleTargetDensity}. The existence probability can be calculated as $p^1_n = \int p(\boldsymbol{x}_n, r_n=1) \mathrm{d} \boldsymbol{x}_n$, where $p(\boldsymbol{x}_n, r_n)$ is given in \eqref{eqn_singleTargetDensity}. If $p^1_n$ exceeds a predefined threshold, the target is declared detected, and state estimate $\hat{\boldsymbol{x}}_n$ is obtained using the minimum mean-square error estimator, i.e., $\hat{\boldsymbol{x}}_n = \int \boldsymbol{x}_n p(\boldsymbol{x}_n) \mathrm{d} \boldsymbol{x}_n$. Hence, our task reduces to computing the density $p(\boldsymbol{x}_n, r_n)$, which is the marginal of the joint PDF in \eqref{eqn_jointPDF}, encompassing all state and association variables. To tractably compute the density $p(\boldsymbol{x}_n, r_n)$, we propose a unified message-passing algorithm, namely TRWBP-MF, to perform inference on the FG shown in Fig. \ref{fig_factorGraph}, which represents the ETT problem.  

\subsection{Joint PDF and FG Representation}
\begin{figure} [tbp]
	\centering
	\includegraphics[width=2.5in]{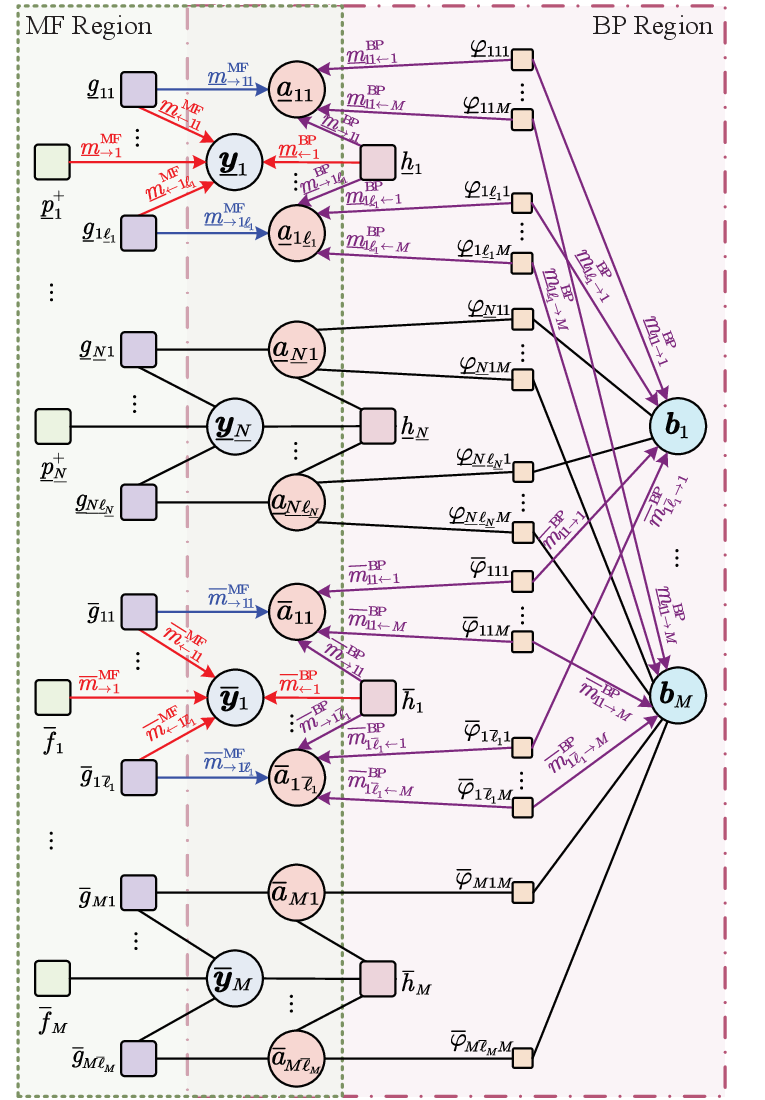}
	\caption{FG representation of the joint PDF in \eqref{eqn_jointPDF}. We use abbreviations for factors and messages, removing their contained variables. For example, $\underline{p}_n^+$ represents $\underline{p}_n^+(\underline{\boldsymbol{y}}_n)$, $\underline{m}^{\mathrm{MF}}_{\to n}$ represents $\underline{m}^{\mathrm{MF}}_{\to n}(\underline{\boldsymbol{y}}_n)$. }
	\label{fig_factorGraph}
\end{figure}

Let $\underline{p}^+_n(\underline{\boldsymbol{y}}_n)$ denote the predicted density of the legacy PT $n$, and $\overline{p}_m(\overline{\boldsymbol{y}}_m)$ denote the prior density of the new PT $m$. Assuming that the number of new PTs follows a Poisson distribution with rate parameter $\lambda_\mathrm{n}$ and that the densities of PTs are independent. Following the derivations in \cite{MW2020} and \cite{MW2021}, the joint PDF of state vector $\boldsymbol{y} = [\underline{\boldsymbol{y}}^T~\overline{\boldsymbol{y}}^T]^T$, the target-oriented association vector $\boldsymbol{a} = [\underline{\boldsymbol{a}}^T~\overline{\boldsymbol{a}}^T]^T$ and the measurement-oriented association vector $\boldsymbol{b}$, conditioned on the measurements $\boldsymbol{z}$, is given by
\begin{equation} \label{eqn_jointPDF}
	\begin{aligned}
		&p(\underline{\boldsymbol{y}}, \overline{\boldsymbol{y}}, \underline{\boldsymbol{a}}, \overline{\boldsymbol{a}}, \boldsymbol{b} \vert \boldsymbol{z}) \\
		\propto& \prod_{n=1}^{\underline{N}} \underline{p}^+_n(\underline{\boldsymbol{y}}_n) \underline{h}_n(\underline{\boldsymbol{y}}_n, \underline{\boldsymbol{a}}_n) \prod_{l=1}^{\underline{\ell}_n} \underline{g}_{nl}(\underline{\boldsymbol{y}}_n, \underline{a}_{nl}; \boldsymbol{z})\\
		\times& \prod_{m=1}^{M} \underline{\varphi}_{nlm}(\underline{a}_{nl}, \boldsymbol{b}_m) \prod_{m=1}^M \overline{f}_m(\overline{\boldsymbol{y}}_m) \overline{h}_m(\overline{\boldsymbol{y}}_m, \overline{\boldsymbol{a}}_m)  \\
        \times& \prod_{l=1}^{\overline{\ell}_{m}} \overline{g}_{ml}(\overline{\boldsymbol{y}}_m, \overline{a}_{ml}; \boldsymbol{z}) \prod_{o=m}^{M} \overline{\varphi}_{mlo}(\overline{a}_{ml}, \boldsymbol{b}_o)   \\
	\end{aligned}
\end{equation}
where the pseudo prior $\overline{f}_m(\overline{\boldsymbol{y}}_m)$ is given by
\begin{equation} \label{eqn_npt_pseudoPrior}
	\overline{f}_m(\overline{\boldsymbol{y}}_m) = 
	\begin{cases}
		\lambda_{\mathrm{n}} \overline{p}_m(\overline{\boldsymbol{x}}_m), & \overline{r}_m = 1 \\
		p_\mathrm{d}(\overline{\boldsymbol{x}}_m), & \overline{r}_m = 0. 
	\end{cases}
\end{equation}   
with prior $\overline{p}_m(\overline{\boldsymbol{x}}_m) = \mathrm{GGIW}(\overline{\boldsymbol{x}}_m; \overline{\boldsymbol{\theta}}_m)$, where the parameter vector $\overline{\boldsymbol{\theta}}_m = (\overline{\alpha}_m, \overline{\beta}_m, \overline{\boldsymbol{\mu}}_m, \overline{\boldsymbol{\Sigma}}_m, \overline{\nu}_m, \overline{\boldsymbol{V}}_m)$. The pseudo detection functions $\underline{h}_n(\underline{\boldsymbol{y}}_n,\underline{\boldsymbol{a}}_n)$ and  $\overline{h}_m(\overline{\boldsymbol{y}}_m,\overline{\boldsymbol{a}}_m)$ are defined as
\begin{equation} 
	\begin{aligned}
		&\underline{h}_n(\underline{\boldsymbol{y}}_n, \underline{\boldsymbol{a}}_n)\\
		\!=\!& 
		\begin{cases}
			\frac{\underline{c}^\gamma_n p_D \underline{\gamma}_n^{\|\underline{\boldsymbol{a}}_n\|_0}\exp(-\underline{\gamma}_n) (\underline{\ell}_n - \|\underline{\boldsymbol{a}}_n\|_0)!}{\underline{\ell}_n!}, &  \underline{r}_n \!=\! 1, \|\underline{\boldsymbol{a}}_n\|_0 \!>\! 0 \\
			\underline{c}^\gamma_n (1-p_D + p_D \exp(-\underline{\gamma}_n)), & \underline{r}_n \!=\! 1, \|\underline{\boldsymbol{a}}_n\|_0 \!=\! 0 \\
			\delta(\|\underline{\boldsymbol{a}}_n\|_0), & \underline{r}_n \!=\! 0 \\
		\end{cases}
	\end{aligned}
\end{equation}
\begin{equation}
    \overline{h}_m(\overline{\boldsymbol{y}}_m, \overline{\boldsymbol{a}}_m) \!=\!
    \begin{cases}
        \frac{ \overline{c}^\gamma_m p_D \overline{\gamma}_m^{\|\overline{\boldsymbol{a}}_m\|_0} \exp(-\overline{\gamma}_m) (\overline{\ell}_m - \|\overline{\boldsymbol{a}}_m\|_0)!} {(1-p_D + p_D \exp(-\overline{\gamma}_m))\overline{\ell}_m!} , &  \overline{r}_m \!=\! 1 \\
        \delta(\|\overline{\boldsymbol{a}}_m\|_0), & \overline{r}_m \!=\! 0.\\
    \end{cases}
\end{equation}
The likelihood ratios $\underline{g}_{nl}(\underline{\boldsymbol{y}}_n, \underline{a}_{nl};\boldsymbol{z})$ and $\overline{g}_{ml}(\overline{\boldsymbol{y}}_m, \overline{a}_{ml}; \boldsymbol{z}) $ are defined as
\begin{align}
	\underline{g}_{nl}(\underline{\boldsymbol{x}}_n, 1, \underline{a}_{nl}; \boldsymbol{z}) &\!=\! 
	\begin{cases}
		\frac{p(\boldsymbol{z}_m | \underline{\boldsymbol{\xi}}_n, \underline{\boldsymbol{e}}_n)} {\kappa_\mathrm{c}}, & \underline{a}_{nl} =m\in\{1,\dots, M\} \\
		1, & \underline{a}_{nl} = 0 
	\end{cases} \nonumber\\
	\underline{g}_{nl}(\underline{\boldsymbol{x}}_n, 0, \underline{a}_{nl}; \boldsymbol{z}) &\!=\! 1,
\end{align}
\begin{align}
	\overline{g}_{ml}(\overline{\boldsymbol{x}}_m, 1, \overline{a}_{ml}; \boldsymbol{z}) &\!=\! 
	\begin{cases}
		\frac{p(\boldsymbol{z}_o | \overline{\boldsymbol{\xi}}_m, \overline{\boldsymbol{e}}_m)} {\kappa_\mathrm{c}}, \!&\! \overline{a}_{ml} \!=\! o \!\in\!\{m,\dots, M\} \\
		0, \!&\! \overline{a}_{ml} \!=\! 0 
	\end{cases}\nonumber \\
	\overline{g}_{ml}(\overline{\boldsymbol{x}}_m, 0, \overline{a}_{ml}; \boldsymbol{z}) &\!=\! 1.
\end{align}
Fig. \ref{fig_factorGraph} shows the FG representation of the joint PDF \eqref{eqn_jointPDF}. A unified message-passing algorithm will be proposed for the inference in this FG. 

\subsection{The Unified TRWBP-MF message-passing Approach} 
We first review the region-based free energy approximation, which is essential to the derivation of the unified TRWBP-MF approach. A region $R= (\mathcal{I}_R, \mathcal{J}_R)$ in a FG consists of subsets of variable and factor indices, $\mathcal{I}_R\subseteq \mathcal{I}$ and $\mathcal{J}_R\subseteq \mathcal{J}$, such that $j\in\mathcal{J}_R$ implies $\mathcal{N}(j)\subseteq\mathcal{I}_R$. Let $\boldsymbol{v}_R = (v_i|i\in\mathcal{I}_R)^T$ represent the collection of variables in region $R$, and let $p_R(\boldsymbol{v}_R)$ and $q_R(\boldsymbol{v}_R)$ denote the exact marginal and the approximate belief for $\boldsymbol{v}_R$, respectively. The free energy $F_R(q_R)$ of region $R$ is defined as $F_R(q_R) = U_R(q_R) - H_R(q_R) $, where 
\begin{equation} \label{eqn_regionAverageEnergyEntropy}
	\begin{aligned}
		U_R(q_R) &=-  \sum_{j\in\mathcal{J}_R}\sum_{\boldsymbol{v}_R} q_R(\boldsymbol{v}_R) \ln f_j(\boldsymbol{v}_j) \\
		H_R(q_R) &=  -\sum_{\boldsymbol{v}_R} q_R(\boldsymbol{v}_R) \ln q_R(\boldsymbol{v}_R)
	\end{aligned}
\end{equation}
are the region average energy and region entropy \cite{YFW2005, EKMBF2013}, respectively. 

For a set $\mathcal{R}$ of regions, the region-based free energy is defined as \cite{YFW2005}
\begin{equation} \label{eqn_regionFreeEnergy}
	F_{\mathcal{R}} = U_{\mathcal{R}} - H_{\mathcal{R}}
\end{equation}
where $U_{\mathcal{R}}\!=\! \sum_{R\in\mathcal{R}}c_RU_R(q_R)$ and $H_{\mathcal{R}}\!=\! \sum_{R\in\mathcal{R}}c_R H_R(q_R)$ are the region-based average energy and region-based approximate entropy, respectively, and $c_R$ are the countering numbers associated with region $R$. A set $\mathcal{R}$ is called valid if its countering numbers ensure that each factor and variable are counted exactly once in the region-based approximation \eqref{eqn_regionFreeEnergy}. The division of the FG into small subregions facilitates the flexibility in selecting different inference approaches (i.e., different approximations to the region entropy $H_R(q_R)$) for individual regions. Each region belief $q_R(\boldsymbol{v}_R)$ must satisfy certain factorization constraint to ensure that the region entropy $H_R(q_R)$ can be appropriately approximated \cite{YFW2005}. 

Now we introduce the unified TRWBP and MF approach based on the region-based free energy approximation \eqref{eqn_regionFreeEnergy}. Suppose that the function $p(\boldsymbol{v})$ can be factorized as
\begin{equation} \label{eqn_joint_pmf_factorization}
	p(\boldsymbol{v}) = \prod_{j\in\mathcal{J}_{\mathrm{BP}}} f_j(\boldsymbol{v}_j) \prod_{j'\in\mathcal{J}_{\mathrm{MF}}} f_{j'}(\boldsymbol{v}_{j'})
\end{equation}
where $\mathcal{J}_{\mathrm{BP}}\cap \mathcal{J}_{\mathrm{MF}} = \emptyset$ and $\mathcal{J}= \mathcal{J}_{\mathrm{BP}}\cup \mathcal{J}_{\mathrm{MF}}$. The graph regions corresponding to the factorizations $\prod_{j\in\mathcal{J}_{\mathrm{BP}}}f_j(\boldsymbol{v}_j)$ and $\prod_{j'\in\mathcal{J}_{\mathrm{MF}}} f_{j'}(\boldsymbol{v}_{j'})$ are referred to as the ``BP region" and ``MF region", respectively. Moreover, we define the following index sets: $\mathcal{I}_{\mathrm{BP}} = \cup_{j\in\mathcal{J}_{\mathrm{BP}}} \mathcal{N}(j)$, $\mathcal{N}_{\mathrm{BP}}(i)= \mathcal{J}_{\mathrm{BP}} \cap \mathcal{N}(i)$, $\mathcal{I}_{\mathrm{MF}} = \cup_{j\in\mathcal{J}_{\mathrm{MF}}} \mathcal{N}(j)$ and $\mathcal{N}_{\mathrm{MF}}(i)= \mathcal{J}_{\mathrm{MF}} \cap \mathcal{N}(i)$. We consider the following regions and their associated counting numbers \cite{EKMBF2013}:
\begin{enumerate} 
	\item one MF region: $R_{\mathrm{MF}}= (\mathcal{I}_{\mathrm{MF}}, \mathcal{J}_{\mathrm{MF}})$ with $c_{R_{\mathrm{MF}}} = 1$; 
	\item small regions: $R_i= (\{i\}, \emptyset)$ with $c_{R_i} = 1-|\mathcal{N}_{\mathrm{BP}}(i)|-\mathbb{I}_{\mathcal{I}_{\mathrm{MF}}}(i)$, for $i\in\mathcal{I}_{\mathrm{BP}}$;
	\item large regions: $R_j \!=\! (\mathcal{N}(j),\{j\})$ with $c_{R_{j}} \!=\! 1$, for $j \!\in \! \mathcal{J}_{\mathrm{BP}}$.
\end{enumerate}
These regions and numbers collectively define a valid set of regions: $\mathcal{R}_{\mathrm{BP}, \mathrm{MF}} = \{ R_{\mathrm{MF}} \} \cup \{R_i| i\in \mathcal{I}_{\mathrm{BP}} \}\cup\{ R_j | j\in \mathcal{J}_{\mathrm{BP}} \}$.

For the regions in the set $\mathcal{R}_{\mathrm{BP}, \mathrm{MF}}$, their region average energy and entropy can be obtained using \eqref{eqn_regionAverageEnergyEntropy}. Substituting their region average energy and entropy, and associated countering numbers into \eqref{eqn_regionFreeEnergy}, we obtain the region-based free energy approximation for the unified TRWBP-MF approach as
\begin{equation} \label{eqn_trwbpmf_regionFreeEneergy}
	\begin{aligned}
		&F_{\mathrm{BP}, \mathrm{MF}} \!=\! \!\sum_{j\!\in\!\mathcal{J}_\mathrm{BP}}\! \rho_j \!\sum_{\boldsymbol{v}_j}\! q_j(\boldsymbol{v}_j) \!\ln\! q_j(\boldsymbol{v}_j) \!-\! \!\sum_{j\!\in\!\mathcal{J}_\mathrm{BP}}\! \!\sum_{\boldsymbol{v}_j}\! q_j(\boldsymbol{v}_j) \!\ln\! f_j(\boldsymbol{v}_j) \\
        &\!-\! \!\sum_{j\in\mathcal{J}_{\mathrm{MF}}}\! \!\sum_{\boldsymbol{v}_j}\! \!\prod_{i\in\mathcal{N}(j)}\! q_i(v_i) \!\ln\! f_j(\boldsymbol{v}_j) \!-\! \!\sum_{i\in\mathcal{I}}\! (\varrho_i \!-\! 1) \!\sum_{v_i} q_i(v_i)\! \ln q_i(v_i)
    \end{aligned}
\end{equation}
where $F_{\mathrm{BP}, \mathrm{MF}} = F_{\mathcal{R}_{\mathrm{BP}, \mathrm{MF}}}$ and $\varrho_i = \sum_{j\in \mathcal{N}(i)} \rho_j$. We provide the derivation of \eqref{eqn_trwbpmf_regionFreeEneergy} in the supplementary material. If the FAPs $\rho_j = 1$ for all $j\in\mathcal{J}$, the above free energy is reduced to the free energy of the merged BP-MF approach in \cite{EKMBF2013}. The beliefs $q_i(v_i)$ and $q_j(\boldsymbol{v}_j)$ should be normalized, and $q_i(v_i)$ must satisfy 
the marginalization constraint 
\begin{equation} \label{eqn_marginalizationConstraint}
	q_i(v_i) = \sum_{\boldsymbol{v}_j \backslash v_i} q_j(\boldsymbol{v}_j), \quad j\in \mathcal{J}_{\mathrm{BP}}, i \in \mathcal{N}(j).
\end{equation}

The stationary points of the constrained free energy in \eqref{eqn_trwbpmf_regionFreeEneergy} can be identified using the Lagrange multiplier theory \cite{BV2004}. Following the derivations in \cite{WJW2005, WPS2011}, we handle the marginalization constraint in \eqref{eqn_marginalizationConstraint} using a Lagrange multiplier $\lambda_{j,i}(v_i)$, and enforce the belief normalization constraints explicitly without additional multipliers. The resulting Lagrangian is given by 
\begin{equation} \label{eqn_trwbpmf_lagrangian}
	\begin{aligned}
		L_{\mathrm{BP}, \mathrm{MF}} =& F_{\mathrm{BP}, \mathrm{MF}} \\
		&\!-\! \!\sum_{j\in \mathcal{J}_{\mathrm{BP}}}\! \sum_{i\in\mathcal{N}(j)}\! \!\sum_{v_i}\! \lambda_{j,i}(v_i) \!\Big(\! q_i(v_i) \!-\! \!\sum_{\boldsymbol{v}_j \!\backslash\! v_i}\! q_j(\boldsymbol{v}_j) \!\Big).\!
	\end{aligned}
\end{equation}
The stationary points of the Lagrangian are obtained by setting its derivatives with respect to the marginals $q_i(v_i)$ and $q_j(\boldsymbol{v}_j)$ equal to zero. The relationship between the stationary points of the Lagrangian and the TRWBP-MF fixed-points equations is established in the following theorem.    

$\textit{Theorem 1:}$ The stationary points of the Lagrangian in \eqref{eqn_trwbpmf_lagrangian} for the unified TRWBP-MF approach must be the points with marginals satisfying 
\begin{align} \label{eqn_beliefs_trwbp_mf}
	q_i(v_i) &= c_i \prod_{j\in\mathcal{N}_\mathrm{BP}(i)} \left( m^{\mathrm{BP}}_{j\to i}(v_i) \right)^{\rho_j}  \prod_{j\in\mathcal{N}_{\mathrm{MF}}(i)} m^{\mathrm{MF}}_{j\to i}(v_i), ~  i \in \mathcal{I} \nonumber \\
	q_j(\boldsymbol{v}_j) &= c_j f^{\frac{1}{\rho_j}}_j(\boldsymbol{v}_j) \prod_{i\in \mathcal{N}(j)} m_{i\to j}(v_i), ~ j \in \mathcal{J}_{\mathrm{BP}}
\end{align} 
where 
\begin{align} \label{eqn_messages_trwbp_mf}
	m_{i\to j}(v_i) =& (m^{\mathrm{BP}}_{j\to i}(v_i))^{- 1} \prod_{j'\in\mathcal{N}_\mathrm{BP}(i)} \left( m^{\mathrm{BP}}_{j'\to i}(v_i) \right)^{\rho_{j'}} \nonumber \\
	& \times \prod_{j''\in\mathcal{N}_{\mathrm{MF}}(i)} m^{\mathrm{MF}}_{j''\to i}(v_i),~ \text{for } j\in\mathcal{J}, i\in\mathcal{N}(j) \nonumber \\
	m^{\mathrm{BP}}_{j\to i}(v_i) =& \sum_{\boldsymbol{v}_j\backslash v_i} f^{\frac{1}{\rho_j}}_j(\boldsymbol{v}_j) \prod_{i'\in \mathcal{N}(j) \backslash i} m_{i'\to j}(v_{i'}), \\
	& \text{for } j\in\mathcal{J}_{\mathrm{BP}}, i\in\mathcal{N}(j) \nonumber\\
	m^{\mathrm{MF}}_{j\to i}(v_i) =& \exp\Bigg( \sum_{\boldsymbol{v}_j \backslash v_i} \prod_{i'\in\mathcal{N}(j)\backslash i} m_{i'\to j}(v_{i'}) \ln f_j(\boldsymbol{v}_j) \Bigg), \nonumber\\
	&\text{for } j\in\mathcal{J}_{\mathrm{MF}}, i\in\mathcal{N}(j) \nonumber
\end{align} 
and vice versa. In \eqref{eqn_beliefs_trwbp_mf}, $c_i$ and $c_j$ are the normalizing constants. Note that the messages $m_{i\to j}(v_i)$ equal to $q_i(v_i)$ for $j\in\mathcal{J}_{\mathrm{MF}}$ and $i\in\mathcal{N}(j)$. We provide the proof of Theorem 1 in the supplementary material. The unified TRWBP-MF approach in \eqref{eqn_beliefs_trwbp_mf} and \eqref{eqn_messages_trwbp_mf} serves as the fundamental inference framework for the proposed ETT algorithm in the subsequent section. This approach inherently requires optimizing the FAPs $\rho_j$ introduced by TRWBP. To simplify the presentation and derivation of the proposed algorithm, we focus on constant FAPs in this work. 

\section{TRWBP-MF for ETT } \label{section_four}
In this section, we describe how to compute the messages between factors and variables, and the beliefs for state variables, for the FG depicted in Fig. \ref{fig_factorGraph}, using the unified TRWBP-MF approach given by \eqref{eqn_beliefs_trwbp_mf} and \eqref{eqn_messages_trwbp_mf}. To implement this unified message-passing algorithm, we split the FG shown in Fig. \ref{fig_factorGraph} into the MF region for state estimation and the BP region for data association. The MF region includes factors $\underline{p}^+_n(\cdot)$, $\overline{f}_m(\cdot)$, $\underline{g}_{nl}(\cdot)$ and $\overline{g}_{ml}(\cdot)$, along with all latent variables in \eqref{eqn_jointPDF} except $\boldsymbol{b}$. The BP region comprises factors $\underline{h}_n(\cdot)$, $\overline{h}_m(\cdot)$, $\underline{\varphi}_{nlm}(\cdot)$ and $\overline{\varphi}_{mlo}(\cdot)$, as well as all latent variables in \eqref{eqn_jointPDF}. 

The messages to be computed are illustrated in Fig. \ref{fig_factorGraph}, with different colors indicating their roles in the proposed algorithm. The procedure of our method is summarized as follows. It begins with computing the predicted densities of legacy PTs, corresponding to the messages sent from factors $\underline{p}^+_n(\cdot)$. Next, the messages marked in blue are calculated to initialize the message iterations in the BP region. The BP messages, highlighted in violet, are then iteratively computed to resolve data association. Once the BP messages converge, the red-marked messages are computed, except those sent from factors $\underline{p}^+_n(\cdot)$ and $\overline{f}_m(\cdot)$. Finally, the beliefs for state variables are computed using all red-marked messages directed to them. 

\subsection{State Prediction}
Suppose that the posterior density $p^-_n(\boldsymbol{x}^-_n,r^-_n)$ of legacy PT $n$ at last step is given by $p^-_n(\boldsymbol{x}^-_n,1) = p^{-,1}_n\mathrm{GGIW}(\boldsymbol{x}^-_n;\boldsymbol{\theta}^-_n)$ for target existence, where $\boldsymbol{\theta}^-_n = (\alpha^-_n, \beta^-_n, \boldsymbol{\mu}^-_n, \boldsymbol{\Sigma}^-_n, \nu^-_n, \boldsymbol{V}^-_n)$, and $p^-_n(\boldsymbol{x}^-_n,0) = p^{-,0}_n p_\mathrm{d}(\boldsymbol{x}^-_n)$ for target nonexistence. We estimate measurement rate $\underline{\gamma}_n$ and extent $\underline{\boldsymbol{e}}_n$ using the heuristic approaches in \cite{FFK2011, LGO2013, GFS2020}. The parameters of $\underline{\gamma}_n$ and $\underline{\boldsymbol{e}}_n$ are adjusted by the factors $\eta$ and $\tau$, respectively, such that the expectations of $\underline{\gamma}_n$ and $\underline{\boldsymbol{e}}_n$ retain through the prediction. Substituting the posterior density and transition density \eqref{eqn_singleTarget_stateTransition} into the Chapman-Kolmogorov equation, the predicted density $\underline{p}^+_n(\underline{\boldsymbol{x}}_n,\underline{r}_n)$ is given by $\underline{p}^+_n(\underline{\boldsymbol{x}}_n,1) = \underline{p}^{+,1}_n \mathrm{GGIW}(\underline{\boldsymbol{x}}_n; \underline{\boldsymbol{\theta}}^+_n) $ for target existence and $\underline{p}^+_n(\underline{\boldsymbol{x}}_n,0)=\underline{p}^{+,0}_n p_\mathrm{d}(\underline{\boldsymbol{x}}_n)$ for target nonexistence, where $\underline{p}^{+,1}_n=p_S p^{-,1}_n$, $\underline{p}^{+,0}_n = p^{-,0}_n + (1-p_S)p^{-,0}_n$, and the components of $\underline{\boldsymbol{\theta}}^+_n = (\underline{\alpha}^+_n, \underline{\beta}^+_n, \underline{\boldsymbol{\mu}}^+_n, \underline{\boldsymbol{\Sigma}}^+_n, \underline{\nu}^+_n, \underline{\boldsymbol{V}}^+_n)$ are given by \citep[Table I]{MJDL2024}.

\subsection{Messages in the MF Region} \label{sec_five_2}
We denote the message from $\underline{p}^+_n(\underline{\boldsymbol{y}}_n)$ to $\underline{\boldsymbol{y}}_n$ as $\underline{m}^{\mathrm{MF}}_{\to n}(\underline{\boldsymbol{y}}_n)$, the message from $\overline{f}_m(\overline{\boldsymbol{y}}_m)$ to $\overline{\boldsymbol{y}}_m$ as $\overline{m}^{\mathrm{MF}}_{\to m}(\overline{\boldsymbol{y}}_m)$. According to $\eqref{eqn_messages_trwbp_mf}$ we get $\underline{m}^{\mathrm{MF}}_{\to n}(\underline{\boldsymbol{y}}_n) = \underline{p}^+_n(\underline{\boldsymbol{y}}_n)$ and $\overline{m}^{\mathrm{MF}}_{\to m}(\overline{\boldsymbol{y}}_m) = \overline{f}_m (\overline{\boldsymbol{y}}_m)$. 
\subsubsection{Messages \texorpdfstring{$\underline{m}^{\mathrm{MF}}_{\to nl}(\underline{a}_{nl})$}{} and \texorpdfstring{$\overline{m}^{\mathrm{MF}}_{\to ml}(\overline{a}_{ml})$}{}}
We represent the message from $\underline{g}_{nl}(\underline{\boldsymbol{y}}_n, \underline{a}_{nl}; \boldsymbol{z})$ to $\underline{a}_{nl}$ as $\underline{m}^{\mathrm{MF}}_{\to nl}(\underline{a}_{nl})$, the message from $\overline{g}_{ml}(\overline{\boldsymbol{y}}_m, \overline{a}_{ml}; \boldsymbol{z})$ to $\overline{a}_{ml}$ as $\overline{m}^{\mathrm{MF}}_{\to ml}(\overline{a}_{ml})$. Using the message update rule in \eqref{eqn_messages_trwbp_mf}, we obtain 
\begin{equation} \label{eqn_mf_legacy_message_y2a}
	\begin{aligned}
		& \underline{m}^{\mathrm{MF}}_{\to nl}(\underline{a}_{nl}) \\
        =& \exp\left( \int \underline{m}^{\mathrm{MF}}_{\to n}(\underline{\boldsymbol{y}}_n) \ln ( \underline{g}_{nl}(\underline{\boldsymbol{y}}_n, \underline{a}_{nl}; \boldsymbol{z}) ) \mathrm{d}\underline{\boldsymbol{y}}_n \right) \\
        =& \begin{cases}
        	\frac{\exp\left(-\frac{1}{2}  \mathbb{E}_{\underline{\boldsymbol{\xi}}_n, \underline{\boldsymbol{E}}_n} \left[  \underline{f}(\underline{\boldsymbol{\xi}}_n, \underline{\boldsymbol{E}}_n, \boldsymbol{z}_{m}) \right] \right)}{\kappa_{\mathrm{c}} (2\pi)^{d_{\boldsymbol{z}}/2} }, & \underline{a}_{nl} = m \in\{1,\dots,M\}  \\
        	1, & \underline{a}_{nl} = 0, 
        \end{cases} \\
	\end{aligned}
\end{equation}
\begin{equation} \label{eqn_mf_new_message_y2a} 
	\begin{aligned}
		& \overline{m}^{\mathrm{MF}}_{\to ml}(\overline{a}_{ml}) \\
        \!=\!& \exp\left( \int \overline{m}^{\mathrm{MF}}_{\to m}(\overline{\boldsymbol{y}}_m) \ln \left( \overline{g}_{ml}(\overline{\boldsymbol{y}}_m, \overline{a}_{ml}; \boldsymbol{z})  \right) \mathrm{d}\boldsymbol{y}_m \right) \\
        \!=\!& \begin{cases}
        	\frac{\lambda_\mathrm{n} \exp\left(\!-\!\frac{1}{2}  \mathbb{E}_{\overline{\boldsymbol{\xi}}_m, \overline{\boldsymbol{E}}_m} \left[ \overline{f}(\overline{\boldsymbol{\xi}}_m, \overline{\boldsymbol{E}}_m, \boldsymbol{z}_o) \right] \right)}{\kappa_{\mathrm{c}} (2\pi)^{d_{\boldsymbol{z}}/2} } , & \overline{a}_{ml} \!=\! o \!\in\!\{m,\dots,M\}  \\
        	0, & \overline{a}_{ml} \!=\! 0. 
        \end{cases} \\
	\end{aligned}
\end{equation}
The function $\underline{f}(\underline{\boldsymbol{\xi}}_n, \underline{\boldsymbol{E}}_n, \boldsymbol{z}_{m})$ in \eqref{eqn_mf_legacy_message_y2a} is given by
\begin{equation} \nonumber
	\begin{aligned}
		\underline{f}(\underline{\boldsymbol{\xi}}_n, \underline{\boldsymbol{E}}_n, \boldsymbol{z}_{m}) =& \ln(|s\underline{\boldsymbol{E}}_n + \boldsymbol{R}|) \\
		&\!+\! (\boldsymbol{z}_{m} \!-\! \boldsymbol{H}\underline{\boldsymbol{\xi}}_n)^T (s\underline{\boldsymbol{E}}_n \!+\! \boldsymbol{R})^{-1} (\boldsymbol{z}_{m} \!-\! \boldsymbol{H}\underline{\boldsymbol{\xi}}_n)
	\end{aligned}
\end{equation}
and $\overline{f}(\overline{\boldsymbol{\xi}}_m, \overline{\boldsymbol{E}}_m, \boldsymbol{z}_o)$ in \eqref{eqn_mf_new_message_y2a} is defined similarly. Next we derive the solution to $\mathbb{E}_{\underline{\boldsymbol{\xi}}_n, \underline{\boldsymbol{E}}_n} [ \underline{f}(\underline{\boldsymbol{\xi}}_n, \underline{\boldsymbol{E}}_n, \boldsymbol{z}_{m}) ]$, while noting that $\mathbb{E}_{\overline{\boldsymbol{\xi}}_m, \overline{\boldsymbol{E}}_m} [ \overline{f}(\overline{\boldsymbol{\xi}}_m, \overline{\boldsymbol{E}}_m, \boldsymbol{z}_o) ]$ can be computed analogously. 

We rewrite the expectation in \eqref{eqn_mf_legacy_message_y2a} as 
\begin{equation} \label{eqn_expectation2XiE}
	\begin{aligned}
		&\mathbb{E}_{\underline{\boldsymbol{\xi}}_n, \underline{\boldsymbol{E}}_n}[\underline{f}(\underline{\boldsymbol{\xi}}_n, \underline{\boldsymbol{E}}_n, \boldsymbol{z}_{m})] = \mathbb{E}_{\underline{\boldsymbol{E}}_n} \left[ \ln(|s\underline{\boldsymbol{E}}_n+\boldsymbol{R}|)\right] \\
		&\!+\! \mathrm{tr}( \mathbb{E}_{\underline{\boldsymbol{\xi}}_n} [ (\boldsymbol{z}_{m} \!-\! \boldsymbol{H}\underline{\boldsymbol{\xi}}_n)(\boldsymbol{z}_{m} \!-\! \boldsymbol{H}\underline{\boldsymbol{\xi}}_n)^T ] \mathbb{E}_{\underline{\boldsymbol{E}}_n} [ (s\underline{\boldsymbol{E}}_n \!+\! \boldsymbol{R})^{-1} ] )
	\end{aligned}
\end{equation}
where the expectation with respect to $\underline{\boldsymbol{\xi}}_n$ is given by \cite{PP2012}
\begin{equation}
	\begin{aligned}
		&\mathbb{E}_{\underline{\boldsymbol{\xi}}_n}[ (\boldsymbol{z}_{m} - \boldsymbol{H}\underline{\boldsymbol{\xi}}_n)(\boldsymbol{z}_{m} - \boldsymbol{H}\underline{\boldsymbol{\xi}}_n)^T ] \\
		=& (\boldsymbol{z}_{m} - \boldsymbol{H}\underline{\boldsymbol{\mu}}^+_n)(\boldsymbol{z}_{m} - \boldsymbol{H}\underline{\boldsymbol{\mu}}^+_n)^T + \boldsymbol{H}\underline{\boldsymbol{\Sigma}}^+_n\boldsymbol{H}^T.
	\end{aligned}
\end{equation}
We assume that $\boldsymbol{R}$ is relatively small compared to $s\underline{\boldsymbol{E}}_n$. This assumption holds for high precision sensors such as lidar. The two expectations with respect to $\underline{\boldsymbol{E}}_n$ in \eqref{eqn_expectation2XiE} can then be approximated as follows. On the one hand, we have
\begin{equation} \label{eqn_expectation_logDet_EPlusR}
	\begin{aligned}
		\mathbb{E}_{\underline{\boldsymbol{E}}_n} \left[ \ln(|s\underline{\boldsymbol{E}}_n+\boldsymbol{R}|)\right] &\approx \mathbb{E}_{\underline{\boldsymbol{E}}_n} \left[ \ln(|s\underline{\boldsymbol{E}}_n|)\right] \\
		&= d_{\boldsymbol{z}}\ln s + \mathbb{E}_{\underline{\boldsymbol{E}}_n} \left[ \ln(|\underline{\boldsymbol{E}}_n|)\right]. \\
    \end{aligned}
\end{equation}
Note that if matrix $\boldsymbol{X}$ follows a IW distribution $\mathrm{IW}(\boldsymbol{X};\nu,\boldsymbol{V})$, then $\boldsymbol{Y} = \boldsymbol{X}^{-1}$ follows the Wishart distribution $\mathrm{W}(\boldsymbol{Y}; \nu-d_{\boldsymbol{Y}}-1, \boldsymbol{V}^{-1})$ \cite{GD1999}, and that $\mathbb{E}_{\boldsymbol{Y}}[\ln(|\boldsymbol{Y}|)]$ is given by \cite{D2023}
\begin{equation} \label{eqn_expectation_logDetX}
	\mathbb{E}_{\boldsymbol{Y}}[\ln(|\boldsymbol{Y}|)] \!=\! \sum^{d_{\boldsymbol{Y}}}_{i=1} \psi\left(\frac{\nu-d_{\boldsymbol{Y}}-i}{2}\right) \!+\! d_{\boldsymbol{Y}}\ln2 \!-\! \ln(|\boldsymbol{V}|)
\end{equation}
where $\psi(\cdot)$ is the digamma function. Using the relationship $\mathbb{E}_{\boldsymbol{X}} \left[ \ln(|\boldsymbol{X}|)\right] = -\mathbb{E}_{\boldsymbol{Y}} \left[ \ln(|\boldsymbol{Y}|)\right]$ and \eqref{eqn_expectation_logDetX}, for target extent $\underline{\boldsymbol{E}}_n$ with IW distribution $\mathrm{IW}(\underline{\boldsymbol{E}}_n; \underline{\nu}^+_n, \underline{\boldsymbol{V}}^+_n)$, we obtain
\begin{equation}
	\begin{aligned}
		\mathbb{E}_{\underline{\boldsymbol{E}}_n} \left[ \ln(|s\underline{\boldsymbol{E}}_n+\boldsymbol{R}|)\right] \approx&  d_{\boldsymbol{z}}\ln s - d_{\boldsymbol{z}}\ln2 + \ln(|\underline{\boldsymbol{V}}^+_n|) \\
		& -\sum^{d_{\boldsymbol{z}}}_{i=1} \psi\left(\frac{\underline{\nu}^+_n-d_{\boldsymbol{z}}-i}{2}\right). 
	\end{aligned}
\end{equation}
Note that the dimension $d_{\boldsymbol{z}}$ of a measurement and the dimension $d_{\boldsymbol{E}}$ of the target extent are identical. On the other hand, we have the following approximation
\begin{equation}
	\begin{aligned}
		\mathbb{E}_{\underline{\boldsymbol{E}}_n}\left[ (s\underline{\boldsymbol{E}}_n + \boldsymbol{R})^{-1} \right] &\approx \mathbb{E}_{\underline{\boldsymbol{E}}_n}\left[ (s\underline{\boldsymbol{E}}_n)^{-1}  \right] \\
		&= \frac{(\underline{\nu}^+_n - d_{\boldsymbol{z}} - 1) (\underline{\boldsymbol{V}}^+_n)^{-1}}{s}. 
	\end{aligned}
\end{equation}
\subsubsection{Messages \texorpdfstring{$\underline{m}^{\mathrm{MF}}_{\leftarrow nl}(\underline{\boldsymbol{y}}_n)$}{} and \texorpdfstring{$\overline{m}^{\mathrm{MF}}_{\leftarrow ml}(\overline{\boldsymbol{y}}_m)$}{}}
We denote the message from $\underline{g}_{nl}(\underline{\boldsymbol{y}}_n, \underline{a}_{nl}; \boldsymbol{z})$ to $\underline{\boldsymbol{y}}_n$ as $\underline{m}^{\mathrm{MF}}_{\leftarrow nl}(\underline{\boldsymbol{y}}_n)$, the message from $\overline{g}_{ml}(\overline{\boldsymbol{y}}_m, \overline{a}_{ml}; \boldsymbol{z})$ to $\overline{\boldsymbol{y}}_m$ as $\overline{m}^{\mathrm{MF}}_{\leftarrow ml}(\overline{\boldsymbol{y}}_m)$. According to \eqref{eqn_messages_trwbp_mf}, we obtain 
\begin{equation} \label{eqn_lpt_message_g2y}
	\begin{aligned}
        \underline{m}^{\mathrm{MF}}_{\leftarrow nl}(\underline{\boldsymbol{y}}_n)
		=& \exp \!\Bigg(\! \sum_{\underline{a}_{nl} = 0}^{M} m_{\underline{a}_{nl}\rightarrow \underline{g}_{nl}}(\underline{a}_{nl}) \ln ( \underline{g}_{nl}(\underline{\boldsymbol{y}}_n, \underline{a}_{nl} ; \boldsymbol{z}) ) \!\Bigg)\! \\
        =& \exp \!\Bigg(\! \sum_{\underline{a}_{nl} = 0}^{M} \underline{q}_{nl}(\underline{a}_{nl}) \ln ( \underline{g}_{nl} (\underline{\boldsymbol{y}}_n, \underline{a}_{nl} ; \boldsymbol{z}) ) \!\Bigg)\! \\
        =& \begin{cases}
        	\prod_{m = 1}^{M} \left( \frac{p(\boldsymbol{z}_m | \underline{\boldsymbol{\xi}}_n, \underline{\boldsymbol{e}}_n)} {\kappa_{\mathrm{c}}} \right)^{\underline{q}_{nl}(m)}, & \underline{r}_n = 1 \\
        	1, & \underline{r}_n = 0
        \end{cases}
	\end{aligned} 
\end{equation}
\begin{equation} \label{eqn_npt_message_g2y}
	\begin{aligned}
        \overline{m}^{\mathrm{MF}}_{\leftarrow ml}(\overline{\boldsymbol{y}}_m) 
		=& \exp\left( \sum_{\overline{a}_{ml}} \overline{q}_{ml}(\overline{a}_{ml}) \ln \left( \overline{g}_{ml}(\overline{\boldsymbol{y}}_m, \overline{a}_{ml} ; \boldsymbol{z})  \right) \right) \\
        = & \begin{cases}
        	0^{\overline{q}_{ml}(0)} \prod_{o = m}^{M} \!\left(\! \frac{p(\boldsymbol{z}_o | \overline{\boldsymbol{\xi}}_m, \overline{\boldsymbol{e}}_m)} {\kappa_{\mathrm{c}} } \!\right)\!^{\overline{q}_{ml}(o)}, \!&\!  \overline{r}_m \!=\! 1 \\
        	1, \!&\! \overline{r}_m \!=\! 0 
        \end{cases}
	\end{aligned} 
\end{equation}
where $\underline{q}_{nl}(\underline{a}_{nl})$ and $\overline{q}_{ml}(\overline{a}_{ml})$ are the beliefs for the association variables $\underline{a}_{nl}$ and $\overline{a}_{ml}$, which will be obtained by \eqref{eqn_belief_a_lpt} and \eqref{eqn_belief_a_npt} after the messages in the BP region converge. It should be noted that the value of the exponentiation $0^{\overline{q}_{ml}(0)}$ in \eqref{eqn_npt_message_g2y} is one because of $\overline{q}_{ml}(0) = 0$ according to \eqref{eqn_belief_a_npt}.  

\subsection{Messages in the BP Region}
\subsubsection{Data association messages}
As shown in Fig. \ref{fig_factorGraph}, the BP region comprises the factors $\underline{h}_n(\cdot)$, $\overline{h}_m(\cdot)$, $\underline{\varphi}_{nlm}(\cdot)$ and $\overline{\varphi}_{mlo}(\cdot)$. We denote the FAPs of the factors $\underline{h}_n(\cdot)$ and $\overline{h}_m(\cdot)$ as $\rho_h$, and the FAPs of the factors $\underline{\varphi}_{nlm}(\cdot)$ and $\overline{\varphi}_{mlo}(\cdot)$ as $\rho_\varphi$. Although the FAPs of these factors can take the same value, as we assume constant FAPs in this paper, assigning distinct FAPs to different types of factors can simplify the derivations of messages in the BP region. For the legacy PT $n$, we denote the messages from $\underline{\varphi}_{nlm}(\underline{a}_{nl}, \boldsymbol{b}_m)$ to $\underline{a}_{nl}$ and $\boldsymbol{b}_m$ as $\underline{m}^{\mathrm{BP}}_{nl \leftarrow m } (\underline{a}_{nl})$ and $\underline{m}^{\mathrm{BP}}_{nl\rightarrow m}(\boldsymbol{b}_m)$, and the message from $\underline{h}_n(\underline{\boldsymbol{y}}_n, \underline{\boldsymbol{a}}_n)$ to $\underline{a}_{nl}$ as $\underline{m}^{\mathrm{BP}}_{\rightarrow nl}(\underline{a}_{nl})$. For the new PT $m$, we denote the messages from $\overline{\varphi}_{mlo}(\overline{a}_{ml}, \boldsymbol{b}_o)$ to $\overline{a}_{ml}$ and $\boldsymbol{b}_o$ as $\overline{m}^{\mathrm{BP}}_{ml \leftarrow o } (\overline{a}_{ml})$ and $\overline{m}^{\mathrm{BP}}_{ml\rightarrow o}(\boldsymbol{b}_o)$, and the message from $\overline{h}_m(\overline{\boldsymbol{y}}_m, \overline{\boldsymbol{a}}_m)$ to $\overline{a}_{ml}$ as $\overline{m}^{\mathrm{BP}}_{\rightarrow ml}(\overline{a}_{ml})$. These messages have to be calculated iteratively since the BP region contains numerous loops. Let $I_{\mathrm{BP}}$ denote the number of BP iterations. According to \eqref{eqn_messages_trwbp_mf}, the messages for the legacy PT $n$ at iteration $\iota\in\{1, \dots, I_{\mathrm{BP}}\}$ are given by
\begin{equation} \label{eqn_lpt_message_b2a}
	\begin{aligned}
		&\underline{m}^{\mathrm{BP},[\iota]}_{nl \leftarrow m } (\underline{a}_{nl}) \!=\! \sum_{\boldsymbol{b}_m \in \mathcal{B}_m } (\underline{\varphi}_{n l m} (\underline{a}_{nl}, \boldsymbol{b}_m ) )^{\frac{1}{\rho_\varphi}} (\underline{m}^{\mathrm{BP},[\iota-1]}_{nl\rightarrow m}(\boldsymbol{b}_m))^{-1} \\
		&\!\times\! \!\prod_{\left(n', l'\right) \in \underline{\mathcal{B}}_m }\! ( \underline{m}^{\mathrm{BP},[\iota\!-\!1]}_{n'l'\rightarrow m}(\boldsymbol{b}_m) )^{\rho_\varphi}  \!\prod_{\left(n'', l''\right) \in \overline{\mathcal{B}}_m }\! ( \overline{m}^{\mathrm{BP},[\iota\!-\!1]}_{n''l'' \rightarrow m}(\boldsymbol{b}_m) )^{\rho_\varphi} 
	\end{aligned}
\end{equation}
\begin{equation} \label{eqn_lpt_message_h2a_exact}
	\begin{aligned} 
		&\underline{m}^{\mathrm{BP},[\iota]}_{\rightarrow nl}(\underline{a}_{nl})  \\
		&\!=\! \sum_{\underline{\boldsymbol{a}}_{nl \sim}} \sum_{\underline{r}_n} \int \left( \underline{h}_n(\underline{\boldsymbol{x}}_n, \underline{r}_n, \underline{\boldsymbol{a}}_n ) \right)^{\frac{1}{\rho_h}} ( \underline{m}^{\mathrm{BP},[\iota]}_{\leftarrow n}(\underline{\boldsymbol{x}}_n, \underline{r}_n) )^{\rho_h - 1}  \\
		&\!\times\! \underline{m}^{\mathrm{MF}}_{\to n}(\underline{\boldsymbol{x}}_n, \underline{r}_n) \!\prod_{l' \in \underline{\mathcal{L}}_n}\! \underline{m}^{\mathrm{MF}}_{\leftarrow nl'}(\underline{\boldsymbol{x}}_n, \underline{r}_n) \mathrm{d} \underline{\boldsymbol{x}}_n \!\prod_{l'' \in \underline{\mathcal{L}}_n \backslash l} \!\underline{m}^{\mathrm{MF}}_{\rightarrow nl''}(\underline{a}_{nl''})  \\
		&\!\times\! ( \underline{m}^{\mathrm{BP},[\iota]}_{\rightarrow nl''}(\underline{a}_{nl''}) )^{\rho_h - 1}  \prod_{m=1}^M ( \underline{m}^{\mathrm{BP},[\iota]}_{nl'' \leftarrow m  } (\underline{a}_{nl''}) )^{\rho_\varphi} 
	\end{aligned}
\end{equation}
\begin{equation} \label{eqn_lpt_message_a2b}
	\begin{aligned} 
		&\underline{m}^{\mathrm{BP},[\iota]}_{nl\rightarrow m}(\boldsymbol{b}_m) = \sum_{\underline{a}_{n l}=0}^M (\underline{\varphi}_{n l m} (\underline{a}_{nl}, \boldsymbol{b}_m ) )^{\frac{1}{\rho_\varphi}} (\underline{m}^{\mathrm{BP},[\iota]}_{nl \leftarrow m } (\underline{a}_{nl}))^{-1} \\
		&\times \underline{m}^{\mathrm{MF}}_{\rightarrow nl}(\underline{a}_{nl}) \underline{m}^{\mathrm{BP},[\iota]}_{\rightarrow nl}(\underline{a}_{nl})  \prod_{m'=1}^{M} \left( \underline{m}^{\mathrm{BP},[\iota]}_{nl \leftarrow m' }(\underline{a}_{nl})  \right)^{\rho_\varphi}
	\end{aligned}
\end{equation}
where $\mathcal{B}_m = \underline{\mathcal{B}}_m \cup \overline{\mathcal{B}}_m\cup{(0,0)}$ represents the set of all possible values of $\boldsymbol{b}_m$ with subsets $\underline{\mathcal{B}}_m = \bigcup_{n=1}^{\underline{N}}\{n\} \times \underline{\mathcal{L}}_n$ and $\overline{\mathcal{B}}_m = \bigcup_{o=\underline{N}+1}^{\underline{N}+m}\{o\} \times \overline{\mathcal{L}}_{o-\underline{N}}$, and $\sum_{\underline{\boldsymbol{a}}_{n l \sim}}$ denotes the summation over $\underline{\boldsymbol{a}}_n$ excluding $\underline{a}_{nl}$. 

Note that the message $\underline{m}^{\mathrm{BP},[\iota]}_{\rightarrow nl}(\underline{a}_{nl})$ in \eqref{eqn_lpt_message_h2a_exact} depends on the messages $\underline{m}^{\mathrm{BP},[\iota]}_{\leftarrow n}(\underline{\boldsymbol{x}}_n, \underline{r}_n)$ in \eqref{eqn_lpt_message_h2y}, $\underline{m}^{\mathrm{MF}}_{\leftarrow nl'}(\underline{\boldsymbol{x}}_n, \underline{r}_n)$ in \eqref{eqn_lpt_message_g2y} and $\underline{m}^{\mathrm{BP},[\iota]}_{\rightarrow nl''}(\underline{a}_{nl''})$, which are currently unavailable. To simplify the computation of $\underline{m}^{\mathrm{BP},[\iota]}_{\rightarrow nl}(\underline{a}_{nl})$, we assume $\rho_h=1$ such that the terms $( \underline{m}^{\mathrm{BP},[\iota]}_{\leftarrow n}(\underline{\boldsymbol{x}}_n, \underline{r}_n) )^{\rho_h - 1}$ and $( \underline{m}^{\mathrm{BP},[\iota]}_{\rightarrow nl''}(\underline{a}_{nl''}) )^{\rho_h - 1}$ in \eqref{eqn_lpt_message_h2a_exact} reduce to one. Moreover, from equation \eqref{eqn_belief_lpt_state} we observe that the term $\underline{m}^{\mathrm{MF}}_{\to n}(\underline{\boldsymbol{x}}_n, \underline{r}_n) \prod_{l' \in \underline{\mathcal{L}}_n} \underline{m}^{ \mathrm{MF}}_{\leftarrow nl'}(\underline{\boldsymbol{x}}_n, \underline{r}_n)$ in \eqref{eqn_lpt_message_h2a_exact} is an approximate posterior of $\underline{\boldsymbol{y}}_n$ without incorporating the message $\underline{m}^{\mathrm{BP}}_{\leftarrow n}(\underline{\boldsymbol{y}}_n)$ given by \eqref{eqn_lpt_message_h2y}, which contains the information for updating the measurement rate. Since $\underline{m}^{\mathrm{MF}}_{\to n}(\underline{\boldsymbol{y}}_n) = \underline{p}^+_n(\underline{\boldsymbol{y}}_n)$ is the predicted density of $\underline{\boldsymbol{y}}_n$, it can be considered a reasonable approximation to $\underline{m}^{\mathrm{MF}}_{\to n}(\underline{\boldsymbol{x}}_n, \underline{r}_n) \prod_{l' \in \underline{\mathcal{L}}_n} \underline{m}^{ \mathrm{MF}}_{\leftarrow nl'}(\underline{\boldsymbol{x}}_n, \underline{r}_n)$. Taking into account the above simplifications, the approximate message for $\underline{m}^{\mathrm{BP},[\iota]}_{\rightarrow nl}(\underline{a}_{nl})$ is obtained as
\begin{equation} \label{eqn_lpt_message_h2a_approx}
	\begin{aligned} 
		\underline{m}^{\mathrm{BP},[\iota]}_{\rightarrow nl}(\underline{a}_{nl}) \approx& \sum_{\underline{\boldsymbol{a}}_{nl \sim}} \sum_{\underline{r}_n} \int \underline{h}_n\left(\underline{\boldsymbol{x}}_n, \underline{r}_n, \underline{\boldsymbol{a}}_n \right) \underline{m}^{\mathrm{MF}}_{\to n}(\underline{\boldsymbol{x}}_n, \underline{r}_n) \mathrm{d} \underline{\boldsymbol{x}}_n \\
		&\times \prod_{l' \in \underline{\mathcal{L}}_n \backslash l} \underline{m}^{\mathrm{MF}}_{\rightarrow nl'}(\underline{a}_{nl'}) \prod_{m=1}^M ( \underline{m}^{\mathrm{BP},[\iota]}_{nl' \leftarrow m  } (\underline{a}_{nl'}) )^{\rho_\varphi}. 
	\end{aligned}
\end{equation}

Similarly, the messages for new PT $m$ are given by
\begin{equation} \label{eqn_npt_message_b2a}
	\begin{aligned}
		&\overline{m}^{\mathrm{BP},[\iota]}_{ml \leftarrow o } (\overline{a}_{ml}) = \sum_{\boldsymbol{b}_o \in \mathcal{B}_o} \left(\overline{\varphi}_{m l o} (\overline{a}_{ml}, \boldsymbol{b}_o ) \right)^{\frac{1}{\rho_\varphi}} (\overline{m}^{\mathrm{BP},[\iota-1]}_{ml\rightarrow o}(\boldsymbol{b}_o) )^{-1} \\
		& \!\times\! \!\prod_{\left(m', l'\right) \in \underline{\mathcal{B}}_o }\! ( \underline{m}^{\mathrm{BP},[\iota-1]}_{m'l'\rightarrow o}(\boldsymbol{b}_o) )^{\rho_\varphi}  \!\prod_{\left(m'', l''\right) \in \overline{\mathcal{B}}_o } \!( \overline{m}^{\mathrm{BP},[\iota-1]}_{m''l''\rightarrow o}(\boldsymbol{b}_o) )^{\rho_\varphi} 
	\end{aligned}
\end{equation}
\begin{equation} \label{eqn_npt_message_h2a_approx}
	\begin{aligned} 
		\overline{m}^{\mathrm{BP},[\iota]}_{\rightarrow ml}\!(\overline{a}_{ml}\!)\! \!\approx\!& \!\sum_{\overline{\boldsymbol{a}}_{m l \sim}}\! \!\sum_{\overline{r}_m}\! \!\int\! \overline{h}_m\!\left(\!\overline{\boldsymbol{x}}_m, \overline{r}_m, \overline{\boldsymbol{a}}_m \!\right)\! \overline{m}^{\mathrm{MF}}_{\to m}(\overline{\boldsymbol{x}}_m, \overline{r}_m) \mathrm{d} \overline{\boldsymbol{x}}_m \\
		& \!\times\! \!\prod_{l' \in \overline{\mathcal{L}}_m \backslash l}\! \overline{m}^{\mathrm{MF}}_{\rightarrow ml'}(\overline{a}_{ml'})  \prod_{o=m}^M ( \overline{m}^{\mathrm{BP},[\iota]}_{ml' \leftarrow o  } (\overline{a}_{ml'})  )^{\rho_\varphi} \\
	\end{aligned}
\end{equation}
\begin{equation} \label{eqn_npt_message_a2b}
	\begin{aligned}
		&\overline{m}^{\mathrm{BP},[\iota]}_{ml\rightarrow o}(\boldsymbol{b}_o) = \sum_{\overline{a}_{m l}} \left(\overline{\varphi}_{m l o} (\overline{a}_{ml}, \boldsymbol{b}_o ) \right)^{\frac{1}{\rho_\varphi}} (\overline{m}^{\mathrm{BP},[\iota]}_{ml \leftarrow o } (\overline{a}_{m l}))^{-1}  \\
		& \times  \overline{m}^{\mathrm{MF}}_{\rightarrow ml}(\overline{a}_{ml}) \overline{m}^{\mathrm{BP},[\iota]}_{\rightarrow ml}(\overline{a}_{ml})  \prod_{o'=m}^{M} ( \overline{m}^{\mathrm{BP},[\iota]}_{ml \leftarrow o' }(\overline{a}_{ml})  )^{\rho_\varphi}.
	\end{aligned}
\end{equation}

\subsubsection{Rescaled data association messages}
Due to the binary constraints $\underline{\varphi}_{nlm}(\underline{a}_{nl}, \boldsymbol{b}_m)$ and $\overline{\varphi}_{mlo}(\overline{a}_{ml}, \boldsymbol{b}_o)$, and the fact that $\underline{h}_n(\underline{\boldsymbol{y}}_n, \underline{\boldsymbol{a}}_n)$ and $\overline{h}_m(\overline{\boldsymbol{y}}_m, \overline{\boldsymbol{a}}_m)$ depend only on the nonzero entries of $\underline{\boldsymbol{a}}_n$ and $\overline{\boldsymbol{a}}_m$, all the messages in \eqref{eqn_lpt_message_b2a}-\eqref{eqn_npt_message_a2b} only comprise of two different values. Specifically, $\underline{m}^{\mathrm{BP},[\iota]}_{nl \leftarrow m } (\underline{a}_{nl})$ takes distinct values for $\underline{a}_{nl} = m$ and $\underline{a}_{nl}\neq m$, $\underline{m}^{\mathrm{BP},[\iota]}_{\rightarrow nl}(\underline{a}_{nl})$ for $\underline{a}_{nl}\neq 0$ and $\underline{a}_{nl}=0$, $\underline{m}^{\mathrm{BP},[\iota]}_{nl\rightarrow m}(\boldsymbol{b}_m)$ for $\boldsymbol{b}_m=[n~l]^T$ and $\boldsymbol{b}_m\neq[n~l]^T$, $\overline{m}^{\mathrm{BP},[\iota]}_{ml \leftarrow o } (\overline{a}_{ml})$ for $\overline{a}_{ml} = o$ and $\overline{a}_{ml} \neq o$, $\overline{m}^{\mathrm{BP},[\iota]}_{\rightarrow ml}(\overline{a}_{ml})$ for $\overline{a}_{ml}\neq 0$ and $\overline{a}_{ml} = 0$, $\overline{m}^{\mathrm{BP},[\iota]}_{ml\rightarrow o}(\boldsymbol{b}_o)$ for $\boldsymbol{b}_o=[\underline{N}+m~l]^T$ and $\boldsymbol{b}_o\neq[\underline{N}+m~l]^T$.

Since BP messages can be rescaled by an arbitrary positive constant without changing the resulting beliefs \cite{KFL2001, EKMBF2013}, we rescale the BP messages in \eqref{eqn_lpt_message_b2a}-\eqref{eqn_npt_message_a2b} as follows. The messages $\underline{m}^{\mathrm{BP},[\iota]}_{nl \leftarrow m } (\underline{a}_{nl})$, $\underline{m}^{\mathrm{BP},[\iota]}_{\rightarrow nl}(\underline{a}_{nl})$ and $\underline{m}^{\mathrm{BP},[\iota]}_{nl\rightarrow m}(\boldsymbol{b}_m)$ are rescaled by dividing their values at $\underline{a}_{nl}\neq m$, $\underline{a}_{nl}=0$ and $\boldsymbol{b}_m\neq[n~l]^T$, respectively. The messages $\overline{m}^{\mathrm{BP},[\iota]}_{ml \leftarrow o } (\overline{a}_{ml})$, $\overline{m}^{\mathrm{BP},[\iota]}_{\rightarrow ml}(\overline{a}_{ml})$ and $\overline{m}^{\mathrm{BP},[\iota]}_{ml\rightarrow o}(\boldsymbol{b}_o) $ are rescaled by dividing their values at $\overline{a}_{ml}\neq o$, $\overline{a}_{ml}=0$ and $\boldsymbol{b}_o\neq[\underline{N}+m~l]^T$, respectively. Thus, all the message in \eqref{eqn_lpt_message_b2a}-\eqref{eqn_npt_message_a2b} can be represented by scalars. 

The rescaled messages for the legacy PT $n$ are given by
\begin{equation} \label{eqn_lpt_scaledMessage_b2a}
	\begin{aligned}
		&\underline{\mathrm{m}}^{\mathrm{BP},[\iota]}_{nl \leftarrow m } \\
		\!=\!& \frac{ ( \underline{\mathrm{m}}^{\mathrm{BP},[\iota-1]}_{nl\rightarrow m} )^{\rho_\varphi-1} }{ 1 \!-\! ( \underline{\mathrm{m}}^{\mathrm{BP},[\iota\!-\!1]}_{nl\rightarrow m} \!)^{\rho_\varphi} \!+\! \!\sum\limits_{n'=1}^{\underline{N}}\! \!\underline{\ell}_{n'}\! ( \underline{\mathrm{m}}^{\mathrm{BP},[\iota\!-\!1]}_{n'l'\rightarrow m} \!)^{\rho_\varphi} \!+\! \!\sum\limits_{o=1}^m\! \!\overline{\ell}_{o}\! ( \overline{\mathrm{m}}^{\mathrm{BP},[\iota\!-\!1]}_{ol''\rightarrow m} \!)^{\rho_\varphi} }
	\end{aligned}
\end{equation}
\begin{equation} \label{eqn_lpt_scaledMessage_h2a}
	\underline{\mathrm{m}}^{\mathrm{BP},[\iota]}_{\rightarrow nl} = \frac{\sum_{\varepsilon = 0}^{\underline{\ell}_n - 1} \underline{f}^1_n(\varepsilon + 1) ( \widetilde{\underline{m}}^{\mathrm{BP},[\iota]}_{\rightarrow nl} )^{\varepsilon} / \varepsilon!}{\sum_{\varepsilon = 0}^{\underline{\ell}_n - 1} \underline{f}^0_n(\varepsilon) ( \widetilde{\underline{m}}^{\mathrm{BP},[\iota]}_{\rightarrow nl} )^{\varepsilon}/\varepsilon!}
\end{equation}
\begin{equation} \label{eqn_lpt_scaledMessage_a2b}
	\underline{\mathrm{m}}^{\mathrm{BP},[\iota]}_{nl\rightarrow m} \!=\! \frac{ ( \underline{\mathrm{m}}^{\mathrm{BP},[\iota]}_{ nl \leftarrow m } )^{\rho_\varphi-1} \underline{m}^{\mathrm{MF}}_{\rightarrow nl}(m) \underline{\mathrm{m}}^{\mathrm{BP},[\iota]}_{\rightarrow nl} } { 1 +  \sum_{\substack{m'=1\\ m'\neq m}}^M \underline{m}^{\mathrm{MF}}_{\rightarrow nl}(m') \underline{\mathrm{m}}^{\mathrm{BP},[\iota]}_{\rightarrow nl} ( \underline{\mathrm{m}}^{\mathrm{BP},[\iota]}_{nl \leftarrow m'} )^{\rho_\varphi} }
\end{equation}
where
\begin{equation} \nonumber 
	\begin{aligned}
		\underline{f}^1_n(\varepsilon) &= \frac{ \underline{c}^\gamma_n p_D \underline{p}^{+,1}_n  (\underline{\beta}^+_n)^{\underline{\alpha}^+_n} \Gamma(\varepsilon + \underline{\alpha}^+_n) }{ \underline{\ell}_n (\underline{\beta}^+_n+1)^{\varepsilon + \underline{\alpha}^+_n}\Gamma(\underline{\alpha}^+_n)} \\
		\underline{f}^0_n(\varepsilon) 
		&= \begin{cases}
			\frac{ \underline{c}^\gamma_n p_D \underline{p}^{+,1}_n (\underline{\ell}_n - \varepsilon) (\underline{\beta}^+_n)^{\underline{\alpha}^+_n} \Gamma(\varepsilon + \underline{\alpha}^+_n) }{ \underline{\ell}_n (\underline{\beta}^+_n+1)^{\varepsilon + \underline{\alpha}^+_n}\Gamma(\underline{\alpha}^+_n)}, & \varepsilon > 0 \\
			\underline{c}^\gamma_n \underline{p}^{+,1}_n \Big(1 - p_D + \frac{p_D (\underline{\beta}^+_n)^{\underline{\alpha}^+_n}}{(\underline{\beta}^+_n + 1)^{\underline{\alpha}^+_n}}  \Big) + \underline{p}^{+,0}_n, & \varepsilon = 0 \\
		\end{cases} \\
		\widetilde{\underline{m}}^{\mathrm{BP},[\iota]}_{\rightarrow nl} &= \sum_{m = 1}^M \underline{m}^{\mathrm{MF}}_{\rightarrow nl}(m) ( \underline{\mathrm{m}}^{\mathrm{BP},[\iota]}_{nl \leftarrow m} )^{\rho_\varphi}
	\end{aligned}
\end{equation}
and $\Gamma(\cdot)$ is the gamma function. 

The rescaled messages for the new PT $m$ are given by
\begin{equation} \label{eqn_npt_scaledMessage_b2a}
	\begin{aligned}
		&\overline{\mathrm{m}}^{\mathrm{BP},[\iota]}_{ml \leftarrow o}  \\
		\!=\!& \frac{ ( \overline{\mathrm{m}}^{\mathrm{BP},[\iota-1]}_{ml\rightarrow o} )^{\rho_\varphi-1} } { 1 \!-\! ( \overline{\mathrm{m}}^{\mathrm{BP},[\iota\!-\! 1]}_{ml\rightarrow o} \!)^{\rho_\varphi} \!+\! \!\sum\limits_{n\!=\!1}^{\underline{N}} \! \!\underline{\ell}_n\! ( \underline{\mathrm{m}}^{\mathrm{BP},[\iota\!-\! 1]}_{nl'\rightarrow o} \!)^{\rho_\varphi} \!+\! \!\sum\limits_{m'\!=\!1}^o \! \!\overline{\ell}_{m'}\! ( \overline{\mathrm{m}}^{\mathrm{BP},[\iota\!-\! 1]}_{m'l''\rightarrow o} \!)^{\rho_\varphi} }
	\end{aligned}
\end{equation}
\begin{equation} \label{eqn_npt_scaledMessage_h2a}
	\overline{\mathrm{m}}^{\mathrm{BP},[\iota]}_{\rightarrow ml} = \frac{ \sum_{\varepsilon = 0}^{\overline{\ell}_m - 1} \overline{f}^1_m(\varepsilon + 1)( \widetilde{\overline{m}}^{\mathrm{BP},[\iota]}_{\rightarrow ml} )^{\varepsilon} / \varepsilon! } { \sum_{\varepsilon = 0}^{\overline{\ell}_m - 1} \overline{f}^0_m(\varepsilon) ( \widetilde{\overline{m}}^{\mathrm{BP},[\iota]}_{\rightarrow ml} )^{\varepsilon} / \varepsilon!}
\end{equation}
\begin{equation} \label{eqn_npt_scaledMessage_a2b}
	\overline{\mathrm{m}}^{\mathrm{BP},[\iota]}_{ml\rightarrow o}= \frac{ ( \overline{\mathrm{m}}^{\mathrm{BP},[\iota]}_{ml \leftarrow o })^{\rho_\varphi-1} \overline{m}^{\mathrm{MF}}_{\rightarrow ml}(o) \overline{\mathrm{m}}^{\mathrm{BP},[\iota]}_{\rightarrow ml} } { \sum_{ \substack{o'=m \\ o'\neq o}}^M \overline{m}^{\mathrm{MF}}_{\rightarrow ml}(o') \overline{\mathrm{m}}^{\mathrm{BP},[\iota]}_{\rightarrow ml} ( \overline{\mathrm{m}}^{\mathrm{BP},[\iota]}_{ml \leftarrow o' } )^{\rho_\varphi} }
\end{equation}
where
\begin{equation} \nonumber
	\begin{aligned}
	\overline{f}^1_m(\varepsilon) &= \frac{ \lambda_{\mathrm{n}} \overline{c}^\gamma_m p_D \overline{\beta}_m^{\overline{\alpha}_m} \Gamma(\varepsilon) }{ \overline{\ell}_m (1-p_D) (\overline{\beta}_m+1)^{\varepsilon + \overline{\alpha}_m}\Gamma(\overline{\alpha}_m)} \\ 
    \overline{f}^0_m(\varepsilon) 
    &= \begin{cases}
    	\frac{ \lambda_{\mathrm{n}} \overline{c}^\gamma_m p_D (\overline{\ell}_m - \varepsilon)\overline{\beta}_m^{\overline{\alpha}_m} \Gamma(\varepsilon) }{ \overline{\ell}_m (1-p_D) (\overline{\beta}_m+1)^{\varepsilon + \overline{\alpha}_m}\Gamma(\overline{\alpha}_m)},& \varepsilon>0 \\
    	1,& \varepsilon = 0
    \end{cases}\\
	\widetilde{\overline{m}}^{\mathrm{BP},[\iota]}_{\rightarrow ml} &= \sum_{o = m}^M \overline{m}^{\mathrm{MF}}_{\rightarrow ml}(o)  ( \overline{\mathrm{m}}^{\mathrm{BP},[\iota]}_{ml \leftarrow o} )^{\rho_\varphi}.
    \end{aligned}
\end{equation}

The complexity of computing the rescaled message in \eqref{eqn_lpt_scaledMessage_b2a}-\eqref{eqn_npt_scaledMessage_a2b} can be further reduced by exploiting the symmetrical structure of the FG in Fig. \ref{fig_factorGraph}. Specifically, these messages are identical for all values of $l$ (e.g., $\underline{\mathrm{m}}^{\mathrm{BP},[\iota]}_{n1 \leftarrow m } = \underline{\mathrm{m}}^{\mathrm{BP},[\iota]}_{n2 \leftarrow m }$), and we only need to compute them once for each target and measurement. We initialize the rescaled message iterations by 
\begin{equation} \nonumber
	\underline{\mathrm{m}}^{\mathrm{BP},[0]}_{nl\rightarrow m} \!=\! \frac{ \underline{m}^{\mathrm{MF}}_{\rightarrow nl}(m) } { 1 \!+\!  \!\sum_{\substack{o=1\\ o\neq m}}^M\! \underline{m}^{\mathrm{MF}}_{\rightarrow nl}(o)}, 
	\overline{\mathrm{m}}^{\mathrm{BP},[\iota]}_{ml\rightarrow o} \!=\! \frac{ \overline{m}^{\mathrm{MF}}_{\rightarrow ml}(o) } { \!\sum_{\substack{o'=m\\ o'\neq o}}^M\! \overline{m}^{\mathrm{MF}}_{\rightarrow ml}(o') }.
\end{equation} 
The detailed derivations of the rescaled messages in \eqref{eqn_lpt_scaledMessage_b2a}-\eqref{eqn_npt_scaledMessage_a2b} are referred to the supplementary material.
\subsubsection{Messages \texorpdfstring{$\underline{m}^{\mathrm{BP}}_{\leftarrow n}(\underline{\boldsymbol{y}}_n)$}{} and \texorpdfstring{$\overline{m}^{\mathrm{BP}}_{\leftarrow m}(\overline{\boldsymbol{y}}_m)$}{}}
After the final iteration of the rescaled data association messages, we proceed to calculate the message $\underline{m}^{\mathrm{BP}}_{\leftarrow n}(\underline{\boldsymbol{y}}_n)$ sent from $\underline{h}_n(\underline{\boldsymbol{y}}_n, \underline{\boldsymbol{a}}_n )$ to $\underline{\boldsymbol{y}}_n$ and the message $\overline{m}^{\mathrm{BP}}_{\leftarrow m}(\overline{\boldsymbol{y}}_m)$ sent from $\overline{h}_m(\overline{\boldsymbol{y}}_m, \overline{\boldsymbol{a}}_m)$ to $\overline{\boldsymbol{y}}_m$. These messages are necessary for computing the state beliefs. According to \eqref{eqn_messages_trwbp_mf}, they are given by 
\begin{equation} \label{eqn_lpt_message_h2y}
	\begin{aligned} 
		\underline{m}^{\mathrm{BP}}_{\leftarrow n}(\underline{\boldsymbol{y}}_n) =& \sum_{\boldsymbol{\underline{a}}_{n}} (\underline{h}_n(\underline{\boldsymbol{y}}_n, \underline{\boldsymbol{a}}_n ))^{\frac{1}{\rho_h}} \prod_{l \in \underline{\mathcal{L}}_n} (\underline{m}^{\mathrm{BP},[I_{\mathrm{BP}}]}_{\rightarrow nl}(\underline{a}_{nl}))^{\rho_h - 1}\\
		&\times \underline{m}^{\mathrm{MF}}_{\rightarrow nl}(\underline{a}_{nl}) \prod_{m=1}^M ( \underline{m}^{\mathrm{BP},[I_{\mathrm{BP}}]}_{nl \leftarrow m } (\underline{a}_{nl}) )^{\rho_\varphi}  \\
		\propto&
		\begin{cases}
			1,& \underline{r}_n = 0, \\
			\sum_{\varepsilon=0}^{\underline{\ell}_n} \underline{f}_n(\varepsilon | \underline{\gamma}_n) ( \widetilde{\underline{m}}^{\mathrm{BP},[I_{\mathrm{BP}}]}_{\rightarrow nl} )^\varepsilon, & \underline{r}_n = 1, \\
		\end{cases}
	\end{aligned}
\end{equation}
\begin{equation}
	\begin{aligned}
		\overline{m}^{\mathrm{BP}}_{\leftarrow m}(\overline{\boldsymbol{y}}_m) \!=\!& \sum_{\overline{\boldsymbol{a}}_{m }} (\overline{h}_n\left(\overline{\boldsymbol{y}}_m, \overline{\boldsymbol{a}}_m \right))^{\frac{1}{\rho_h}}  \!\prod_{l \in \overline{\mathcal{L}}_m }\! (\overline{m}^{\mathrm{BP},[I_{\mathrm{BP}}]}_{\rightarrow ml}(\overline{a}_{ml}))^{\rho_h-1}  \\
		&\times \overline{m}^{\mathrm{MF}}_{\rightarrow ml}(\overline{a}_{ml}) \prod_{o=m}^M ( \overline{m}^{\mathrm{BP},[I_{\mathrm{BP}}]}_{ml \leftarrow o} (\overline{a}_{ml}))^{\rho_\varphi}   \\
		\!\propto\!&
		\begin{cases}
			0,& \overline{r}_m = 0 \\
			\sum_{\varepsilon=1}^{\overline{\ell}_m} \overline{f}_m(\varepsilon | \overline{\gamma}_m) ( \widetilde{\overline{m}}^{\mathrm{BP},[I_{\mathrm{BP}}]}_{\rightarrow ml} )^\varepsilon, & \overline{r}_m = 1 \\
		\end{cases} 
	\end{aligned}
\end{equation}
note that $\rho_h=1$ and
\begin{equation} \nonumber
	\begin{aligned}
		\underline{f}_n(\varepsilon | \underline{\gamma}_n) &= 
		\begin{cases}
			\frac{\underline{c}^\gamma_n p_D \underline{\gamma}_n^{\varepsilon} \exp(-\underline{\gamma}_n)}{\varepsilon!} , &  \varepsilon > 0 \\
			\underline{c}^\gamma_n (1-p_D + p_D \exp(-\underline{\gamma}_n)), & \varepsilon = 0 \\
		\end{cases}\\
		\overline{f}_m(\varepsilon | \overline{\gamma}_m) &= \frac{ \overline{c}^\gamma_m p_D \overline{\gamma}_m^{\varepsilon} \exp(-\overline{\gamma}_m) } {(1-p_D + p_D \exp(-\overline{\gamma}_m))\varepsilon!}.
	\end{aligned}
\end{equation}
We also provide the derivations of these messages in the supplementary material.  

\subsection{Beliefs}
Once all messages in the MF and BP regions are obtained, the beliefs $\underline{q}_{nl}(\underline{a}_{nl})$ and $\overline{q}_{ml}(\overline{a}_{ml})$, for association variables $\underline{a}_{nl}$ and $\overline{a}_{ml}$, as well as the beliefs $\underline{q}_n(\underline{\boldsymbol{y}}_n)$ and $\overline{q}_m(\overline{\boldsymbol{y}}_m)$, for state variables $\underline{\boldsymbol{y}}_n$ and $\overline{\boldsymbol{y}}_m$, can be computed according to \eqref{eqn_beliefs_trwbp_mf}. The beliefs $\underline{q}_{nl}(\underline{a}_{nl})$ and $\overline{q}_{ml}(\overline{a}_{ml})$ are given by
\begin{equation} \label{eqn_belief_a_lpt}
	\begin{aligned} 
		&\underline{q}_{nl}(\underline{a}_{nl}) \\
		\!\propto & \underline{m}^{\mathrm{MF}}_{\rightarrow nl}(\underline{a}_{nl}) ( \underline{m}^{\mathrm{BP},[I_{\mathrm{BP}}]}_{\rightarrow nl}(\underline{a}_{nl}) )^{\rho_h} \prod_{m=1}^M ( \underline{m}^{\mathrm{BP},[I_{\mathrm{BP}}]}_{nl\leftarrow m}(\underline{a}_{nl}) )^{\rho_\varphi}\\
		\!\propto &
		\begin{cases}
			1, \!&\! \underline{a}_{nl} \!=\! 0 \\
			\underline{m}^{\mathrm{MF}}_{\rightarrow nl}(m)  \underline{\mathrm{m}}^{\mathrm{BP},[I_{\mathrm{BP}}]}_{\rightarrow nl} ( \underline{\mathrm{m}}^{\mathrm{BP},[I_{\mathrm{BP}}]}_{nl\leftarrow m} )^{\rho_\varphi}, \!&\! \underline{a}_{nl} \!=\! m\!\in\! \!\{\! 1, \!\dots\!, M \!\}\!
		\end{cases}
	\end{aligned}
\end{equation}
\begin{equation} \label{eqn_belief_a_npt}
	\begin{aligned} 
		&\overline{q}_{ml}(\overline{a}_{ml}) \\
		\!\propto& \overline{m}^{\mathrm{MF}}_{\rightarrow ml}(\overline{a}_{ml}) ( \overline{m}^{\mathrm{BP},[I_{\mathrm{BP}}]}_{\rightarrow ml}(\overline{a}_{ml}) )^{\rho_h} \prod_{o=m}^M ( \overline{m}^{\mathrm{BP},[I_{\mathrm{BP}}]}_{ml\leftarrow o}(\overline{a}_{ml}) )^{\rho_\varphi} \\
		\!\propto & 
		\begin{cases}
			0, \!&\! \overline{a}_{ml} \!=\! 0 \\
			\overline{m}^{\mathrm{MF}}_{\rightarrow ml}(o)  \overline{\mathrm{m}}^{\mathrm{BP},[I_{\mathrm{BP}}]}_{\rightarrow ml} ( \overline{\mathrm{m}}^{\mathrm{BP},[I_{\mathrm{BP}}]}_{ml\leftarrow o} )^{\rho_\varphi}, \!&\! \overline{a}_{ml} \!=\! o\!\in\!\!\{\!m, \!\dots\!, M \!\}\!.
		\end{cases}
	\end{aligned}
\end{equation}

The belief $\underline{q}_n(\underline{\boldsymbol{y}}_n)$ for the legacy PT $n$ is given by 
\begin{equation} \label{eqn_belief_lpt_state}
	\begin{aligned}
		\underline{q}_n(\underline{\boldsymbol{y}}_n) &\propto  \underline{m}^{\mathrm{MF}}_{\rightarrow n}(\underline{\boldsymbol{y}}_n) ( \underline{m}^{\mathrm{BP}}_{\leftarrow n}(\underline{\boldsymbol{y}}_n) )^{\rho_h} \prod_{l=1}^{\underline{\ell}_n} \underline{m}^{\mathrm{MF}}_{\leftarrow nl}(\underline{\boldsymbol{y}}_n) \\
		&\propto \begin{cases}
			\underline{p}^{+,0}_n p_{\mathrm{d}}(\underline{\boldsymbol{x}}_n),& \underline{r}_n = 0\\
            \underline{p}^{+,1}_n \underline{c}_n  \underline{p}_n(\underline{\gamma}_n) \underline{p}_n(\underline{\boldsymbol{\xi}}_n, \underline{\boldsymbol{e}}_n, \boldsymbol{z} ), & \underline{r}_n = 1
		\end{cases} 
	\end{aligned}
\end{equation}
where $\underline{c}_n = \underline{c}^\gamma_n(1-p_D + \sum_{\varepsilon=0}^{\underline{\ell}_n} \underline{w}_n^\varepsilon) (\kappa_{\mathrm{c}})^{-\sum_{m=1}^M \underline{q}_{nl}(m)}$ and 
\begin{equation} \label{eqn_lpt_gammaPosterior}
	\begin{aligned}
		\underline{p}_n(\underline{\gamma}_n) \propto& \underline{c}^\gamma_n(1 - p_D)\mathrm{G}(\underline{\gamma}_n; \underline{\alpha}^+_n, \underline{\beta}^+_n) \\
		&+ \sum_{\varepsilon=0}^{\underline{\ell}_n} \underline{w}_n^\varepsilon \mathrm{G}(\underline{\gamma}_n; \underline{\alpha}^+_n + \varepsilon, \underline{\beta}^+_n + 1)  
	\end{aligned}
\end{equation}
\begin{equation}
	\underline{w}_n^\varepsilon = \frac{p_D (\underline{\beta}^+_n)^{\underline{\alpha}^+_n} \Gamma(\underline{\alpha}^+_n + \varepsilon) ( \widetilde{\underline{m}}^{\mathrm{BP},[I_{\mathrm{BP}}]}_{\rightarrow nl} )^\varepsilon }{ (\underline{\beta}^+_n+1)^{\underline{\alpha}^+_n+\varepsilon} \Gamma(\underline{\alpha}^+_n) \varepsilon! } 
\end{equation}
\begin{equation} \label{eqn_lpt_joinPosteriorPDF}
	\underline{p}_n(\underline{\boldsymbol{\xi}}_n, \underline{\boldsymbol{e}}_n, \boldsymbol{z}) \!=\! \underline{p}^+_n(\underline{\boldsymbol{\xi}}_n) \underline{p}^+_n(\underline{\boldsymbol{e}}_n) \!\prod_{m=1}^M\! p(\boldsymbol{z}_m| \underline{\boldsymbol{\xi}}_n, \underline{\boldsymbol{e}}_n)
\end{equation}
\begin{equation} \label{eqn_lpt_PowerOfMeasLikelihood}
	p(\boldsymbol{z}_m| \underline{\boldsymbol{\xi}}_n, \underline{\boldsymbol{e}}_n) = ( \mathrm{N}(\boldsymbol{z}_m; \boldsymbol{H} \underline{\boldsymbol{\xi}}_n, s\underline{\boldsymbol{E}}_n + \boldsymbol{R}) )^{\underline{\ell}_n\underline{q}_{nl}(m)}.
\end{equation}

Moreover, the belief $\overline{q}_m(\overline{\boldsymbol{y}}_m)$ for the new PT $m$ is given by 
\begin{equation} \label{eqn_belief_npt_state}
	\begin{aligned}
		\overline{q}_m(\overline{\boldsymbol{y}}_m) &\!\propto\!  \overline{m}^{\mathrm{MF}}_{\rightarrow m}(\overline{\boldsymbol{y}}_m) \left( \overline{m}^{\mathrm{BP}}_{\leftarrow m}(\overline{\boldsymbol{y}}_m) \right)^{\rho_h} \!\prod_{l=1}^{\overline{\ell}_m}\! \overline{m}^{\mathrm{MF}}_{\leftarrow ml}(\overline{\boldsymbol{y}}_m) \\
		&\!\propto\! \begin{cases}
			0, & \overline{r}_m = 0\\
            \overline{p}_m(\overline{\gamma}_m) \overline{p}_m(\overline{\boldsymbol{\xi}}_m, \overline{\boldsymbol{e}}_m, \boldsymbol{z} ), & \overline{r}_m = 1
		\end{cases} 
	\end{aligned}
\end{equation}
where
\begin{equation} \label{eqn_npt_gammaPosterior}
	\overline{p}_m(\overline{\gamma}_m) \propto \sum_{\varepsilon=1}^{\overline{\ell}_m} \overline{w}_m^\varepsilon \mathrm{G}(\overline{\gamma}_m; \overline{\alpha}_m + \varepsilon, \overline{\beta}_m + 1)  
\end{equation}
\begin{equation}
	\overline{w}_m^\varepsilon \propto \frac{ \Gamma(\overline{\alpha}_m+\varepsilon) ( \widetilde{\underline{m}}^{\mathrm{BP},[I_{\mathrm{BP}}]}_{\rightarrow nq} )^\varepsilon }{ (\overline{\beta}_m+1)^{\overline{\alpha}_m+\varepsilon} \varepsilon! } 
\end{equation}
\begin{equation} \label{eqn_npt_joinPosteriorPDF}
	\overline{p}_m(\overline{\boldsymbol{\xi}}_m, \overline{\boldsymbol{e}}_m, \boldsymbol{z} ) = \overline{p}_m(\overline{\boldsymbol{\xi}}_m) \overline{p}_m(\overline{\boldsymbol{e}}_m) \!\prod_{o=m}^M\! p(\boldsymbol{z}_o| \overline{\boldsymbol{\xi}}_m, \overline{\boldsymbol{e}}_m)
\end{equation}
\begin{equation} \label{eqn_npt_PowerOfMeasLikelihood}
	p(\boldsymbol{z}_o| \overline{\boldsymbol{\xi}}_m, \overline{\boldsymbol{e}}_m) = ( \mathrm{N}(\boldsymbol{z}_o; \boldsymbol{H} \overline{\boldsymbol{\xi}}_m, s\overline{\boldsymbol{E}}_m + \boldsymbol{R}) )^{\overline{\ell}_m\overline{q}_{ml}(o)}.
\end{equation}
The detailed derivations of the beliefs $\underline{q}_n(\underline{\boldsymbol{y}}_n)$ and $\overline{q}_m(\overline{\boldsymbol{y}}_m)$ can be found in the supplementary material. Note that the densities $\underline{p}_n(\underline{\gamma}_n)$ and $\overline{p}_m(\overline{\gamma}_m)$ in \eqref{eqn_lpt_gammaPosterior} and \eqref{eqn_npt_gammaPosterior} are normalized and they are mixtures of gamma distributions. We apply the truncation approach in \cite{GO2012C} to merge these mixtures into a single gamma. Furthermore, due to the additions and the exponentiations in \eqref{eqn_lpt_PowerOfMeasLikelihood} and \eqref{eqn_npt_PowerOfMeasLikelihood}, no closed-form solutions exist for the joint densities $\underline{p}_n(\underline{\boldsymbol{\xi}}_n, \underline{\boldsymbol{e}}_n, \boldsymbol{z})$ and $\overline{p}_m(\overline{\boldsymbol{\xi}}_m, \overline{\boldsymbol{e}}_m, \boldsymbol{z} )$ in \eqref{eqn_lpt_joinPosteriorPDF} and \eqref{eqn_npt_joinPosteriorPDF}. To overcome this challenge, we apply the MF approach in \eqref{eqn_mf_KLD} and \eqref{eqn_mf_posteriorUpdate} to derive approximate solutions.

\subsection{MF Approximation of the State Beliefs}
Note that the densities $\underline{p}_n(\underline{\boldsymbol{\xi}}_n, \underline{\boldsymbol{e}}_n, \boldsymbol{z})$ and $\overline{p}_m(\overline{\boldsymbol{\xi}}_m, \overline{\boldsymbol{e}}_m, \boldsymbol{z})$ in \eqref{eqn_lpt_joinPosteriorPDF} and \eqref{eqn_npt_joinPosteriorPDF} have similar structures, hence similar solutions. Here, we only provide the solution to $\underline{p}_n(\underline{\boldsymbol{\xi}}_n, \underline{\boldsymbol{e}}_n, \boldsymbol{z})$ following the derivations in \cite{O2012}. Let $\boldsymbol{\zeta} = [\boldsymbol{\zeta}_1^T \cdots \boldsymbol{\zeta}_M^T]^T$ be the measurement sources (i.e., noise-free measurements) corresponding to the noisy measurements $\boldsymbol{z}$, and let $\omega_m = \underline{\ell}_n\underline{p}_{nl}(m)$. Then the density $p(\boldsymbol{z}_m| \underline{\boldsymbol{\xi}}_n, \underline{\boldsymbol{e}}_n)$ in \eqref{eqn_lpt_PowerOfMeasLikelihood} can be reformulated as 
\begin{align} \label{eqn_lpt_PowerOfMeasLikelihoodReformulated}
	&p(\boldsymbol{z}_m| \underline{\boldsymbol{\xi}}_n, \underline{\boldsymbol{e}}_n) \nonumber \\
	=& C^{\omega_m}_{s\underline{\boldsymbol{E}}_n + \boldsymbol{R}} \mathrm{N}\left(\boldsymbol{z}_m; \boldsymbol{H} \underline{\boldsymbol{\xi}}_n, \frac{s\underline{\boldsymbol{E}}_n + \boldsymbol{R}}{\omega_m}\right)  \\
	=& C^{\omega_m}_{s\underline{\boldsymbol{E}}_n + \boldsymbol{R}} \int \mathrm{N}\left(\boldsymbol{z}_m; \boldsymbol{\zeta}_m, \frac{\boldsymbol{R}}{\omega_m}\right) \mathrm{N}\left(\boldsymbol{\zeta}_m; \boldsymbol{H} \underline{\boldsymbol{\xi}}_n, \frac{s\underline{\boldsymbol{E}}_n}{\omega_m}\right) \mathrm{d}\boldsymbol{\zeta}_m \nonumber
\end{align}
where $C^\omega_{\boldsymbol{X}} = |2\pi\boldsymbol{X}\omega^{-1}|^{\frac{1}{2}} / |2\pi\boldsymbol{X}|^{\frac{\omega}{2}}$. We define 
\begin{equation} 
	\begin{aligned}
		p(\boldsymbol{\zeta}_m, \boldsymbol{z}_m | \underline{\boldsymbol{\xi}}_n, \underline{\boldsymbol{e}}_n) &= C^{\omega_m}_{s\underline{\boldsymbol{E}}_n + \boldsymbol{R}} p(\boldsymbol{z}_m | \boldsymbol{\zeta}_m) p(\boldsymbol{\zeta}_m |\underline{\boldsymbol{\xi}}_n, \underline{\boldsymbol{e}}_n)  \\
		p(\boldsymbol{z}_m | \boldsymbol{\zeta}_m) &= \mathrm{N}(\boldsymbol{z}_m; \boldsymbol{\zeta}_m, \boldsymbol{R}/\omega_m) \\
		p(\boldsymbol{\zeta}_m |\underline{\boldsymbol{\xi}}_n, \underline{\boldsymbol{e}}_n) &= \mathrm{N}(\boldsymbol{\zeta}_m; \boldsymbol{H} \underline{\boldsymbol{\xi}}_n, s\underline{\boldsymbol{E}}_n/ \omega_m)\\
	\end{aligned}
\end{equation}
such that $p(\boldsymbol{z}_m| \underline{\boldsymbol{\xi}}_n, \underline{\boldsymbol{e}}_n)$ can be obtained by integrating out $\boldsymbol{\zeta}_m$ from $p(\boldsymbol{\zeta}_m, \boldsymbol{z}_m | \underline{\boldsymbol{\xi}}_n, \underline{\boldsymbol{e}}_n)$ according to \eqref{eqn_lpt_PowerOfMeasLikelihoodReformulated}. 

Let $p(\underline{\boldsymbol{\xi}}_n, \underline{\boldsymbol{e}}_n) = \underline{p}^+_n(\underline{\boldsymbol{\xi}}_n) \underline{p}^+_n(\underline{\boldsymbol{e}}_n) $, we have
\begin{align} \label{eqn_lpt_joinPosteriorPDF_zeta}
	&p(\underline{\boldsymbol{\xi}}_n, \underline{\boldsymbol{e}}_n, \boldsymbol{\zeta}, \boldsymbol{z}) \nonumber \\
	=& p(\boldsymbol{\zeta}, \boldsymbol{z} | \underline{\boldsymbol{\xi}}_n, \underline{\boldsymbol{e}}_n) p(\underline{\boldsymbol{\xi}}_n, \underline{\boldsymbol{e}}_n) \nonumber\\
	=& p(\underline{\boldsymbol{\xi}}_n, \underline{\boldsymbol{e}}_n) \prod_{m=1}^M p(\boldsymbol{\zeta}_m, \boldsymbol{z}_m | \underline{\boldsymbol{\xi}}_n, \underline{\boldsymbol{e}}_n) \\
	=& \underline{p}^+_n(\underline{\boldsymbol{\xi}}_n) \underline{p}^+_n(\underline{\boldsymbol{e}}_n) \prod_{m=1}^M C^{\omega_m}_{s\underline{\boldsymbol{E}}_n + \boldsymbol{R}} p(\boldsymbol{z}_m | \boldsymbol{\zeta}_m) p(\boldsymbol{\zeta}_m |\underline{\boldsymbol{\xi}}_n, \underline{\boldsymbol{e}}_n) \nonumber
\end{align}
Using the MF approximation, we approximate the joint density $p(\underline{\boldsymbol{\xi}}_n, \underline{\boldsymbol{e}}_n, \boldsymbol{\zeta})$ of the latent variables as $q_{\boldsymbol{\xi}}(\underline{\boldsymbol{\xi}}_n) q_{\boldsymbol{e}}(\underline{\boldsymbol{e}}_n) q_{\boldsymbol{\zeta}}(\boldsymbol{\zeta})$. The optimal estimates of $q_{\boldsymbol{\xi}}(\cdot)$, $q_{\boldsymbol{e}}(\cdot)$ and $q_{\boldsymbol{\zeta}}(\cdot)$ can be obtained by substituting \eqref{eqn_lpt_joinPosteriorPDF_zeta} into the variational equation \eqref{eqn_mf_posteriorUpdate}. Since the latent variables $\underline{\boldsymbol{\xi}}_n$, $\underline{\boldsymbol{e}}_n$ and $\boldsymbol{\zeta}$ are coupled in \eqref{eqn_lpt_joinPosteriorPDF_zeta}, we resort to fixed-point iterations \cite{BMN2006, TLG2008} to solve \eqref{eqn_mf_posteriorUpdate}. The approximate densities at the $\iota$th iteration are given by
\begin{equation} \label{eqn_mf_posteriors}
	\begin{aligned}
		q^{[\iota]}_{\boldsymbol{\xi}}(\underline{\boldsymbol{\xi}}_n) &= \mathrm{N}(\underline{\boldsymbol{\xi}}_n; \underline{\boldsymbol{\mu}}_n^{[\iota]}, \underline{\boldsymbol{\Sigma}}_n^{[\iota]}) \\
		q^{[\iota]}_{\boldsymbol{e}}(\underline{\boldsymbol{e}}_n) &= \mathrm{IW}(\underline{\boldsymbol{E}}_n; \underline{\nu}^{[\iota]}_n, \underline{\boldsymbol{V}}^{[\iota]}_n) \\
		q^{[\iota]}_{\boldsymbol{\zeta}}(\boldsymbol{\zeta}) &= \prod_{m=1}^M \mathrm{N}(\boldsymbol{\zeta}_m; \widetilde{\boldsymbol{\mu}}_m^{[\iota]}, \widetilde{\boldsymbol{\Sigma}}_m^{[\iota]})
	\end{aligned}
\end{equation}
with parameters
\begin{equation} \nonumber
	\begin{aligned}
		\underline{\boldsymbol{\mu}}_n^{[\iota]} =& \underline{\boldsymbol{\Sigma}}_n^{[\iota]} \bigg( (\underline{\boldsymbol{\Sigma}}_n^+)^{-1}\underline{\boldsymbol{\mu}}^+_n + \frac{1}{s} \boldsymbol{H}^T \mathbb{E}_{\underline{\boldsymbol{E}}_n}[\underline{\boldsymbol{E}}_n^{-1}] \sum_{m=1}^M \omega_m \widetilde{\boldsymbol{\mu}}_m^{[\iota]} \bigg)\\
		\underline{\boldsymbol{\Sigma}}_n^{[\iota]} =& \Big( (\underline{\boldsymbol{\Sigma}}^+_n)^{-1} + \frac{1}{s}\Omega \boldsymbol{H}^T \mathbb{E}_{\underline{\boldsymbol{E}}_n}[\underline{\boldsymbol{E}}_n^{-1}]\boldsymbol{H} \Big)^{-1} \\
		\underline{\nu}^{[\iota]}_n =& \underline{\nu}^+_n + \Omega \\
		\underline{\boldsymbol{V}}^{[\iota]}_n =& \underline{\boldsymbol{V}}^+_n + \frac{1}{s} \sum_{m=1}^M \omega_m \Big( \boldsymbol{H}\underline{\boldsymbol{\Sigma}}_n^{[\iota]}\boldsymbol{H}^T + \widetilde{\boldsymbol{\Sigma}}_m^{[\iota]} \Big. \\
		&\Big. + (\widetilde{\boldsymbol{\mu}}_m^{[\iota]} - \boldsymbol{H}\underline{\boldsymbol{\mu}}_n^{[\iota]})(\widetilde{\boldsymbol{\mu}}_m^{[\iota]} - \boldsymbol{H}\underline{\boldsymbol{\mu}}_n^{[\iota]})^T \Big)  \\
		\widetilde{\boldsymbol{\mu}}_m^{[\iota]} =& \omega_m \widetilde{\boldsymbol{\Sigma}}_m^{[\iota]} \Big( \frac{1}{s} \mathbb{E}_{\underline{\boldsymbol{E}}_n}[\underline{\boldsymbol{E}}_n^{-1}] \boldsymbol{H} \widetilde{\boldsymbol{\mu}}_m^{[\iota]} + \boldsymbol{R}^{-1}\boldsymbol{z}_m \Big) \\
		\widetilde{\boldsymbol{\Sigma}}_m^{[\iota]} =& \frac{1}{\omega_m}\Big( \frac{1}{s} \mathbb{E}_{\underline{\boldsymbol{E}}_n}[\underline{\boldsymbol{E}}_n^{-1}] + \boldsymbol{R}^{-1} \Big)^{-1}
	\end{aligned}
\end{equation}
where $\mathbb{E}_{\underline{\boldsymbol{E}}_n}[\underline{\boldsymbol{E}}_n^{-1}] = (\underline{\nu}^{[\iota-1]}_n - d_{\boldsymbol{z}} - 1)(\underline{\boldsymbol{V}}^{[\iota-1]}_n)^{-1}$ and $\Omega = \sum_{m=1}^M\omega_m$. The derivation of \eqref{eqn_mf_posteriors} refers to the supplementary material. The iterations in \eqref{eqn_mf_posteriors} are initialized by $\underline{\boldsymbol{\mu}}_n^{[0]} = \underline{\boldsymbol{\mu}}^+_n$, $\underline{\boldsymbol{\Sigma}}_n^{[0]} = \underline{\boldsymbol{\Sigma}}^+_n$, $\underline{\nu}^{[0]}_n = \underline{\nu}^+_n$, $\underline{\boldsymbol{V}}^{[0]}_n = \underline{\boldsymbol{V}}^+_n$, $\widetilde{\boldsymbol{\mu}}_m^{[0]} = \boldsymbol{z}_m$ and $\widetilde{\boldsymbol{\Sigma}}_m^{[0]}= \mathbb{E}_{\underline{\boldsymbol{E}}_n} [s\underline{\boldsymbol{E}}_n]$. After the final iteration $\iota = I_{\mathrm{MF}}$, we obtain
\begin{equation} \label{eqn_mf_finalPosterior}
	q_{\boldsymbol{\xi}}(\underline{\boldsymbol{\xi}}_n) = \mathrm{N}(\underline{\boldsymbol{\xi}}_n; \underline{\boldsymbol{\mu}}_n, \underline{\boldsymbol{\Sigma}}_n), ~ q_{\boldsymbol{e}}(\underline{\boldsymbol{e}}_n) = \mathrm{IW}(\underline{\boldsymbol{E}}_n; \underline{\nu}_n, \underline{\boldsymbol{V}}_n)
\end{equation}
and $q_{\boldsymbol{\zeta}}(\boldsymbol{\zeta}) =\prod_{m=1}^M \mathrm{N}(\boldsymbol{\zeta}_m; \widetilde{\boldsymbol{\mu}}_m, \widetilde{\boldsymbol{\Sigma}}_m) $, where $\underline{\boldsymbol{\mu}}_n \!=\! \underline{\boldsymbol{\mu}}_n^{[I_{\mathrm{MF}}]}$, $\underline{\boldsymbol{\Sigma}}_n \!=\! \underline{\boldsymbol{\Sigma}}_n^{[I_{\mathrm{MF}}]}$, $\underline{\nu}_n = \underline{\nu}^{[I_{\mathrm{MF}}]}_n$, $\underline{\boldsymbol{V}}_n = \underline{\boldsymbol{V}}^{[I_{\mathrm{MF}}]}_n$, $\widetilde{\boldsymbol{\mu}}_m = \widetilde{\boldsymbol{\mu}}_m^{[I_{\mathrm{MF}}]}$ and $\widetilde{\boldsymbol{\Sigma}}_m = \widetilde{\boldsymbol{\Sigma}}_m^{[I_{\mathrm{MF}}]}$.

The density $\underline{p}_n(\underline{\boldsymbol{\xi}}_n, \underline{\boldsymbol{e}}_n, \boldsymbol{z})$ in \eqref{eqn_lpt_joinPosteriorPDF} can be approximated by
\begin{equation} \label{eqn_lpt_posteriorApprox}
	\underline{p}_n(\underline{\boldsymbol{\xi}}_n, \underline{\boldsymbol{e}}_n, \boldsymbol{z}) \!=\! \underline{p}_n(\underline{\boldsymbol{\xi}}_n, \underline{\boldsymbol{e}}_n | \boldsymbol{z}) p(\boldsymbol{z}) \!\approx\! q_{\boldsymbol{\xi}}(\underline{\boldsymbol{\xi}}_n) q_{\boldsymbol{e}}(\underline{\boldsymbol{e}}_n) p(\boldsymbol{z}) 
\end{equation}
where $p(\boldsymbol{z})$ is the predictive likelihood and is necessary for computing the posterior existence probability. The variational inference provides an approximation to $p(\boldsymbol{z})$ by taking exponential of the evidence lower bound \cite{BMN2006}.  Due to space limitation, we present the formula of $p(\boldsymbol{z})$ in the supplementary material. 

Inserting \eqref{eqn_lpt_posteriorApprox} into \eqref{eqn_belief_lpt_state}, we obtain the belief $\underline{q}_n(\underline{\boldsymbol{y}}_n)$ for the legacy PT $n$ as 
\begin{equation} \label{eqn_posterior_lpt}
	\underline{q}_n(\underline{\boldsymbol{y}}_n) = 
	\begin{cases}
		\underline{p}^0_n p_{\mathrm{d}}(\underline{\boldsymbol{x}}_n),& \underline{r}_n = 0\\
		\underline{p}^1_n \underline{p}_n(\underline{\gamma}_n) q_n(\underline{\boldsymbol{\xi}}_n) q_n(\underline{\boldsymbol{e}}_n), & \underline{r}_n = 1
	\end{cases}
\end{equation}
where $\underline{p}_n(\underline{\gamma}_n)$ is given by \eqref{eqn_lpt_gammaPosterior} (note that we have to merge the gamma mixture in $\eqref{eqn_lpt_gammaPosterior}$ into a single gamma for the posterior GGIW distribution), the densities $q_n(\underline{\boldsymbol{\xi}}_n)$ and $q_n(\underline{\boldsymbol{e}}_n)$ are given by \eqref{eqn_mf_finalPosterior}, and the posterior target nonexistence and existence probabilities are as follows
\begin{equation} \nonumber
	\underline{p}^0_n = \frac{\underline{p}^{+, 0}_n}{\underline{p}^{+, 0}_n + \underline{p}^{+, 1}_n \underline{c}_n p(\boldsymbol{z})}, \quad \underline{p}^1_n = \frac{\underline{p}^{+, 1}_n \underline{c}_n p(\boldsymbol{z})}{\underline{p}^{+, 0}_n + \underline{p}^{+, 1}_n \underline{c}_n p(\boldsymbol{z})}.
\end{equation}
Target detection and state estimation for legacy PT $n$ can now be performed based on the posterior density $\underline{q}_n(\underline{\boldsymbol{y}}_n)$ in \eqref{eqn_posterior_lpt}.  
\section{Simulation} \label{section_five}
In this section, we evaluate the performance of the
proposed TRWBP-MF approach by simulations and comparisons with state-of-the-art algorithms.
\vspace{-0.2cm}
\subsection{Scenario setting}
\begin{figure}[tbp]
	\centering
	\subfigure{
	  \includegraphics[width=1.6in]{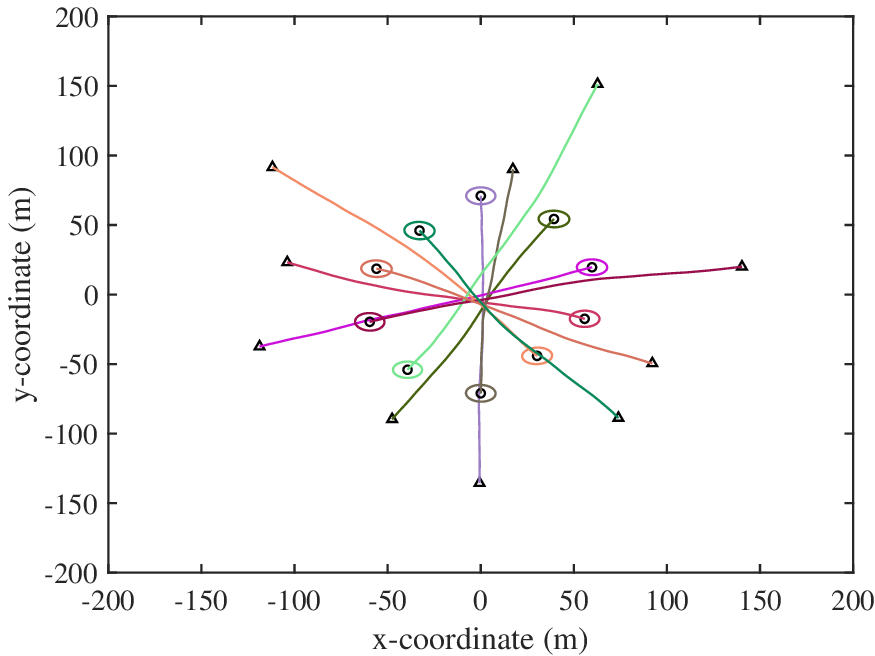}
	}
	\subfigure{
	  \includegraphics[width=1.6in]{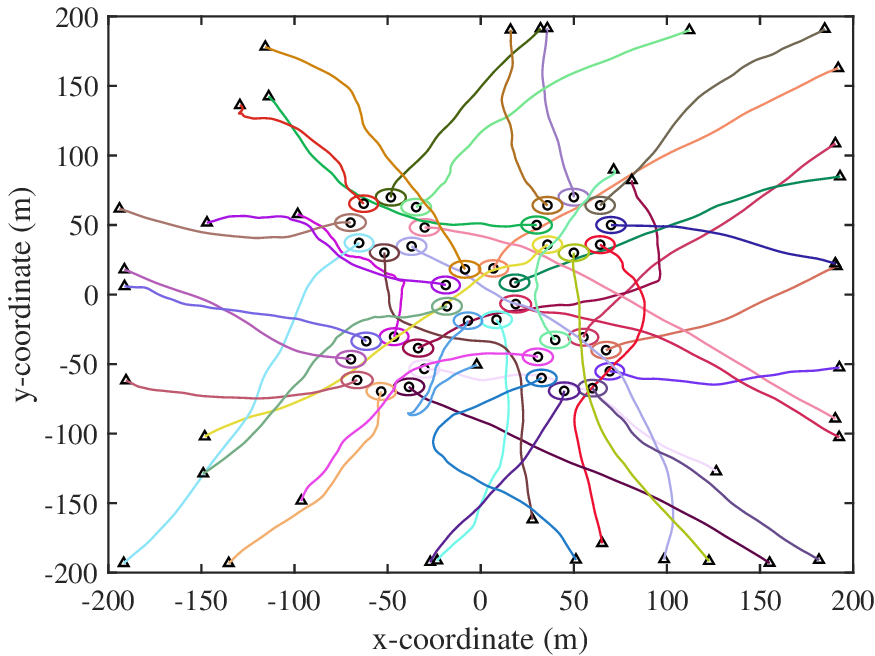}
	}
	\caption{True tracks for scenario 1 (left) and scenario 2 (right). Circles, triangles and ellipses represent starting positions, stopping positions and extents.} 
	\label{fig_groundTruth}
\end{figure}

We consider two simulated scenarios: one adapted from \cite{MW2021}, and the other designed for this paper. Targets move on a two-dimensional SR defined as [-200, 200]$\times$[-200, 200] (in m). Fig. \ref{fig_groundTruth} shows the simulated trajectories for the two scenarios. In the first scenario, ten targets move toward the SR center from a circle of 75m about the SR center, with initial positions uniformly generated on the circle and initial velocities of 10m/s. This scenario is challenging because data association becomes extremely difficult when targets cross near the origin. In the second scenario, forty targets move toward the boundary of the SR from the circles of 20m about five locations, i.e., $[\pm 50 ~ \pm 50]^T$ and the origin, with initial velocities of 5m/s. This scenario involves a large number of targets and is designed to demonstrate the scalability of the proposed algorithm.

The common settings for both scenarios are as follows. The kinematic state of each target, denoted as $\boldsymbol{\xi}_n^k=[\boldsymbol{p}_n^k~ \dot{\boldsymbol{p}}_n^k]^T$, consists of target position $\boldsymbol{p}_n^k\in\mathbb{R}^2$ and velocity $\dot{\boldsymbol{p}}_n^k\in\mathbb{R}^2$. Each target follows the nearly constant velocity (NCV) motion model with the transition matrix $\boldsymbol{F}$ and process noise covariance $\boldsymbol{Q}$ defined as 
\begin{equation}\nonumber
	\boldsymbol{F} =
	\begin{bmatrix}
	  \boldsymbol{I}_2 & T\boldsymbol{I}_2 \\
	  \boldsymbol{0}_2 & \boldsymbol{I}_2 \\
	\end{bmatrix}, ~
	\boldsymbol{Q} = \sigma^2_a\boldsymbol{G}\boldsymbol{G}^T, ~
	\boldsymbol{G} =
	\begin{bmatrix}
	  \frac{T^2}{2}\boldsymbol{I}_2\\
	  T \boldsymbol{I}_2\\
	\end{bmatrix}
\end{equation}
where $T=0.2$s is the sampling period, $\boldsymbol{I}_n$ is the $n\times n$ identity matrix, and $\sigma_a = 1 \text{m/s}^2$ is the acceleration standard deviation. The duration of each scenario is 100 time steps. The initial target extents are sampled from the IW distribution $\mathrm{IW}(\boldsymbol{E}; 1000, 994\times\text{diag}([8^2~6^2]^T))$ such that the means of their long and short semi-axes lengths are 8m and 6m, respectively. The initial measurement rate of each target is 10. We assume that the target extents and measurement rates remain unchanged over time. Each target has a survival probability $p_S=0.99$, and different detection probabilities $p_D\in\{0.85, 0.95\}$ are tested. The parameters for the measurement likelihood in \eqref{eqn_singleMeasLikelihood} are set as $\boldsymbol{H} = [\boldsymbol{I}_2~\boldsymbol{0}_2]$, $s = 1/4$ and $\boldsymbol{R} = \boldsymbol{I}_2$. Finally, the Poisson rate $\lambda_\mathrm{c}$ of clutter measurements is set to 10. 

\vspace{-0.4cm}
\subsection{Performance comparison}
The proposed TRWBP-MF approach is compared with the max-sum algorithm (MSA) \cite{MJDL2024} and the PMBM filter \cite{GFS2020} both with GGIW implementation, and the BP-based PMB (PMB-BP) filter using particle implementation \cite{XGMWGS2023}. The performance of these algorithms is measured by the generalized optimal subpattern assignment (GOSPA) metric \cite{RGS2017} with parameters $p=1$, $c=20$ and $\alpha = 2$. We use the Gaussian Wasserstein distance \cite{GS1984} as the base distance for GOSPA. The results are averaged over 100 Monte Carlo runs.

For the TRWBP-MF approach, new tracks are initialized  following the procedure in \cite{MW2021}, which consists of message censoring, and measurement clustering and reordering. Each cluster represent a new track with prior $ \overline{p}(\overline{\boldsymbol{x}}) = \mathrm{GGIW}(\overline{\boldsymbol{x}}, \overline{\boldsymbol{\theta}})$, where the initial parameter vector $\overline{\boldsymbol{\theta}} = (\overline{\alpha}, \overline{\beta}, \overline{\boldsymbol{\mu}}, \overline{\boldsymbol{\Sigma}}, \overline{\nu}, \overline{\boldsymbol{V}})$ is given by: $\overline{\alpha} = 100$, $\overline{\beta} = 10$, $\overline{\boldsymbol{\mu}} = [\boldsymbol{z}_{\mathrm{c}}^T ~0~0]^T$ with $\boldsymbol{z}_{\mathrm{c}}$ being the central measurement of a cluster, $\overline{\boldsymbol{\Sigma}} = \text{diag}([10^2~10^2~5^2~5^2]^T)$, $\overline{\nu} = 1000$ and $\overline{\boldsymbol{V}} = 994\times\text{diag}([8^2~6^2]^T)$. Additionally, the following parameters are used: the Poisson new PT birth rate $\lambda_\mathrm{n} = 0.01$, the factor $\eta=1.1$ and the constant $\tau=50$ for the prediction of measurement rate and extent, the number of BP and MF message iterations $I_\mathrm{BP} = 30$ and $I_\mathrm{MF} = 20$, the target detection threshold 0.5, the track pruning threshold $10^{-5}$, and the upper bound $\ell$ of the truncated Poisson distribution $\mathrm{P}_{\mathrm{t}}(l|\lambda)$ introduced in Section \ref{section_two} is obtained from the conventional Poisson distribution such that $\sum_{l=\ell+1}^{\infty}\mathrm{P}(l|\lambda)\leq 10^{-5}$. The parameters of MSA are identical to those of TRWBP-MF, except that the number of message iterations is 20.

For the PMBM filter, the gating threshold is set to 13.8. Measurements are partitioned using the DBSCAN algorithm \cite{EKSX1996} with different distance values equally spaced between 0.1 and 20, at intervals of 0.1, and a maximum number of 20 assignments is preserved for each partition. The pruning thresholds for Bernoulli components and global hypotheses are set to $10^{-3}$ and $10^{-2}$. The birth distribution of PMBM is the same as that of TRWBP-MF except $\overline{\boldsymbol{\mu}} = [0~0~0~0]^T$ and $\overline{\boldsymbol{\Sigma}} = \text{diag}([4000~4000~25~25]^T)$, ensuring that most of the SR is within three standard deviations of the mean. 

The PMB-BP filter assumes that the measurement rate is known, and it only estimates the kinematic state and extent. It utilizes the method in \cite{MW2021} to initialize new tracks. For the particle implementation of PMB-BP, we set the number of particles to $\{2000, 5000\}$, and the resulting realizations are denoted as PMB-BP-2000 and PMB-BP-5000. New extent samples are drawn form $\mathrm{W}(\underline{\boldsymbol{E}}^k_n;\nu,\boldsymbol{E}^{k-1}_n/\nu)$ with $\nu = 5000$, and new kinematic samples are drawn from the NCV motion model. Moreover, the number of message iterations is 3 and the
threshold for pruning unlikely tracks is $10^{-3}$.

\begin{figure}[tbp]
	\centering
	\includegraphics[width=2.8in]{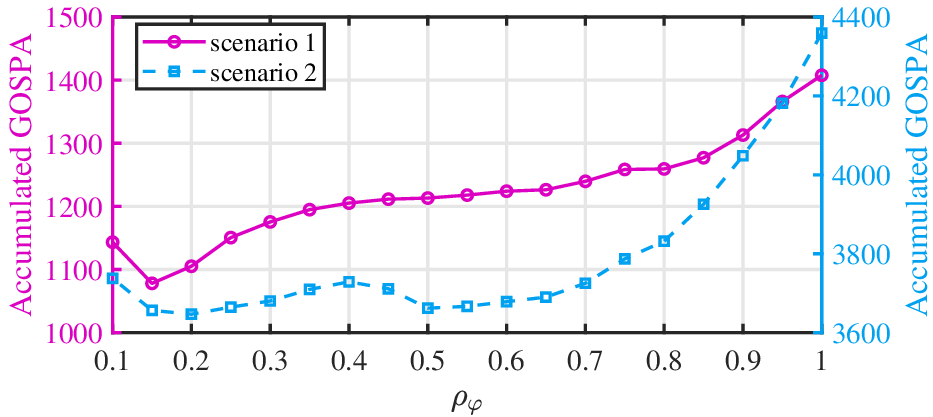}
	\caption{Accumulated GOSPA with respect to FAP $\rho_\varphi$ given $p_D=0.95$. TRWBP corresponds to $\rho_\varphi < 1$ and BP corresponds to $\rho_\varphi = 1$.} 
	\label{fig_gospaVsRho}
\end{figure} 

We first evaluate the performance of TRWBP-MF regarding the FAP $\rho_\varphi$, which takes uniform values from 0.1 to 1, with an interval of 0.05. Fig. \ref{fig_gospaVsRho} shows the accumulated GOSPA errors for the two scenarios under $p_D = 0.95$. The results show that TRWBP-MF performs better with $\rho_\varphi < 1$ compared to $\rho_\varphi = 1$ for both scenarios, indicating that TRWBP ($\rho_\varphi < 1$) is superior to BP ($\rho_\varphi = 1$) for solving the data association problem in this paper. The minimum GOSPA errors are achieved at $\rho_\varphi = 0.15$ and $\rho_\varphi = 0.2$ for the first and second scenarios, respectively. For $0.1\leq\rho_\varphi\leq 0.7$, the performance of TRWBP-MF is comparable and satisfactory, suggesting that this region provides an appropriate choice for the FAP $\rho_\varphi$.  

\begin{figure}[tbp]
	\centering
	\subfigure{
	  \includegraphics[width=1.6in]{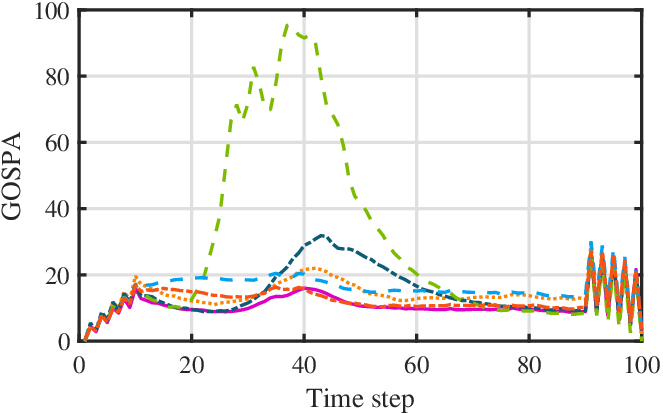}
	} \vspace{-0.3cm}
	\subfigure{
	  \includegraphics[width=1.6in]{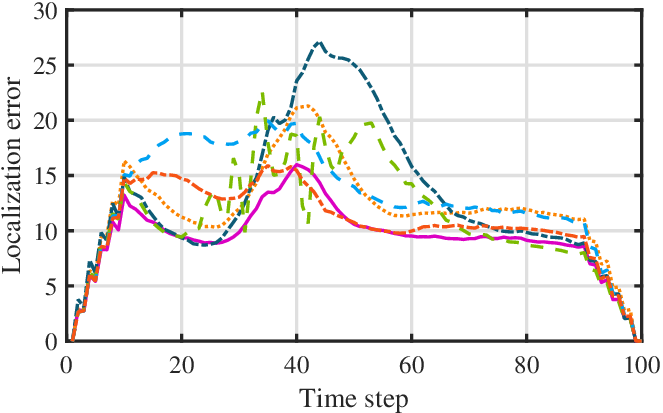}
	} \vspace{-0.3cm}
	\subfigure{
	  \includegraphics[width=1.6in]{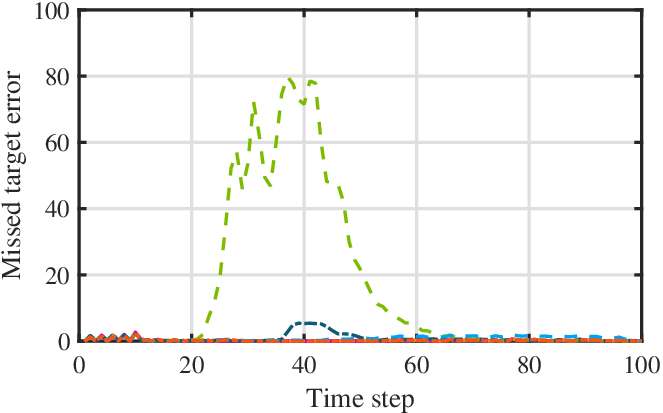}
	}
	\subfigure{
	  \includegraphics[width=1.6in]{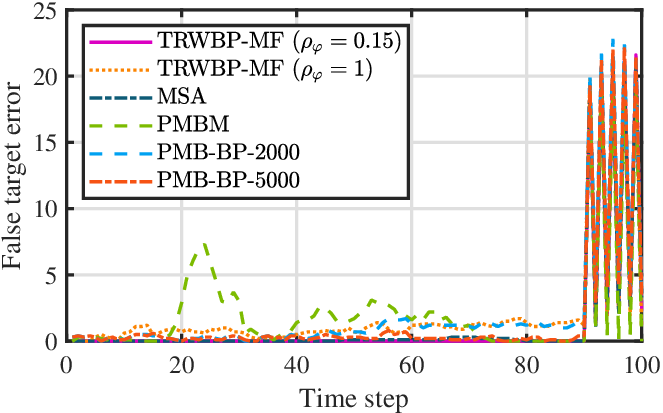}
	}
	\caption{GOSPA errors and their components for scenario 1 given $p_D = 0.95$.} 
	\label{fig_gospaS1}
\end{figure}

\begin{figure}[tbp]
	\centering
	\subfigure{
	  \includegraphics[width=1.6in]{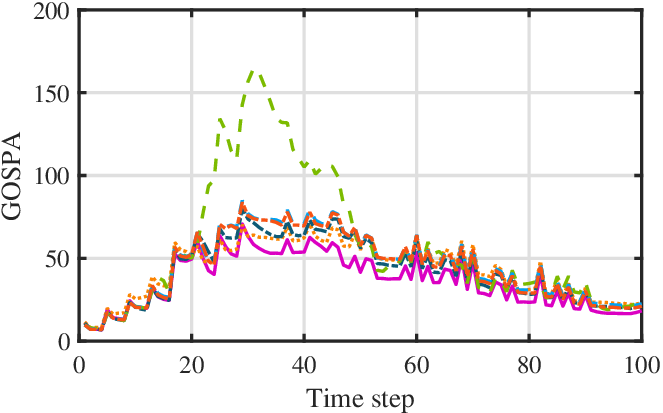}
	} \vspace{-0.3cm}
	\subfigure{
	  \includegraphics[width=1.6in]{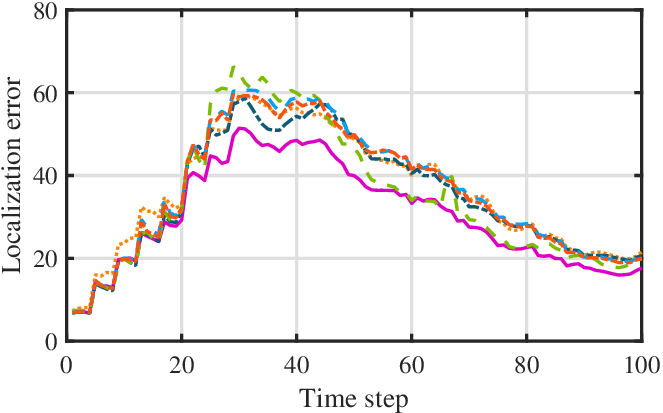}
	} \vspace{-0.3cm}
	\subfigure{
	  \includegraphics[width=1.6in]{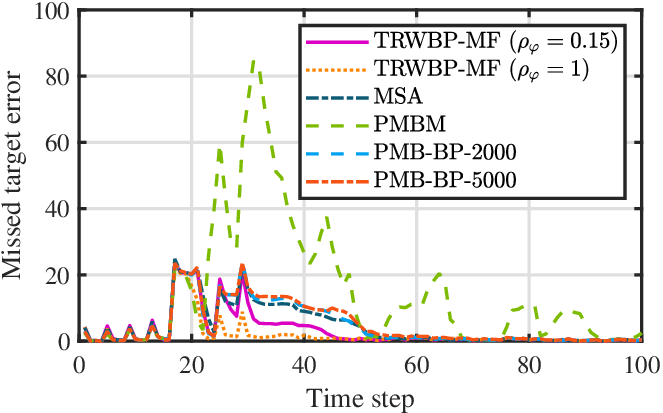}
	}
	\subfigure{
	  \includegraphics[width=1.6in]{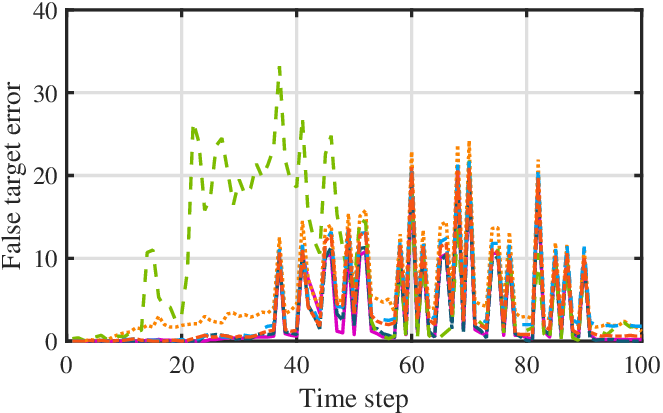}
	}
	\caption{GOSPA errors and their components for scenario 2 given $p_D = 0.95$.} 
	\label{fig_gospaS2}
\end{figure}

Under $p_D = 0.95$ and $\rho_\varphi = 0.15$, Figs. \ref{fig_gospaS1} and \ref{fig_gospaS2} show the GOSPA errors and their components for the two scenarios. For scenario 1, PMBM exhibits a significant increase in missed target error when targets intersect around step $k=40$. This degradation results from the poor performance of the distance-based clustering algorithm which struggles to appropriately group closely-spaced measurements. As all target originated measurements tend to be grouped into a single cluster, most target are updated with no measurement and declared to be undetected. In contrast, the probabilistic association-based TRWBP-MF, MSA and PMB-BP, which avoid measurement clustering, are more robust, though their localization errors increase due to higher uncertainty in data association. At the intersection region, the localization errors of TRWBP-MF and MSA increase more sharply than that of particle-based PMB-BP because plenty of particles overcome the association uncertainty better than a single GGIW component. Compared to MSA, TRWBP-MF significantly reduces the localization error when targets intersect, especially for $\rho_\varphi = 0.15$. For scenario 2, TRWBP-MF with $\rho_\varphi = 0.15$ outperforms the other algorithms in terms of GOSPA and localization errors. TRWBP-MF, MSA and PMB-BP are comparable regarding missed and false target errors, whereas PMBM experiences significantly larger missed and false target errors at steps when some targets are in close proximity. Furthermore, the localization performance of TRWBP-MF improves as FAP $\rho_\varphi$ decreases from 1 to 0.15 across both scenarios. This highlights the critical role of $\rho_\varphi$ in enhancing state estimation for the unified TRWBP-MF approach, warranting further investigation into its optimization rather than setting it as a constant.

Table \ref{tab_statistics} summarizes the accumulated GOSPA errors and their components, as well as runtimes (with MATLAB implementations on an Intel Core i9-10900F CPU) over all time steps for $p_D\in\{0.85, 0.95\}$. Overall, TRWBP-MF with $\rho_\varphi = 0.15$ outperforms MSA, PMBM and PMB-BP in terms of GOSPA and localization error, as well as computational efficiency. The localization and false target errors of TRWBP-MF decrease as $\rho_\varphi$ varies from 1 to 0.15. For PMB-BP, employing more particles reduces localization error but comes at the price of higher computational load. TRWBP-MF significantly reduces computational time in comparison to PMBM and PMB-BP for scenario 2, demonstrating its superior scalability for a large number of targets.

\begin{table*}[tbp]
	\renewcommand{\arraystretch}{1.1}
	\centering
	\caption{Mean GOSPA Error and Runtime in Seconds (Both Summed Over All the Time Steps)}
	\label{tab_statistics}
	\resizebox{\linewidth}{!}
	{
		\begin{tabular}{cc|ccccc|ccccc} \hline\hline
			\multirow{2}{*}{$p_D$} & \multirow{2}{*}{Algorithm} & \multicolumn{5}{c|}{Scenario 1} & \multicolumn{5}{c}{Scenario 2}\\ \cline{3-12}
			& & GOSPA & Localization & Missed & False & Runtime & GOSPA & Localization & Missed & False & Runtime \\ \hline
			\multirow{6}{*}{0.95} & TRWBP-MF ($\rho_\varphi=0.15$) & \textbf{1078.0} & \textbf{942.1} & 23.0 & \textbf{112.9} & 5.4 & \textbf{3656.4} & \textbf{3030.4} & 321.1 & 304.9 & \textbf{12.4} \\
			& TRWBP-MF ($\rho_\varphi=1$) & 1407.7 & 1193.0 & 26.7 & 188.0 & \textbf{4.1} & 4359.4 & 3625.3 & \textbf{189.0} & 545.1 & 13.8 \\
			& MSA & 1491.8 & 1289.0 & 83.9 & 118.9 & 6.9 & 4259.3 & 3505.4 & 449.4 & \textbf{304.5} & 77.0 \\
			& PMBM & 2905.5 & 1132.8 & 1554.2 & 218.5 & 76.9 & 5781.1 & 3447.9 & 1499.1 & 834.1 & 2030.8  \\
			& PMB-BP-2000 & 1577.9 & 1300.3 & 83.3 & 194.3 & 46.8 &  4535.7 & 3654.7 & 464.0 & 417.0 & 454.7\\ 
			& PMB-BP-5000 & 1238.8 & 1071.1 & \textbf{21.2} & 146.5 & 114.7 & 4453.4 & 3592.1 & 495.8 & 365.5 & 1379.9  \\
			\hline
			\multirow{6}{*}{0.85} & TRWBP-MF ($\rho_\varphi=0.15$) & \textbf{1341.7} & \textbf{1086.3} & \textbf{35.6} & 219.8 & 5.0 & \textbf{4181.8} & \textbf{3259.0} & 381.7 & \textbf{541.1} & 10.9 \\
			& TRWBP-MF ($\rho_\varphi=1$) & 1744.6 & 1266.3 & 42.4 & 435.9 & \textbf{4.8} & 5653.0 & 4034.7 & \textbf{266.4} & 1351.9 & \textbf{10.4} \\
			& MSA & 1585.3 & 1290.6 & 79.6 & 215.1 & 6.3 & 4691.2 & 3597.9 & 529.9 & 563.4 & 65.5 \\
			& PMBM & 2964.0 & 1165.7 & 1606.8 & \textbf{191.5} & 113.3 & 6353.1 & 3796.7 & 1120.5 & 1435.9 & 2998.5  \\
			& PMB-BP-2000 & 1642.4 & 1304.9 & 52.1 & 285.4 & 41.9 & 5207.4 & 3898.7 & 514.3 & 794.4 & 412.0\\ 
			& PMB-BP-5000 & 1484.2 & 1169.1 & 40.6 & 274.5 & 100.5  & 5199.0 & 3857.1 & 542.6 & 799.3 & 1387.4 \\
			\hline\hline
		\end{tabular}
	}
\end{table*} 

\section{Conclusion} \label{section_six}
In this paper, we present a unified TRWBP-MF approach for tracking extended targets, which combines the TRWBP and MF approximation using the region-based free energy approximation. With GGIW prior for the joint density of measurement rate, kinematic state and extent, and under linear Gaussian assumptions of the target models, a closed-form tracker is derived using the unified TRWBP-MF approach. The TRWBP provides a probabilistic solution to data association that avoids measurement clustering, offering improved performance than ordinary BP, particularly for tracking closely-spaced and large-scale extended targets. Although the proposed algorithm is implemented with constant FAPs without optimization, simulation results from two challenging scenarios demonstrate that it outperforms the MSA, and the PMBM and PMB-BP filters, while significantly reducing computational load. Future work will focus on optimizing the FAPs within the unified TRWBP-MF approach to further enhance tracking performance.

\begingroup
\footnotesize 
\bibliographystyle{IEEEtran}
\balance
\bibliography{reference.bib}
\endgroup

\newpage 

\onecolumn  
\begin{center}

	{\linespread{1.5}\selectfont \huge \text{Unifying Tree-Reweighted Belief Propagation and} \text{ Mean Field for Tracking Extended Targets: } \text{Supplementary Material} }
	
	\text{{Weizhen Ma, Zhongliang Jing, Peng Dong, Henry Leung}}
\end{center}

\setcounter{section}{0}  
\setcounter{equation}{0}  
\setcounter{page}{1}
This manuscript provides detailed derivations for some important equations in the manuscript ``Unifying Tree-Reweighted Belief Propagation and Mean Field for Tracking Extended Targets'' \cite{MJDL} by the same authors. The notations follow these in the original manuscript.

\subsection{The free energy \texorpdfstring{$F_{\mathrm{BP, MF}}$}{}}
For the valid region set $\mathcal{R}_{\mathrm{BP}, \mathrm{MF}} = \{ R_{\mathrm{MF}} \} \cup \{R_i| i\in \mathcal{I}_{\mathrm{BP}} \}\cup\{ R_j | j\in \mathcal{J}_{\mathrm{BP}} \}$ defined in \cite{MJDL}, we first compute the region entropy and region average energy for each region according to equation (14) in \cite{MJDL}. For region $R_{\mathrm{MF}}$, its region average energy $U_{R_{\mathrm{MF}}}(q_{R_{\mathrm{MF}}})$ and entropy $H_{R_{\mathrm{MF}}}(q_{R_{\mathrm{MF}}})$ are given by
\begin{equation} \label{eqn_RMF_energyEntropy}
	U_{R_{\mathrm{MF}}}(q_{R_{\mathrm{MF}}}) = -\sum_{j\in\mathcal{J}_{\mathrm{MF}}} \sum_{\boldsymbol{v}_j} \prod_{i\in\mathcal{N}(j)}q_i(v_i) \ln f_j(\boldsymbol{v}_j), \quad
	H_{R_{\mathrm{MF}}}(q_{R_{\mathrm{MF}}}) = -\sum_{i\in\mathcal{I}_{\mathrm{MF}}} \sum_{v_i} q_i(v_i) \ln q_i(v_i) 
\end{equation}
where the MF factorization $q(\boldsymbol{v}_j) \!=\! \prod_{i\in\mathcal{N}(j)} q_i(v_i)$ has been utilized for $U_{R_{\mathrm{MF}}}(q_{R_{\mathrm{MF}}})$. For the region $R_i$, $i\in \mathcal{I}_{\mathrm{BP}}$, its region average energy $U_{R_i}(q_{R_i})$ and entropy $H_{R_i}(q_{R_i})$ are given by
\begin{equation} \label{eqn_Ri_energyEntropy}
	U_{R_i}(q_{R_i}) = 0, \quad H_{R_i}(q_{R_i}) = -\sum_{v_i} q_i(v_i) \ln q_i(v_i).
\end{equation}
For the region $R_j$, $j\in \mathcal{J}_{\mathrm{BP}}$, its region average energy $U_{R_j}(q_{R_j})$ is given by 
\begin{equation} \label{eqn_Rj_energy} 
	U_{R_j}(q_{R_j}) = -\sum_{\boldsymbol{v}_j} q_j(\boldsymbol{v}_j) \ln f_j(\boldsymbol{v}_j). 
\end{equation}
The region entropy $H_{R_j}(q_{R_j})$ of $R_j$ is approximated as follows. For a function $p(\boldsymbol{v})=\prod_{j\in\mathcal{J}}f_j(\boldsymbol{v}_j)$, a convex upper bound on the entropy $H(q)$ of the approximate densities $q(\cdot)$ is given by \cite{WPS2011}
\begin{equation} \label{eqn_trwbp_entropyBound}
	H(q;\boldsymbol{\rho}) = -\sum_{i\in\mathcal{I}} \sum_{v_i} q_i(v_i) \ln q_i(v_i) - \sum_{j\in\mathcal{J}}\rho_j \sum_{\boldsymbol{v}_j}q_j(\boldsymbol{v}_j)\ln\frac{q_j(\boldsymbol{v}_j)}{\prod_{l\in \mathcal{N}(j)}q_l(v_l)}
\end{equation}
where $\boldsymbol{\rho} = (\rho_j|j\in\mathcal{J})^T$ is the factor appearance probability vector. Exploiting the marginalization constraint $q_i(v_i) = \sum_{\boldsymbol{v}_j\backslash v_i}q_j(\boldsymbol{v}_j)$ for $i\in\mathcal{N}(j)$, we have
\begin{equation} \label{eqn_partMutualInfo}
	\sum_{\boldsymbol{v}_j}q_j(\boldsymbol{v}_j)\ln \prod_{i\in \mathcal{N}(j)} q_i(v_i) \!=\! \sum_{\boldsymbol{v}_j} \sum_{i\in \mathcal{N}(j)} q_j(\boldsymbol{v}_j)\ln q_i(v_i) \!=\! \sum_{i\in \mathcal{N}(j)} \sum_{\boldsymbol{v}_j \backslash v_i}  q_j(\boldsymbol{v}_j)\ln q_i(v_i) \!=\! \sum_{i\in \mathcal{N}(j)} \sum_{ v_i} q_i(v_i)\ln q_i(v_i). 
\end{equation}
Using \eqref{eqn_partMutualInfo} in \eqref{eqn_trwbp_entropyBound}, we obtain 
\begin{equation}
	\begin{aligned}
		H(q;\boldsymbol{\rho}) &= -\sum_{j\in\mathcal{J}}\rho_j \sum_{\boldsymbol{v}_j}q_j(\boldsymbol{v}_j)\ln q_j(\boldsymbol{v}_j) -\sum_{i\in\mathcal{I}} \sum_{v_i} q_i(v_i) \ln q_i(v_i) + \sum_{j\in\mathcal{J}}\rho_j \sum_{i\in \mathcal{N}(j)} \sum_{ v_i} q_i(v_i)\ln q_i(v_i) \\
		&= -\sum_{j\in\mathcal{J}}\rho_j \sum_{\boldsymbol{v}_j}q_j(\boldsymbol{v}_j)\ln q_j(\boldsymbol{v}_j) -\sum_{i\in\mathcal{I}} \sum_{v_i} q_i(v_i) \ln q_i(v_i) + \sum_{i\in\mathcal{I}} \sum_{j\in \mathcal{N}(i)} \rho_j \sum_{ v_i} q_i(v_i)\ln q_i(v_i) \\
		&= -\sum_{j\in\mathcal{J}}\rho_j \sum_{\boldsymbol{v}_j}q_j(\boldsymbol{v}_j)\ln q_j(\boldsymbol{v}_j) + \sum_{i\in\mathcal{I}} (\varrho_i - 1) \sum_{ v_i} q_i(v_i)\ln q_i(v_i) 
	\end{aligned}
\end{equation} 
where $\varrho_i = \sum_{j\in \mathcal{N}(i)} \rho_j$. For region $R_j$, a convex upper bound on the region entropy $H_{R_j}(q_{R_j})$ is then given by
\begin{equation} \label{eqn_Rj_entropy} 
	H_{R_j}(q_{R_j}, \boldsymbol{\rho}) = - \rho_j \sum_{\boldsymbol{v}_j} q_j(\boldsymbol{v}_j) \ln q_j(\boldsymbol{v}_j) + \sum_{i\in\mathcal{N}(j)}(\rho_j - 1) \sum_{v_i} q_i(v_i) \ln q_i(v_i) .
\end{equation}
which will be used as an approximation to $H_{R_j}(q_{R_j})$. 

For the regions $R_{\mathrm{MF}}$, $R_i$ and $R_j$ in the set $\mathcal{R}_{\mathrm{BP}, \mathrm{MF}}$, substituting their region average energy and entropy in \eqref{eqn_RMF_energyEntropy}-\eqref{eqn_Rj_energy}, and \eqref{eqn_Rj_entropy}, and their associated countering numbers $c_{R_{\mathrm{MF}}} = 1$, $c_{R_i} = 1-|\mathcal{N}_{\mathrm{BP}}(i)|-\mathbb{I}_{\mathcal{I}_{\mathrm{MF}}}(i)$ and $c_{R_{j}} = 1$ as presented in \cite{MJDL} into the region-based free energy $F_{\mathcal{R}} = U_{\mathcal{R}} - H_{\mathcal{R}}$, i.e., equation (15) in \cite{MJDL}, we obtain the region-based free energy approximation for the unified TRWBP-MF approach as
\begin{equation}
	\begin{aligned}
		F_{\mathcal{R}_{\mathrm{BP}, \mathrm{MF}}} =& - \sum_{j\in\mathcal{J}_{\mathrm{MF}}} \sum_{\boldsymbol{v}_j} \prod_{i\in\mathcal{N}(j)}q_i(v_i) \ln f_j(\boldsymbol{v}_j) + \sum_{i\in\mathcal{I}_{\mathrm{MF}}} \sum_{v_i} q_i(v_i) \ln q_i(v_i) \\
		& + \sum_{i\in\mathcal{I}_{\mathrm{BP}}} (1-|\mathcal{N}_{\mathrm{BP}}(i)|-\mathbb{I}_{\mathcal{I}_{\mathrm{MF}}}(i)) \sum_{v_i} q_i(v_i) \ln q_i(v_i) + \sum_{j\in \mathcal{J}_{\mathrm{BP}}} \rho_j \sum_{\boldsymbol{v}_j} q_j(\boldsymbol{v}_j) \ln q_j(\boldsymbol{v}_j) \\
		& + \sum_{j\in \mathcal{J}_{\mathrm{BP}}}(1 - \rho_j ) \sum_{i\in\mathcal{N}(j)} \sum_{v_i} q_i(v_i) \ln q_i(v_i) - \sum_{j\in\mathcal{J}_{\mathrm{BP}}} \sum_{\boldsymbol{v}_j} q_j(\boldsymbol{v}_j) \ln f_j(\boldsymbol{v}_j)  \\
		=& \sum_{j\in \mathcal{J}_{\mathrm{BP}}} \rho_j \sum_{\boldsymbol{v}_j} q_j(\boldsymbol{v}_j) \ln q_j(\boldsymbol{v}_j) -\sum_{j\in\mathcal{J}_{\mathrm{BP}}} \sum_{\boldsymbol{v}_j} q_j(\boldsymbol{v}_j) \ln f_j(\boldsymbol{v}_j) - \sum_{j\in\mathcal{J}_{\mathrm{MF}}} \sum_{\boldsymbol{v}_j} \prod_{i\in\mathcal{N}(j)}q_i(v_i) \ln f_j(\boldsymbol{v}_j)\\
		& + \sum_{i\in\mathcal{I}_{\mathrm{MF}}} \sum_{v_i} q_i(v_i) \ln q_i(v_i) + \sum_{i\in\mathcal{I}_{\mathrm{BP}}} (1-|\mathcal{N}_{\mathrm{BP}}(i)|-\mathbb{I}_{\mathcal{I}_{\mathrm{MF}}}(i)) \sum_{v_i} q_i(v_i) \ln q_i(v_i) \\
		& + \sum_{i\in\mathcal{I}_{\mathrm{BP}}} (|\mathcal{N}_{\mathrm{BP}}(i)| - \varrho_i) \sum_{v_i} q_i(v_i) \ln q_i(v_i) \\
		=& \sum_{j\in \mathcal{J}_{\mathrm{BP}}} \rho_j \sum_{\boldsymbol{v}_j} q_j(\boldsymbol{v}_j) \ln q_j(\boldsymbol{v}_j) -\sum_{j\in\mathcal{J}_{\mathrm{BP}}} \sum_{\boldsymbol{v}_j} q_j(\boldsymbol{v}_j) \ln f_j(\boldsymbol{v}_j) - \sum_{j\in\mathcal{J}_{\mathrm{MF}}} \sum_{\boldsymbol{v}_j} \prod_{i\in\mathcal{N}(j)}q_i(v_i) \ln f_j(\boldsymbol{v}_j)\\
		& + \sum_{i\in\mathcal{I}_{\mathrm{MF}}} \sum_{v_i} q_i(v_i) \ln q_i(v_i) + \sum_{i\in\mathcal{I}_{\mathrm{BP}}} (1 - \mathbb{I}_{\mathcal{I}_{\mathrm{MF}}}(i) - \varrho_i) \sum_{v_i} q_i(v_i) \ln q_i(v_i) \\
		=& \sum_{j\in \mathcal{J}_{\mathrm{BP}}} \rho_j \sum_{\boldsymbol{v}_j} q_j(\boldsymbol{v}_j) \ln q_j(\boldsymbol{v}_j) -\sum_{j\in\mathcal{J}_{\mathrm{BP}}} \sum_{\boldsymbol{v}_j} q_j(\boldsymbol{v}_j) \ln f_j(\boldsymbol{v}_j) - \sum_{j\in\mathcal{J}_{\mathrm{MF}}} \sum_{\boldsymbol{v}_j} \prod_{i\in\mathcal{N}(j)}q_i(v_i) \ln f_j(\boldsymbol{v}_j)\\
		& - \sum_{i\in\mathcal{I}} (\varrho_i - 1) \sum_{v_i} q_i(v_i) \ln q_i(v_i).
	\end{aligned}
\end{equation}

\subsection{Theorem 1}
We present the proof of Theorem 1 in \cite{MJDL} as follows. The Lagrangian for the unified TRWBP and mean field (MF) approach, referred to as TRWBP-MF, is given by \cite{MJDL}
\begin{equation} \label{eqn_lagrangian}
	\begin{aligned}
		L_{\mathrm{BP}, \mathrm{MF}} =& \sum_{j\in\mathcal{J}_\mathrm{BP}} \rho_j \sum_{\boldsymbol{v}_j} q_j(\boldsymbol{v}_j) \ln q_j(\boldsymbol{v}_j) - \sum_{j\in\mathcal{J}_\mathrm{BP}} \sum_{\boldsymbol{v}_j} q_j(\boldsymbol{v}_j) \ln f_j(\boldsymbol{v}_j) - \sum_{j\in\mathcal{J}_{\mathrm{MF}}}\sum_{\boldsymbol{v}_j}\prod_{i\in\mathcal{N}(j)} q_i(v_i) \ln f_j(\boldsymbol{v}_j) \\
		& - \sum_{i\in\mathcal{I}} (\varrho_i - 1) \sum_{v_i} q_i(v_i) \ln q_i(v_i) - \sum_{j\in \mathcal{J}_{\mathrm{BP}}} \sum_{i\in\mathcal{N}(j)} \sum_{v_i} \lambda_{j,i}(v_i) \Big( q_i(v_i) - \sum_{\boldsymbol{v}_j \backslash v_i} q_j(\boldsymbol{v}_j) \Big).
	\end{aligned}
\end{equation}
The stationary points of the Lagrangian in $\eqref{eqn_lagrangian}$ are then obtained by setting the derivatives of $L_{\mathrm{BP}, \mathrm{MF}}$ with respect to the beliefs $q_i(v_i)$ and $q_j(\boldsymbol{v}_j)$ equal to zero. These derivatives are given by
\begin{equation}
	\begin{aligned}
		\frac{\partial L_{\mathrm{BP}, \mathrm{MF}}}{\partial q_i(v_i)} &= \ln q_i(v_i) - \sum_{j\in\mathcal{N}_{\mathrm{MF}}(i)} \sum_{\boldsymbol{v}_j \backslash v_i} \prod_{l\in\mathcal{N}(j)\backslash i} q_l(v_l) \ln f_{j}(\boldsymbol{v}_j) + \sum_{j\in\mathcal{N}_\mathrm{BP}(i)} \lambda_{j,i}(v_i) + c_i \\
		\frac{\partial L_{\mathrm{BP}, \mathrm{MF}}}{\partial q_j(\boldsymbol{v}_j)} &= \rho_j\ln q_j(\boldsymbol{v}_j) - \rho_j \ln \prod_{i\in \mathcal{N}(j) }q_i(v_i) - \ln f_j(\boldsymbol{v}_j)-\sum_{i\in \mathcal{N}(j) } \lambda_{j,i}(v_i) + c_j
	\end{aligned}
\end{equation}
where $c_i$ and $c_j$ are constants. 
Setting the above derivatives to zero and taking exponentiations, we get the following fixed-point equations for the stationary points:
\begin{equation} \label{eqn_fixedPoint_qi}
	q_i(v_i) = c_i \prod_{j\in\mathcal{N}_\mathrm{BP}(i)} \exp \left(-\lambda_{j,i}(v_i)\right) \prod_{j\in\mathcal{N}_{\mathrm{MF}}(i)} \exp \left( \sum_{\boldsymbol{v}_j \backslash v_i} \prod_{l\in\mathcal{N}(j) \backslash i} q_l(v_l) \ln f_j(\boldsymbol{v}_j) \right) 
\end{equation}
\begin{equation} \label{eqn_fixedPoint_qj}
	q_j(\boldsymbol{v}_j) = c_j f^{\frac{1}{\rho_j}}_j(\boldsymbol{v}_j) \prod_{i\in \mathcal{N}(j) } q_i(v_i) \exp\left(\frac{\lambda_{j,i}(v_i)}{\rho_j}\right).
\end{equation}

By introducing the following messages 
\begin{equation}
	\begin{aligned}
		m^{\mathrm{BP}}_{j\to i}(v_i) &= \exp\left(-\frac{\lambda_{j,i}(v_i)}{\rho_j}\right), \quad \text{for } j\in\mathcal{J}_{\mathrm{BP}}, i\in\mathcal{N}(j) \\
		m^{\mathrm{MF}}_{j\to i}(v_i) &= \exp\left( \sum_{\boldsymbol{v}_j \backslash v_i} \prod_{l\in\mathcal{N}(j)\backslash i} q_l(x_l) \ln f_j(\boldsymbol{v}_j) \right),  \quad \text{for } j\in\mathcal{J}_{\mathrm{MF}}, i\in\mathcal{N}(j)
	\end{aligned}
\end{equation}
we can rewrite \eqref{eqn_fixedPoint_qi} and \eqref{eqn_fixedPoint_qj} as
\begin{equation} \label{eqn_trwbpmf_belief_qi}
	q_i(v_i) = c_i \prod_{j\in\mathcal{N}_\mathrm{BP}(i)} \left( m^{\mathrm{BP}}_{j\to i}(v_i) \right)^{\rho_j}  \prod_{j\in\mathcal{N}_{\mathrm{MF}}(i)} m^{\mathrm{MF}}_{j\to i}(v_i), \quad \text{for } i\in\mathcal{I}
\end{equation}
\begin{equation} \label{eqn_fixedPoint_qj_intermediate}
	q_j(\boldsymbol{v}_j) = c_j f^{\frac{1}{\rho_j}}_j(\boldsymbol{v}_j) \prod_{i\in \mathcal{N}(j)} \frac{q_i(v_i) }{m^{\mathrm{BP}}_{j\to i}(v_i)}, \quad \text{for } j\in\mathcal{J}_{\mathrm{BP}}.
\end{equation}

Next we define the messages 
\begin{equation} \label{eqn_trwbpmf_message_i2j}
	m_{i\to j}(v_i) = (m^{\mathrm{BP}}_{j\to i}(v_i))^{-1} \prod_{l\in\mathcal{N}_\mathrm{BP}(i)} \left( m^{\mathrm{BP}}_{l\to i}(v_i) \right)^{\rho_l} \prod_{l\in\mathcal{N}_{\mathrm{MF}}(i)} m^{\mathrm{MF}}_{l\to i}(v_i)
\end{equation}
for all $j\in\mathcal{J}$, $i\in\mathcal{N}(j)$. We can rewrite $q_j(\boldsymbol{v}_j)$ in \eqref{eqn_fixedPoint_qj_intermediate} as
\begin{equation} \label{eqn_trwbpmf_belief_qj}
	q_j(\boldsymbol{v}_j) = c_j f^{\frac{1}{\rho_j}}_j(\boldsymbol{v}_j) \prod_{i\in \mathcal{N}(j)} m_{i\to j}(v_i), \quad \text{for } j\in\mathcal{J}_{\mathrm{BP}}.
\end{equation}
Using the marginalization constraint (16) in \cite{MJDL}, and the above formula, we further obtain
\begin{equation} \label{eqn_n_times_m}
	\begin{aligned}
		m_{i\to j}(x_i) m^{\mathrm{BP}}_{j\to i}(v_i) &= \prod_{j\in\mathcal{N}_\mathrm{BP}(i)} \left( m^{\mathrm{BP}}_{j\to i}(v_i) \right)^{\rho_j}  \prod_{j\in\mathcal{N}_{\mathrm{MF}}(i)} m^{\mathrm{MF}}_{j\to i}(v_i) \\
		&= \frac{1}{c_i} q_i(v_i) \\
        &= \frac{1}{c_i} \sum_{\boldsymbol{v}_j \backslash v_i} q_j(\boldsymbol{v}_j) \\
        &= \frac{c_j}{c_i} \sum_{\boldsymbol{v}_j \backslash v_i} f^{\frac{1}{\rho_j}}_j(\boldsymbol{v}_j) \prod_{l\in \mathcal{N}(j)} m_{l\to j}(x_l)
	\end{aligned}
\end{equation}
for all $ j\in\mathcal{J}_{\mathrm{BP}}$, $i\in\mathcal{N}(j)$. Dividing both sides of \eqref{eqn_n_times_m} by $m_{i\to j}(v_i)$ gives                
\begin{equation} \label{eqn_trwbpmf_message_bp}
	m^{\mathrm{BP}}_{j\to i}(v_i) =  \sum_{\boldsymbol{v}_j \backslash v_i} f^{\frac{1}{\rho_j}}_j(\boldsymbol{v}_j) \prod_{l\in \mathcal{N}(j) \backslash i} m_{l\to j}(v_l)
\end{equation}
where we have dropped the constant $c_j/c_i$ in \eqref{eqn_n_times_m} as BP messages can be rescaled by an arbitrary positive constant without changing the beliefs. Noting that $m_{i\to j}(v_i) = q_i(v_i)$ for all  $ j\in\mathcal{J}_{\mathrm{MF}}$ and $i\in\mathcal{N}(j)$, we can write the messages $m^{\mathrm{MF}}_{j\to i}(v_i)$ as
\begin{equation} \label{eqn_trwbpmf_message_mf}
	m^{\mathrm{MF}}_{j\to i}(v_i) = \exp\left( \sum_{\boldsymbol{v}_j \backslash v_i} \prod_{l\in\mathcal{N}(j)\backslash i} m_{l\to j}(v_l) \ln f_j(\boldsymbol{v}_j) \right)
\end{equation}
for all $j\in\mathcal{J}_{\mathrm{MF}}$, $i\in\mathcal{N}(j)$. Note that \eqref{eqn_trwbpmf_belief_qi} and \eqref{eqn_trwbpmf_belief_qj} are equivalent to the equation (20) in \cite{MJDL}, and \eqref{eqn_trwbpmf_message_i2j}, \eqref{eqn_trwbpmf_message_bp} and \eqref{eqn_trwbpmf_message_mf} are equivalent to the equation (21) in \cite{MJDL}. This completes the proof that stationary points of the Lagrangian in (20) of \cite{MJDL} are fixed points satisfying (21) of \cite{MJDL}. Since the above derivation steps are reversible, this also completes the proof of Theorem 1 in \cite{MJDL}.  

\subsection{The rescaled data association messages}
The derivations of the rescaled data association messages (39)-(44) in \cite{MJDL} are as follows. The data association messages $\underline{m}^{\mathrm{BP},[\iota]}_{nl \leftarrow m } (\underline{a}_{nl})$, $\underline{m}^{\mathrm{BP},[\iota]}_{\to nl}(\underline{a}_{nl})$, $\underline{m}^{\mathrm{BP},[\iota]}_{nl\to m}(\boldsymbol{b}_m)$, $\overline{m}^{\mathrm{BP},[\iota]}_{ml \leftarrow o } (\overline{a}_{ml})$, $\overline{m}^{\mathrm{BP},[\iota]}_{\to ml}(\overline{a}_{ml})$ and $\overline{m}^{\mathrm{BP},[\iota]}_{ml\to o}(\boldsymbol{b}_o)$ are given by equations (32)-(38) in \cite{MJDL}. Sine each message only comprises of two distinct values, let
\begin{equation} \label{eqn_distinctValues}
	\begin{aligned}
		\underline{m}^{\mathrm{BP},[\iota],1}_{nl \leftarrow m } &= \underline{m}^{\mathrm{BP},[\iota]}_{nl \leftarrow m } (\underline{a}_{nl} = m), \quad 
		\underline{m}^{\mathrm{BP},[\iota],0}_{nl \leftarrow m } = \underline{m}^{\mathrm{BP},[\iota]}_{nl \leftarrow m } (\underline{a}_{nl} \neq m) \\
		\underline{m}^{\mathrm{BP}, [\iota], 1}_{\to nl} &= \underline{m}^{\mathrm{BP},[\iota]}_{\to nl}(\underline{a}_{nl} \neq 0), \quad 
		\underline{m}^{\mathrm{BP}, [\iota], 0}_{\to nl} = \underline{m}^{\mathrm{BP},[\iota]}_{\to nl}(\underline{a}_{nl} = 0) \\
		\underline{m}^{\mathrm{BP},[\iota], 1}_{nl\to m} &= \underline{m}^{\mathrm{BP},[\iota]}_{nl\to m}(\boldsymbol{b}_m = [n~l]^T), \quad 
		\underline{m}^{\mathrm{BP},[\iota], 0}_{nl\to m} = \underline{m}^{\mathrm{BP},[\iota]}_{nl\to m}(\boldsymbol{b}_m \neq [n~l]^T) \\
		\overline{m}^{\mathrm{BP},[\iota], 1}_{ml \leftarrow o } &= \overline{m}^{\mathrm{BP},[\iota]}_{ml \leftarrow o } (\overline{a}_{ml} = o), \quad
		\overline{m}^{\mathrm{BP},[\iota], 0}_{ml \leftarrow o } = \overline{m}^{\mathrm{BP},[\iota]}_{ml \leftarrow o } (\overline{a}_{ml} \neq o)\\
		\overline{m}^{\mathrm{BP},[\iota], 1}_{\to ml} &= \overline{m}^{\mathrm{BP},[\iota]}_{\to ml}(\overline{a}_{ml} \neq 0), \quad
		\overline{m}^{\mathrm{BP},[\iota], 0}_{\to ml} = \overline{m}^{\mathrm{BP},[\iota]}_{\to ml}(\overline{a}_{ml} = 0) \\
		\overline{m}^{\mathrm{BP},[\iota], 1}_{ml\to o} &= \overline{m}^{\mathrm{BP},[\iota]}_{ml\to o}(\boldsymbol{b}_o = [\underline{N}+m~l]^T), \quad 
		\overline{m}^{\mathrm{BP},[\iota], 0}_{ml\to o} = \overline{m}^{\mathrm{BP},[\iota]}_{ml\to o}(\boldsymbol{b}_o \neq [\underline{N}+m~l]^T).
	\end{aligned}
\end{equation}

We rescale the messages of legacy PTs as follows
\begin{equation} \label{eqn_message_normalization_b2a_legacy}
	\underline{\mathrm{m}}^{\mathrm{BP},[\iota]}_{nl \leftarrow m }(\underline{a}_{nl}) = \frac{ \underline{m}^{\mathrm{BP},[\iota]}_{nl \leftarrow m } (\underline{a}_{nl}) } { \underline{m}^{\mathrm{BP},[\iota],0}_{nl \leftarrow m } } 
\end{equation}
\begin{equation} \label{eqn_message_normalization_h2a_legacy}
	\underline{\mathrm{m}}^{\mathrm{BP},[\iota]}_{\to nl}(\underline{a}_{nl}) = \frac{ \underline{m}^{\mathrm{BP},[\iota]}_{\to nl}(\underline{a}_{nl}) } { \underline{m}^{\mathrm{BP},[\iota], 0}_{\to nl} }
\end{equation}
\begin{equation} \label{eqn_message_normalization_a2b_legacy}
	\underline{\mathrm{m}}^{\mathrm{BP},[\iota]}_{nl\to m}(\boldsymbol{b}_m) = \frac{ \underline{m}^{\mathrm{BP},[\iota]}_{nl\to m}(\boldsymbol{b}_m) } { \underline{m}^{\mathrm{BP},[\iota], 0}_{nl\to m} }
\end{equation}
then we have $\underline{\mathrm{m}}^{\mathrm{BP},[\iota]}_{nl \leftarrow m }(\underline{a}_{nl} \neq m) = 1$, $\underline{\mathrm{m}}^{\mathrm{BP},[\iota]}_{\to nl}(\underline{a}_{nl} = 0) = 1$ and $\underline{\mathrm{m}}^{\mathrm{BP},[\iota]}_{nl\to m}(\boldsymbol{b}_m \neq [n~l]^T) = 1$. Thus, we only have to compute the following messages: $\underline{\mathrm{m}}^{\mathrm{BP},[\iota]}_{nl \leftarrow m } = \underline{\mathrm{m}}^{\mathrm{BP},[\iota]}_{nl \leftarrow m }(\underline{a}_{nl}=m)$, $\underline{\mathrm{m}}^{\mathrm{BP},[\iota]}_{\to nl} = \underline{\mathrm{m}}^{\mathrm{BP},[\iota]}_{\to nl}(\underline{a}_{nl} \neq 0)$ and $\underline{\mathrm{m}}^{\mathrm{BP},[\iota]}_{nl\to m} = \underline{\mathrm{m}}^{\mathrm{BP},[\iota]}_{nl\to m}(\boldsymbol{b}_m = [n~l]^T)$. We also rescale the messages of new PTs as follows
\begin{equation} \label{eqn_message_normalization_b2a_new}
	\overline{\mathrm{m}}^{\mathrm{BP},[\iota]}_{ml \leftarrow o } (\overline{a}_{ml}) = \frac{ \overline{m}^{\mathrm{BP},[\iota]}_{ml \leftarrow o } (\overline{a}_{ml}) } { \overline{m}^{\mathrm{BP},[\iota], 0}_{ml \leftarrow o } }
\end{equation}
\begin{equation} \label{eqn_message_normalization_h2a_new}
	\overline{\mathrm{m}}^{\mathrm{BP},[\iota]}_{\to ml}(\overline{a}_{ml}) = \frac{ \overline{m}^{\mathrm{BP},[\iota]}_{\to ml}(\overline{a}_{ml}) }{ \overline{m}^{\mathrm{BP},[\iota],0}_{\to ml} }
\end{equation}
\begin{equation} \label{eqn_message_normalization_a2b_new}
	\overline{\mathrm{m}}^{\mathrm{BP},[\iota]}_{ml\to o}(\boldsymbol{b}_o) = \frac{ \overline{m}^{\mathrm{BP},[\iota]}_{ml\to o}(\boldsymbol{b}_o) } { \overline{m}^{\mathrm{BP},[\iota],0}_{ml\to o} }
\end{equation}
then we have $\overline{\mathrm{m}}^{\mathrm{BP},[\iota]}_{ml \leftarrow o }(\overline{a}_{ml} \neq o) = 1$, $\overline{\mathrm{m}}^{\mathrm{BP},[\iota]}_{\to ml}(\overline{a}_{ml} = 0) = 1$ and $\overline{\mathrm{m}}^{\mathrm{BP},[\iota]}_{ml\to o}(\boldsymbol{b}_o \neq [\underline{N}+m~l]^T) = 1$, and we only have to compute the following messages: $\overline{\mathrm{m}}^{\mathrm{BP},[\iota]}_{ml \leftarrow o } = \overline{\mathrm{m}}^{\mathrm{BP},[\iota]}_{ml \leftarrow o } (\overline{a}_{ml} = o)$, $\overline{\mathrm{m}}^{\mathrm{BP},[\iota]}_{\to ml} = \overline{\mathrm{m}}^{\mathrm{BP},[\iota]}_{\to ml}(\overline{a}_{ml} \neq 0)$ and $\overline{\mathrm{m}}^{\mathrm{BP},[\iota]}_{ml\to o} = \overline{\mathrm{m}}^{\mathrm{BP},[\iota]}_{ml\to o}(\boldsymbol{b}_o = [\underline{N}+m~l]^T)$. 

\subsubsection{The rescaled message \texorpdfstring{$\underline{\mathrm{m}}^{\mathrm{BP},[\iota]}_{nl \leftarrow m }$}{}}
According to equation (32) in \cite{MJDL}, the two distinct values in \eqref{eqn_distinctValues} for message $\underline{m}^{\mathrm{BP},[\iota]}_{nl \leftarrow m } (\underline{a}_{nl})$ are given by 
\begin{equation} \label{eqn_unnormalized_message_b2a_legacy}
	\begin{aligned}
		\underline{m}^{\mathrm{BP},[\iota],1}_{nl \leftarrow m } &=  \left( \underline{m}^{\mathrm{BP},[\iota-1], 1}_{nl\to m} \right)^{\rho_{\varphi}-1} \prod_{\left(n', l'\right) \in \underline{\mathcal{B}}_m \backslash (n, l)} \left( \underline{m}^{\mathrm{BP},[\iota-1], 0}_{n'l'\to m} \right)^{\rho_{\varphi}}  \prod_{\left(n'', l''\right) \in \overline{\mathcal{B}}_m } \left( \overline{m}^{\mathrm{BP},[\iota-1],0}_{n''l''\to m} \right)^{\rho_{\varphi}}  \\
		\underline{m}^{\mathrm{BP},[\iota],0}_{nl \leftarrow m } &= \left( \underline{m}^{\mathrm{BP},[\iota-1], 0}_{nl\to m} \right)^{\rho_{\varphi}-1} \sum_{\boldsymbol{b}_m \in \mathcal{B}_m \backslash (n, l)}  \prod_{\left(n', l'\right) \in \underline{\mathcal{B}}_m \backslash (n, l)} \left( \underline{m}^{\mathrm{BP},[\iota-1]}_{n'l'\to m} (\boldsymbol{b}_m)  \right)^{\rho_{\varphi}} \prod_{\left(n'', l''\right) \in \overline{\mathcal{B}}_m } \left( \overline{m}^{\mathrm{BP},[\iota-1]}_{n''l'' \to m}(\boldsymbol{b}_m)  \right)^{\rho_{\varphi}}.
	\end{aligned}
\end{equation}
Substituting \eqref{eqn_unnormalized_message_b2a_legacy} into \eqref{eqn_message_normalization_b2a_legacy}, and using \eqref{eqn_message_normalization_a2b_legacy} and \eqref{eqn_message_normalization_a2b_new}, we obtain
\begin{equation}
	\begin{aligned}
		&\underline{\mathrm{m}}^{\mathrm{BP},[\iota]}_{nl \leftarrow m } \\
		=& \frac{ \left( \underline{m}^{\mathrm{BP},[\iota-1], 1}_{nl\to m} \right)^{\rho_{\varphi}-1} \prod_{\left(n', l'\right) \in \underline{\mathcal{B}}_m \backslash (n, l)} \left( \underline{m}^{\mathrm{BP},[\iota-1], 0}_{n'l'\to m} \right)^{\rho_{\varphi}}  \prod_{\left(n'', l''\right) \in \overline{\mathcal{B}}_m } \left( \overline{m}^{\mathrm{BP},[\iota-1],0}_{n''l''\to m} \right)^{\rho_{\varphi}} } { \left( \underline{m}^{\mathrm{BP},[\iota-1], 0}_{nl\to m} \right)^{\rho_{\varphi}-1} \sum_{\boldsymbol{b}_m \in \mathcal{B}_m \backslash (n, l)}  \prod_{\left(n'l'\right) \in \underline{\mathcal{B}}_m \backslash (n, l)} \left( \underline{m}^{\mathrm{BP},[\iota-1]}_{n'l'\to m} (\boldsymbol{b}_m)  \right)^{\rho_{\varphi}} \prod_{\left(n'', l''\right) \in \overline{\mathcal{B}}_m } \left( \overline{m}^{\mathrm{BP},[\iota-1]}_{n''l''\to m}(\boldsymbol{b}_m)  \right)^{\rho_{\varphi}} } \\
		=& \frac{ \left( \underline{\mathrm{m}}^{\mathrm{BP},[\iota-1]}_{nl\to m} \right)^{\rho_{\varphi}-1} \prod_{\left(n', l'\right) \in \underline{\mathcal{B}}_m \backslash (n, l)} \left( \underline{m}^{\mathrm{BP},[\iota-1], 0}_{n'l'\to m} \right)^{\rho_{\varphi}}  \prod_{\left(n'', l''\right) \in \overline{\mathcal{B}}_m } \left( \overline{m}^{\mathrm{BP},[\iota-1],0}_{n''l''\to m} \right)^{\rho_{\varphi}}} { \sum_{\boldsymbol{b}_m \in \mathcal{B}_m \backslash\{(n, l)\}}  \prod_{\left(n', l'\right) \in \underline{\mathcal{B}}_m \backslash (n, l)} \left( \underline{\mathrm{m}}^{\mathrm{BP},[\iota-1]}_{n'l'\to m} (\boldsymbol{b}_m) \underline{m}^{\mathrm{BP},[\iota-1],0}_{n'l'\to m} \right)^{\rho_{\varphi}} \prod_{\left(n'', l''\right) \in \overline{\mathcal{B}}_m } \left( \overline{\mathrm{m}}^{\mathrm{BP},[\iota-1]}_{n''l''\to m}(\boldsymbol{b}_m) \overline{m}^{\mathrm{BP},[\iota-1],0}_{n''l''\to m} \right)^{\rho_{\varphi}} } \\
		=& \frac{ \left( \underline{\mathrm{m}}^{\mathrm{BP},[\iota-1]}_{nl\to m} \right)^{\rho_{\varphi}-1} }{ 1 + \sum_{\boldsymbol{b}_m \in \underline{\mathcal{B}}_m \cup \overline{\mathcal{B}}_m  \backslash (n, l)}  \prod_{\left(n', l'\right) \in \underline{\mathcal{B}}_m \backslash (n, l)} \left( \underline{\mathrm{m}}^{\mathrm{BP},[\iota-1]}_{n'l'\to m} (\boldsymbol{b}_m)  \right)^{\rho_{\varphi}} \prod_{\left(n'', l''\right) \in \overline{\mathcal{B}}_m } \left( \overline{\mathrm{m}}^{\mathrm{BP},[\iota-1]}_{n''l''\to m}(\boldsymbol{b}_m)  \right)^{\rho_{\varphi}} } \\
		=& \frac{ \left( \underline{\mathrm{m}}^{\mathrm{BP},[\iota-1]}_{nl\to m} \right)^{\rho_{\varphi}-1} }{ 1 - \left( \underline{\mathrm{m}}^{\mathrm{BP},[\iota-1]}_{nl\to m} \right)^{\rho_{\varphi}} + \sum_{n'=1}^{\underline{N}} \underline{\ell}_{n'} \left( \underline{\mathrm{m}}^{\mathrm{BP},[\iota-1]}_{n'l'\to m} \right)^{\rho_{\varphi}} + \sum_{o=1}^m \overline{\ell}_{o} \left( \overline{\mathrm{m}}^{\mathrm{BP},[\iota-1]}_{ol''\to m} \right)^{\rho_{\varphi}} }.
	\end{aligned}
\end{equation}
Note that $\underline{\mathrm{m}}^{\mathrm{BP},[\iota-1]}_{nl\to m}\left(\boldsymbol{b}_m \neq[n~l]^T\right) = 1$, $\overline{\mathrm{m}}^{\mathrm{BP},[\iota-1]}_{ml\to o}\left(\boldsymbol{b}_o \neq[\underline{N}+m~l]^T\right) = 1$, $\underline{\mathrm{m}}^{\mathrm{BP},[\iota-1]}_{nl\to m}$ are identical for $l\in\underline{\mathcal{L}}_n$, $\overline{\mathrm{m}}^{\mathrm{BP},[\iota-1]}_{ol''\to m}$ are identical for $l''\in\overline{\mathcal{L}}_{o}$, and the term 1 in the denominator corresponds to $\boldsymbol{b}_m = [0~0]^T$. 
\subsubsection{The rescaled message \texorpdfstring{$\underline{\mathrm{m}}^{\mathrm{BP},[\iota]}_{\to nl}$}{}}
The derivation of $\underline{\mathrm{m}}^{\mathrm{BP},[\iota]}_{\to nl}$ follows the the procedure in \cite{MW2020}. According to equation (35) in \cite{MJDL}, the message $\underline{m}^{\mathrm{BP},[\iota]}_{\to nl}(\underline{a}_{nl})$ is given by
\begin{equation} \label{eqn_lpt_message_h2a_1}
	\underline{m}^{\mathrm{BP},[\iota]}_{\rightarrow nl}(\underline{a}_{nl}) \approx \sum_{\underline{\boldsymbol{a}}_{nl \sim}} \sum_{\underline{r}_n} \int \underline{h}_n\left(\underline{\boldsymbol{x}}_n, \underline{r}_n, \underline{\boldsymbol{a}}_n \right) \underline{m}^{\mathrm{MF}}_{\to n}(\underline{\boldsymbol{x}}_n, \underline{r}_n) \mathrm{d} \underline{\boldsymbol{x}}_n \prod_{l' \in \underline{\mathcal{L}}_n \backslash l} \underline{m}^{\mathrm{MF}}_{\rightarrow nl'}(\underline{a}_{nl'}) \prod_{m=1}^M ( \underline{m}^{\mathrm{BP},[\iota]}_{nl' \leftarrow m  } (\underline{a}_{nl'}) )^{\rho_\varphi}. 
\end{equation}
Using $\underline{h}_n\left(\underline{\boldsymbol{x}}_n, \underline{r}_n, \underline{\boldsymbol{a}}_n \right)$ and $\underline{m}^{\mathrm{MF}}_{\to n}(\underline{\boldsymbol{x}}_n, \underline{r}_n) = \underline{p}^+_n(\underline{\boldsymbol{x}}_n, \underline{r}_n)$ from \cite{MJDL}, it is straightforward to obtain
\begin{equation} \label{eqn_lpt_integralWRTx}
	\begin{aligned}
		&\sum_{\underline{r}_n} \int \left( \underline{h}_n\left(\underline{\boldsymbol{x}}_n, \underline{r}_n, \underline{\boldsymbol{a}}_n \right) \right)  \underline{m}^{\mathrm{MF}}_{\to n}(\underline{\boldsymbol{x}}_n, \underline{r}_n) \mathrm{d} \boldsymbol{x}_n \\
		=& \begin{cases}
			\frac{1}{\underline{\ell}_n!} \underline{p}^{+,1}_n \underline{c}^{\gamma}_n p_D (\underline{\ell}_n - \|\underline{\boldsymbol{a}}_n\|_0)! \int  \underline{\gamma}_n^{\|\underline{\boldsymbol{a}}_n\|_0}\exp(-\underline{\gamma}_n) \underline{p}^+_n(\underline{\gamma}_n) \mathrm{d} \underline{\gamma}_n, & \|\underline{\boldsymbol{a}}_n\|_0 > 0 \\
			\underline{p}^{+,1}_n \underline{c}^{\gamma}_n \left(1 - p_D + p_D \int \exp(-\underline{\gamma}_n) \underline{p}^+_n(\underline{\gamma}_n) \mathrm{d} \underline{\gamma}_n \right) + \underline{p}^{+,0}_n, & \|\underline{\boldsymbol{a}}_n\|_0 = 0 \\
		\end{cases} \\
		=& \underline{p}_n(\|\underline{\boldsymbol{a}}_n\|_0).
	\end{aligned}
\end{equation} 

Since $\underline{p}^+_n(\underline{\gamma}_n) = \mathrm{G}(\underline{\gamma}_n; \underline{\alpha}^+_n, \underline{\beta}^+_n)$ is the gamma distribution, the integral in the above equation can be computed as
\begin{equation} \label{eqn_et_legacy_integral}
	\begin{aligned}
		\int \underline{\gamma}_n^{\varepsilon} \exp(-\underline{\gamma}_n) \underline{p}^+_n(\underline{\gamma}_n) \mathrm{d}\underline{\gamma}_n =& \int \underline{\gamma}_n^{\varepsilon} \exp(-\underline{\gamma}_n) \frac{(\underline{\beta}^+_n)^{\underline{\alpha}^+_n}}{\Gamma(\underline{\alpha}^+_n)} \underline{\gamma}_n^{\underline{\alpha}^+_n-1} \exp(-\underline{\beta}^+_n \underline{\gamma}_n) \mathrm{d}\underline{\gamma}_n  \\
		=&\frac{(\underline{\beta}^+_n)^{\underline{\alpha}^+_n}}{\Gamma(\underline{\alpha}^+_n)} \int \underline{\gamma}_n^{\varepsilon+\underline{\alpha}^+_n-1} \exp(-(\underline{\beta}^+_n + 1)\underline{\gamma}_n) \mathrm{d}\underline{\gamma}_n \\
		=& \frac{(\underline{\beta}^+_n)^{\underline{\alpha}^+_n} (\underline{\beta}^+_n+1)^{\varepsilon + \underline{\alpha}^+_n} \Gamma(\varepsilon + \underline{\alpha}^+_n) }{\Gamma(\underline{\alpha}^+_n) (\underline{\beta}^+_n+1)^{\varepsilon + \underline{\alpha}^+_n} \Gamma(\varepsilon + \underline{\alpha}^+_n)}\int \underline{\gamma}_n^{\varepsilon+\underline{\alpha}^+_n-1} \exp(-(\underline{\beta}^+_n + 1)\underline{\gamma}_n) \mathrm{d}\underline{\gamma}_n \\
		=& \frac{ (\underline{\beta}^+_n)^{\underline{\alpha}^+_n} \Gamma(\varepsilon + \underline{\alpha}^+_n) }{(\underline{\beta}^+_n+1)^{\varepsilon + \underline{\alpha}^+_n}\Gamma(\underline{\alpha}^+_n)} \int \mathrm{G}(\underline{\gamma}_n; \varepsilon + \underline{\alpha}^+_n, \underline{\beta}^+_n + 1)\mathrm{d}\underline{\gamma}_n \\
		=& \frac{ (\underline{\beta}^+_n)^{\underline{\alpha}^+_n} \Gamma(\varepsilon + \underline{\alpha}^+_n) }{(\underline{\beta}^+_n+1)^{\varepsilon + \underline{\alpha}^+_n}\Gamma(\underline{\alpha}^+_n)} .
	\end{aligned}
\end{equation}
Using \eqref{eqn_et_legacy_integral} in \eqref{eqn_lpt_integralWRTx}, $\underline{p}_n(\|\underline{\boldsymbol{a}}_n\|_0)$ can be rewritten as
\begin{equation}
	\begin{aligned}
		\underline{p}_n(\|\underline{\boldsymbol{a}}_n\|_0)
		&= \begin{cases}
			\frac{\underline{p}^{+,1}_n \underline{c}^\gamma_n p_D(\underline{\ell}_n - \|\underline{\boldsymbol{a}}_n\|_0)!}{\underline{\ell}_n!} \frac{ (\underline{\beta}^+_n)^{\underline{\alpha}^+_n} \Gamma(\|\underline{\boldsymbol{a}}_n\|_0 + \underline{\alpha}^+_n) }{(\underline{\beta}^+_n+1)^{\|\underline{\boldsymbol{a}}_n\|_0 + \underline{\alpha}^+_n}\Gamma(\underline{\alpha}^+_n)}, & \|\underline{\boldsymbol{a}}_n\|_0 > 0 \\
			\underline{p}^{+,1}_n \underline{c}^\gamma_n \left(1 - p_D + p_D \left( \frac{\underline{\beta}^+_n}{\underline{\beta}^+_n + 1} \right)^{\underline{\alpha}^+_n} \right) + \underline{p}^{+,0}_n, & \|\underline{\boldsymbol{a}}_n\|_0 = 0 .
		\end{cases}
	\end{aligned}
\end{equation}
Substituting $\underline{p}_n(\|\underline{\boldsymbol{a}}_n\|_0)$ into \eqref{eqn_lpt_message_h2a_1}, we have 
\begin{equation}
	\underline{m}^{\mathrm{BP},[\iota]}_{\rightarrow nl}(\underline{a}_{nl}) = \sum_{\boldsymbol{\underline{a}}_{n l \sim}} \underline{p}_n(\|\underline{\boldsymbol{a}}_n\|_0)  \prod_{l' \in \underline{\mathcal{L}}_n \backslash l} \underline{m}^{\mathrm{MF}}_{\rightarrow nl'}(\underline{a}_{nl'})  \prod_{m=1}^M \left( \underline{m}^{\mathrm{BP},[\iota]}_{nl' \leftarrow m  } (\underline{a}_{nl'})  \right)^{\rho_\varphi}.
\end{equation}
Using \eqref{eqn_message_normalization_b2a_legacy} in the above equation, we obtain
\begin{equation} \label{eqn_lpt_message_h2a_2}
    \begin{aligned} 
        \underline{m}^{\mathrm{BP},[\iota]}_{\rightarrow nl}(\underline{a}_{nl}) =& \sum_{\boldsymbol{\underline{a}}_{n l \sim}} \underline{p}_n(\|\underline{\boldsymbol{a}}_n\|_0)  \prod_{l' \in \underline{\mathcal{L}}_n \backslash l} \underline{m}^{\mathrm{MF}}_{\rightarrow nl'}(\underline{a}_{nl'})  \prod_{m=1}^M \left( \underline{\mathrm{m}}^{\mathrm{BP},[\iota]}_{nl' \leftarrow m} (\underline{a}_{nl'}) \underline{m}^{\mathrm{BP},[\iota], 0}_{nl' \leftarrow m}  \right)^{\rho_\varphi} \\ 
        =& C_1 \sum_{\boldsymbol{\underline{a}}_{n l \sim}} \underline{p}_n(\|\underline{\boldsymbol{a}}_n\|_0)  \prod_{l' \in \underline{\mathcal{L}}_n \backslash l} \underline{m}^{\mathrm{MF}}_{\rightarrow nl'}(\underline{a}_{nl'}) \left( \underline{\mathrm{m}}^{\mathrm{BP},[\iota]}_{nl' \leftarrow \underline{a}_{nl'}}  \right)^{\rho_\varphi} \\ 
	\end{aligned}
\end{equation}
where the constant $C_1 = \prod_{l' \in \underline{\mathcal{L}}_n \backslash l}   \prod_{m=1}^M \left(  \underline{m}^{\mathrm{BP},[\iota], 0}_{nl' \leftarrow m} \right)^{\rho_\varphi} $ and we introduce $\underline{\mathrm{m}}^{\mathrm{BP},[\iota]}_{nl' \leftarrow 0} = 1$.  

We introduce a binary vector $\underline{\boldsymbol{c}}_n = [\underline{c}_{n1},\dots,\underline{c}_{n\underline{\ell}_n}]^T$, with entries $\underline{c}_{nl}\in\{0,1\}$, $l\in\underline{\mathcal{L}}_n$, and define a indicator function
\begin{equation} \label{eqn_lpt_indicator1}
	\Phi_1(\underline{a}_{nl},\underline{c}_{nl})=
	\begin{cases}
		1, & \underline{c}_{nl}=1, \underline{a}_{nl}>0 \text{ or } \underline{c}_{nl}=0, \underline{a}_{nl}=0 \\
		0, & \text{otherwise.}
	\end{cases}
\end{equation}
Then, $\underline{m}^{\mathrm{BP},[\iota]}_{\rightarrow nl}(\underline{a}_{nl})$ in \eqref{eqn_lpt_message_h2a_2} can be rewritten as 
\begin{equation} \label{eqn_lpt_message_h2a_3}
    \begin{aligned} 
        \underline{m}^{\mathrm{BP},[\iota]}_{\rightarrow nl}(\underline{a}_{nl}) =& C_1 \sum_{\underline{\boldsymbol{c}}_n} \sum_{\boldsymbol{\underline{a}}_{n l \sim}} \underline{p}_n(\|\underline{\boldsymbol{a}}_n\|_0) \bigg(\prod_{l'\in \underline{\mathcal{L}}_n} \Phi_1(\underline{a}_{nl'}, \underline{c}_{nl'})\bigg) \prod_{l' \in \underline{\mathcal{L}}_n \backslash l} \underline{m}^{\mathrm{MF}}_{\rightarrow nl'}(\underline{a}_{nl'}) \left( \underline{\mathrm{m}}^{\mathrm{BP},[\iota]}_{nl' \leftarrow \underline{a}_{nl'}}  \right)^{\rho_\varphi} \\ 
        =& C_1 \sum_{\underline{c}_{nl}} \Phi_1(\underline{a}_{nl}, \underline{c}_{nl}) \sum_{\underline{\boldsymbol{c}}_{nl\sim}} \underline{p}_n(\|\underline{\boldsymbol{c}}_n\|_0) \prod_{l' \in \underline{\mathcal{L}}_n \backslash l} \sum_{\underline{a}_{nl'} = 0}^M \Phi_1(\underline{a}_{nl'}, \underline{c}_{nl'}) \underline{m}^{\mathrm{MF}}_{\rightarrow nl'}(\underline{a}_{nl'}) \left( \underline{\mathrm{m}}^{\mathrm{BP},[\iota]}_{nl' \leftarrow \underline{a}_{nl'}}  \right)^{\rho_\varphi} \\
        =& C_1 \sum_{\underline{c}_{nl}} \Phi_1(\underline{a}_{nl}, \underline{c}_{nl}) \sum_{\underline{\boldsymbol{c}}_{nl\sim}} \underline{p}_n(\|\underline{\boldsymbol{c}}_n\|_0) \prod_{l' \in \underline{\mathcal{L}}_n \backslash l} \left( \widetilde{\underline{m}}^{\mathrm{BP},[\iota]}_{\rightarrow nl'} \right)^{\underline{c}_{nl'}}
	\end{aligned}
\end{equation}
where $\widetilde{\underline{m}}^{\mathrm{BP},[\iota]}_{\rightarrow nl'} = \sum_{m = 1}^M \underline{m}^{\mathrm{MF}}_{\rightarrow nl'}(m) \left( \underline{\mathrm{m}}^{\mathrm{BP},[\iota]}_{nl' \leftarrow m}  \right)^{\rho_\varphi}$, and note that
\begin{equation} \nonumber
	\sum_{\underline{a}_{nl'} = 0}^M \Phi_1(\underline{a}_{nl'}, 0) \underline{m}^{\mathrm{MF}}_{\rightarrow nl'}(\underline{a}_{nl'}) \left( \underline{\mathrm{m}}^{\mathrm{BP},[\iota]}_{nl' \leftarrow \underline{a}_{nl'}}  \right)^{\rho_\varphi} = 1. 
\end{equation}

We introduce a second indicator function as 
\begin{equation} \label{eqn_lpt_indicator2}
    \Phi_2(\underline{\boldsymbol{c}}_{nl\sim}, \varepsilon ) = 
    \begin{cases}
        1, & \|\underline{\boldsymbol{c}}_{nl\sim}\|_0 = \varepsilon\\
        0, & \text{otherwise}.
    \end{cases}
\end{equation}
Using this function in \eqref{eqn_lpt_message_h2a_3}, we obtain
\begin{equation}
    \underline{m}^{\mathrm{BP},[\iota]}_{\rightarrow nl}(\underline{a}_{nl}) = C_1 \sum_{\underline{c}_{nl}} \Phi_1(\underline{a}_{nl}, \underline{c}_{nl}) \sum_{\varepsilon = 0}^{\underline{\ell}_n - 1} \underline{p}_n(\varepsilon + \underline{c}_{nl}) \sum_{\underline{\boldsymbol{c}}_{nl\sim}} \Phi_2(\underline{\boldsymbol{c}}_{nl\sim}, \varepsilon ) \prod_{l' \in \underline{\mathcal{L}}_n \backslash l} \left( \widetilde{\underline{m}}^{\mathrm{BP},[\iota]}_{\rightarrow nl'} \right)^{\underline{c}_{nl'}} .
\end{equation}
Considering that for each $\varepsilon$ there are $C^{\underline{\ell}_n - 1}_\varepsilon$ different vectors $\underline{\boldsymbol{c}}_{nl\sim}$ for which $\Phi_{2}(\underline{\boldsymbol{c}}_{nl\sim},\varepsilon)$ is one and that the messages $\widetilde{\underline{m}}^{\mathrm{BP},[\iota]}_{\rightarrow nl'}$ are identical for all $l'\in\underline{\mathcal{L}}_n$. Thus we obtain
\begin{equation}
	\begin{aligned}
		\underline{m}^{\mathrm{BP},[\iota]}_{\rightarrow nl}(\underline{a}_{nl}) &= C_1 \sum_{\underline{c}_{nl}} \Phi_1(\underline{a}_{nl}, \underline{c}_{nl}) \sum_{\varepsilon = 0}^{\underline{\ell}_n - 1} \frac{(\underline{\ell}_n - 1)! \underline{p}_n(\varepsilon + \underline{c}_{nl})}{\varepsilon!(\underline{\ell}_n - \varepsilon - 1)!}  \prod_{l' \in \underline{\mathcal{L}}_n \backslash l} \left( \widetilde{\underline{m}}^{\mathrm{BP},[\iota]}_{\rightarrow nl'} \right)^{\underline{c}_{nl'}} \\
		&= \begin{cases}
			C_1 \sum_{\varepsilon = 0}^{\underline{\ell}_n - 1} \frac{\underline{f}^1_n(\varepsilon + 1)}{\varepsilon!} \left( \widetilde{\underline{m}}^{\mathrm{BP},[\iota]}_{\rightarrow nl'} \right)^{\varepsilon},& \underline{a}_{nl} \in\{1,\dots,M\} \\
			C_1 \sum_{\varepsilon = 0}^{\underline{\ell}_n - 1} \frac{\underline{f}^0_n(\varepsilon)}{\varepsilon!} \left( \widetilde{\underline{m}}^{\mathrm{BP},[\iota]}_{\rightarrow nl'} \right)^{\varepsilon},& \underline{a}_{nl} = 0
		\end{cases}
    \end{aligned}
\end{equation}
where
\begin{equation}
	\begin{aligned}
		\underline{f}^1_n(\varepsilon) &= \frac{ \underline{c}^\gamma_n p_D \underline{p}^{+,1}_n (\underline{\beta}^+_n)^{\underline{\alpha}^+_n} \Gamma(\varepsilon + \underline{\alpha}^+_n) }{\underline{\ell}_n (\underline{\beta}^+_n+1)^{\varepsilon + \underline{\alpha}^+_n}\Gamma(\underline{\alpha}^+_n)} \\
		\underline{f}^0_n(\varepsilon) 
		&= \begin{cases}
			\frac{ \underline{c}^\gamma_n p_D  \underline{p}^{+,1}_n (\underline{\ell}_n - \varepsilon)(\underline{\beta}^+_n)^{\underline{\alpha}^+_n} \Gamma(\varepsilon + \underline{\alpha}^+_n) }{\underline{\ell}_n(\underline{\beta}^+_n+1)^{\varepsilon + \underline{\alpha}^+_n}\Gamma(\underline{\alpha}^+_n)}, & \varepsilon > 0 \\
			\underline{p}^{+,1}_n\underline{c}^\gamma_n \left(1 - p_D + \frac{p_D (\underline{\beta}^+_n)^{\underline{\alpha}^+_n}}{(\underline{\beta}^+_n + 1)^{\underline{\alpha}^+_n}}  \right) + \underline{p}^{+,0}_n, & \varepsilon = 0.
		\end{cases}
	\end{aligned}
\end{equation}
Finally, the two distinct values in \eqref{eqn_distinctValues} for $\underline{m}^{\mathrm{BP},[\iota]}_{\rightarrow nl}(\underline{a}_{nl})$ are given by
\begin{equation}
	\begin{aligned}
        \underline{m}^{\mathrm{BP}, [\iota], 1}_{\rightarrow nl} &= C_1 \sum_{\varepsilon = 0}^{\underline{\ell}_n - 1} \frac{1}{\varepsilon!} \underline{f}^1_n(\varepsilon + 1) \left( \widetilde{\underline{m}}^{\mathrm{BP},[\iota]}_{\rightarrow nl'} \right)^{\varepsilon} \\
        \underline{m}^{\mathrm{BP}, [\iota], 0}_{\rightarrow nl} &= C_1 \sum_{\varepsilon = 0}^{\underline{\ell}_n - 1} \frac{1}{\varepsilon!}  \underline{f}^0_n(\varepsilon) \left( \widetilde{\underline{m}}^{\mathrm{BP},[\iota]}_{\rightarrow nl'} \right)^{\varepsilon}
    \end{aligned}
\end{equation}
and the rescaled message $\underline{\mathrm{m}}^{\mathrm{BP},[\iota]}_{\rightarrow nl}$ is given by 
\begin{equation}
	\underline{\mathrm{m}}^{\mathrm{BP},[\iota]}_{\rightarrow nl} = \frac{\underline{m}^{\mathrm{BP}, [\iota], 1}_{\rightarrow nl}}{\underline{m}^{\mathrm{BP}, [\iota], 0}_{\rightarrow nl}} = \frac{\sum_{\varepsilon = 0}^{\underline{\ell}_n - 1} \underline{f}^1_n(\varepsilon + 1) \left( \widetilde{\underline{m}}^{\mathrm{BP},[\iota]}_{\rightarrow nl'} \right)^{\varepsilon} / \varepsilon!}{\sum_{\varepsilon = 0}^{\underline{\ell}_n - 1} \underline{f}^0_n(\varepsilon) \left( \widetilde{\underline{m}}^{\mathrm{BP},[\iota]}_{\rightarrow nl'} \right)^{\varepsilon} / \varepsilon!}.
\end{equation}

\subsubsection{The rescaled message \texorpdfstring{$\underline{\mathrm{m}}^{\mathrm{BP},[\iota]}_{nl\to m}$}{}}
According to equation (34) in \cite{MJDL}, the two distinct values in \eqref{eqn_distinctValues} for message $\underline{m}^{\mathrm{BP},[\iota]}_{nl\to m} (\underline{\boldsymbol{b}}_m)$ are given by 
\begin{equation} \label{eqn_unnormalized_message_a2b_legacy}
	\begin{aligned}
		\underline{m}^{\mathrm{BP},[\iota], 1}_{nl\rightarrow m} &= \left( \underline{m}^{\mathrm{BP},[\iota], 1}_{ nl \leftarrow m } \right)^{\rho_\varphi-1} \underline{m}^{\mathrm{MF}}_{\rightarrow nl}(m)  \underline{m}^{\mathrm{BP},[\iota]}_{\rightarrow nl}(m)  \prod_{m'=1, m'\neq m}^M \left( \underline{m}^{\mathrm{BP},[\iota], 0}_{ nl \leftarrow m' } \right)^{\rho_\varphi} \\
		\underline{m}^{\mathrm{BP},[\iota], 0}_{nl\rightarrow m} &= \left( \underline{m}^{\mathrm{BP},[\iota], 0}_{nl \leftarrow m } \right)^{\rho_\varphi-1} \sum_{\underline{a}_{n l}=0, \underline{a}_{n l}\neq m}^M \underline{m}^{\mathrm{MF}}_{\rightarrow nl}(\underline{a}_{n l}) \underline{m}^{\mathrm{BP},[\iota]}_{\rightarrow nl} (\underline{a}_{n l}) \prod_{m'=1, m'\neq m}^M \left( \underline{m}^{\mathrm{BP},[\iota]}_{nl \leftarrow m' }(\underline{a}_{n l}) \right)^{\rho_\varphi} .
	\end{aligned}
\end{equation}
Substituting \eqref{eqn_unnormalized_message_a2b_legacy} into \eqref{eqn_message_normalization_a2b_legacy}, and using \eqref{eqn_message_normalization_b2a_legacy} and \eqref{eqn_message_normalization_h2a_legacy}, we obtain 
\begin{equation}
	\begin{aligned}
		\underline{\mathrm{m}}^{\mathrm{BP},[\iota]}_{nl\rightarrow m} &= \frac{ \left( \underline{m}^{\mathrm{BP},[\iota], 1}_{ nl \leftarrow m } \right)^{\rho_\varphi-1} \underline{m}^{\mathrm{MF}}_{\rightarrow nl}(m)  \underline{m}^{\mathrm{BP},[\iota]}_{\rightarrow nl}(m)  \prod_{m'=1, m'\neq m}^M \left( \underline{m}^{\mathrm{BP},[\iota], 0}_{ nl \leftarrow m' } \right)^{\rho_\varphi} } { \left( \underline{m}^{\mathrm{BP},[\iota], 0}_{nl \leftarrow m } \right)^{\rho_\varphi-1} \sum_{\underline{a}_{n l}=0, \underline{a}_{n l}\neq m}^M \underline{m}^{\mathrm{MF}}_{\rightarrow nl}(\underline{a}_{n l}) \underline{m}^{\mathrm{BP},[\iota]}_{\rightarrow nl} (\underline{a}_{n l}) \prod_{m'=1, m'\neq m}^M \left( \underline{m}^{\mathrm{BP},[\iota]}_{nl \leftarrow m'}(\underline{a}_{n l}) \right)^{\rho_\varphi} } \\
		&= \frac{ \left( \underline{\mathrm{m}}^{\mathrm{BP},[\iota]}_{ nl \leftarrow m } \right)^{\rho_\varphi-1} \underline{m}^{\mathrm{MF}}_{\rightarrow nl}(m) \underline{\mathrm{m}}^{\mathrm{BP},[\iota]}_{\rightarrow nl} \underline{m}^{\mathrm{BP},[\iota],0}_{\rightarrow nl} \prod_{m'=1, m'\neq m}^M \left( \underline{m}^{\mathrm{BP},[\iota], 0}_{ nl \leftarrow m' } \right)^{\rho_\varphi} } { \sum_{\underline{a}_{n l}=0, \underline{a}_{n l}\neq m}^M \underline{m}^{\mathrm{MF}}_{\rightarrow nl}(\underline{a}_{n l}) \underline{\mathrm{m}}^{\mathrm{BP},[\iota]}_{\rightarrow nl}(\underline{a}_{n l}) \underline{m}^{\mathrm{BP},[\iota],0}_{\rightarrow nl} \prod_{m'=1, m'\neq m}^M \left( \underline{\mathrm{m}}^{\mathrm{BP},[\iota]}_{nl \leftarrow m'}(\underline{a}_{n l}) \underline{m}^{\mathrm{BP},[\iota],0}_{nl \leftarrow m'} \right)^{\rho_\varphi} } \\
		&= \frac{ \left( \underline{\mathrm{m}}^{\mathrm{BP},[\iota]}_{ nl \leftarrow m } \right)^{\rho_\varphi-1} \underline{m}^{\mathrm{MF}}_{\rightarrow nl}(m) \underline{\mathrm{m}}^{\mathrm{BP},[\iota]}_{\rightarrow nl} } { 1 + \sum_{\underline{a}_{n l}=1, \underline{a}_{n l}\neq m}^M \underline{m}^{\mathrm{MF}}_{\rightarrow nl}(\underline{a}_{n l}) \underline{\mathrm{m}}^{\mathrm{BP},[\iota]}_{\rightarrow nl}(\underline{a}_{n l})  \prod_{m'=1, m'\neq m}^M \left( \underline{\mathrm{m}}^{\mathrm{BP},[\iota]}_{nl \leftarrow m'}(\underline{a}_{n l}) \right)^{\rho_\varphi} } \\
		&= \frac{ \left( \underline{\mathrm{m}}^{\mathrm{BP},[\iota]}_{ nl \leftarrow m } \right)^{\rho_\varphi-1} \underline{m}^{\mathrm{MF}}_{\rightarrow nl}(m) \underline{\mathrm{m}}^{\mathrm{BP},[\iota]}_{\rightarrow nl} } { 1 +  \sum_{m'=1, m'\neq m}^M \underline{m}^{\mathrm{MF}}_{\rightarrow nl}(m') \underline{\mathrm{m}}^{\mathrm{BP},[\iota]}_{\rightarrow nl} \left( \underline{\mathrm{m}}^{\mathrm{BP},[\iota]}_{nl \leftarrow m'} \right)^{\rho_\varphi} }.
    \end{aligned}
\end{equation}
Note that $\underline{m}^{\mathrm{MF}}_{\rightarrow nl}(0)= 1$ and $\underline{\mathrm{m}}^{\mathrm{BP},[\iota]}_{nl\leftarrow m}(\underline{a}_{nl}\neq m) = 1$ and the term 1 in the denominator corresponds to $\underline{a}_{nl} = 0$.  

\subsubsection{The rescaled message \texorpdfstring{$\overline{\mathrm{m}}^{\mathrm{BP},[\iota]}_{ml \leftarrow o }$}{}}
According to equation (36) in \cite{MJDL}, the two distinct values in \eqref{eqn_distinctValues} for message $\overline{\mathrm{m}}^{\mathrm{BP},[\iota]}_{ml \leftarrow o } (\overline{a}_{ml})$ are given by
\begin{equation} \label{eqn_unnormalized_message_b2a_new}
    \begin{aligned}
        \overline{m}^{\mathrm{BP},[\iota], 1}_{ml \leftarrow o } &= \left( \overline{m}^{\mathrm{BP},[\iota-1], 1}_{ml\rightarrow o} \right)^{\rho_\varphi-1}  \prod_{\left(n, l'\right) \in \underline{\mathcal{B}}_m } \left( \underline{m}^{\mathrm{BP},[\iota-1],0}_{nl'\rightarrow o} \right)^{\rho_\varphi}  \prod_{\left(m', l''\right) \in \overline{\mathcal{B}}_m \backslash (m, l) } \left( \overline{m}^{\mathrm{BP},[\iota-1],0}_{m'l''\rightarrow o} \right)^{\rho_\varphi} \\
        \overline{m}^{\mathrm{BP},[\iota], 0}_{ml \leftarrow o } &= \left( \overline{m}^{\mathrm{BP},[\iota-1], 0}_{ml\rightarrow o} \right)^{\rho_\varphi-1} \sum_{\boldsymbol{b}_o \in \mathcal{B}_o \backslash (m,l)}  \prod_{\left(n, l'\right) \in \underline{\mathcal{B}}_m } \left( \underline{m}^{\mathrm{BP},[\iota-1]}_{nl'\rightarrow o}(\boldsymbol{b}_o) \right)^{\rho_\varphi}  \!\prod_{\left(m', l''\right) \in \overline{\mathcal{B}}_m \backslash (m,l)} \!\left( \overline{m}^{\mathrm{BP},[\iota-1]}_{m'l''\rightarrow o}(\boldsymbol{b}_o) \right)^{\rho_\varphi}. 
	\end{aligned}
\end{equation}
Substituting \eqref{eqn_unnormalized_message_b2a_new} into \eqref{eqn_message_normalization_b2a_new} and using \eqref{eqn_message_normalization_a2b_new}, we obtain
\begin{equation}
	\begin{aligned}
		&\overline{\mathrm{m}}^{\mathrm{BP},[\iota]}_{ml \leftarrow o} \\
		=& \frac{ \left( \overline{m}^{\mathrm{BP},[\iota-1], 1}_{ml\rightarrow o} \right)^{\rho_\varphi-1}  \prod_{\left(n, l'\right) \in \underline{\mathcal{B}}_m } \left( \underline{m}^{\mathrm{BP},[\iota-1],0}_{nl'\rightarrow o} \right)^{\rho_\varphi}  \prod_{\left(m', l''\right) \in \overline{\mathcal{B}}_m \backslash (m,l) } \left( \overline{m}^{\mathrm{BP},[\iota-1],0}_{m'l''\rightarrow o} \right)^{\rho_\varphi} } { \left( \overline{m}^{\mathrm{BP},[\iota-1], 0}_{ml\rightarrow o} \right)^{\rho_\varphi} \sum_{\boldsymbol{b}_o \in \mathcal{B}_o \backslash (m,l)} \prod_{\left(n, l'\right) \in \underline{\mathcal{B}}_m } \left( \underline{m}^{\mathrm{BP},[\iota-1]}_{nl'\rightarrow o}(\boldsymbol{b}_o) \right)^{\rho_\varphi}  \prod_{\left(m', l''\right) \in \overline{\mathcal{B}}_m \backslash (m,l)} \left( \overline{m}^{\mathrm{BP},[\iota-1]}_{m'l''\rightarrow o}(\boldsymbol{b}_o) \right)^{\rho_\varphi} } \\
		=& \frac{ \left( \overline{\mathrm{m}}^{\mathrm{BP},[\iota-1]}_{ml\rightarrow o} \right)^{\rho_\varphi-1} \prod_{\left(n, l'\right) \in \underline{\mathcal{B}}_m } \left( \underline{m}^{\mathrm{BP},[\iota-1],0}_{nl'\rightarrow o} \right)^{\rho_\varphi}  \prod_{\left(m', l''\right) \in \overline{\mathcal{B}}_m \backslash (m,l) } \left( \overline{m}^{\mathrm{BP},[\iota-1],0}_{m'l''\rightarrow o} \right)^{\rho_\varphi} } { \sum_{\boldsymbol{b}_o \in \mathcal{B}_o \backslash (m,l)} \prod_{\left(n, l'\right) \in \underline{\mathcal{B}}_m } \left( \underline{\mathrm{m}}^{\mathrm{BP},[\iota-1]}_{nl'\rightarrow o}(\boldsymbol{b}_o) \underline{m}^{\mathrm{BP},[\iota-1],0}_{nl'\rightarrow o} \right)^{\rho_\varphi}  \prod_{\left(m', l''\right) \in \overline{\mathcal{B}}_m \backslash (m,l)} \left( \overline{\mathrm{m}}^{\mathrm{BP},[\iota-1]}_{m'l''\rightarrow o}(\boldsymbol{b}_o) \overline{m}^{\mathrm{BP},[\iota-1],0}_{m'l''\rightarrow o}\right)^{\rho_\varphi} } \\
		=& \frac{ \left( \overline{\mathrm{m}}^{\mathrm{BP},[\iota-1]}_{ml\rightarrow o} \right)^{\rho_\varphi-1} } { 1 + \sum_{\boldsymbol{b}_o \in \underline{\mathcal{B}}_o\cup \overline{\mathcal{B}}_o \backslash (m,q)}  \prod_{\left(n, l'\right) \in \underline{\mathcal{B}}_m } \left( \underline{\mathrm{m}}^{\mathrm{BP},[\iota-1]}_{nl'\rightarrow o}(\boldsymbol{b}_o)  \right)^{\rho_\varphi}  \prod_{\left(m', l''\right) \in \overline{\mathcal{B}}_m \backslash (m,l)} \left( \overline{\mathrm{m}}^{\mathrm{BP},[\iota-1]}_{m'l''\rightarrow o}(\boldsymbol{b}_o) \right)^{\rho_\varphi} } \\
		=& \frac{ \left( \overline{\mathrm{m}}^{\mathrm{BP},[\iota-1]}_{ml\rightarrow o} \right)^{\rho_\varphi-1} } { 1 - \left( \overline{\mathrm{m}}^{\mathrm{BP},[\iota-1]}_{ml\rightarrow o} \right)^{\rho_\varphi}+ \sum_{n=1}^{\underline{N}} \underline{\ell}_n \left( \underline{\mathrm{m}}^{\mathrm{BP},[\iota-1]}_{nl'\rightarrow o} \right)^{\rho_\varphi} + \sum_{m'=1}^o \overline{\ell}_{m'} \left( \overline{\mathrm{m}}^{\mathrm{BP},[\iota-1]}_{m'l''\rightarrow o} \right)^{\rho_\varphi} }
	\end{aligned}
\end{equation}
where the term 1 in the denominator corresponds to $\boldsymbol{b}_o = [0~0]^T$, and note that $\underline{\mathrm{m}}^{\mathrm{BP},[\iota-1]}_{nl'\rightarrow o}$ are identical for all $l'\in\underline{\mathcal{L}}_n$, and $\overline{\mathrm{m}}^{\mathrm{BP},[\iota-1]}_{m'l''\rightarrow o}$ are identical for all $l''\in\overline{\mathcal{L}}_{m'}$. 

\subsubsection{The rescaled message \texorpdfstring{$\overline{\mathrm{m}}^{\mathrm{BP},[\iota]}_{\to ml}$}{}}
According to equation (37) in \cite{MJDL}, the message $\overline{m}^{\mathrm{BP},[\iota]}_{\rightarrow ml}(\overline{a}_{ml})$ is given by
\begin{equation} \label{eqn_npt_message_h2a_1}
	\overline{m}^{\mathrm{BP},[\iota]}_{\rightarrow ml}\!(\overline{a}_{ml}\!)\! \approx \!\sum_{\overline{\boldsymbol{a}}_{m l \sim}}\! \!\sum_{\overline{r}_m}\! \!\int\! \overline{h}_m\left(\overline{\boldsymbol{x}}_m, \overline{r}_m, \overline{\boldsymbol{a}}_m \right) \overline{m}^{\mathrm{MF}}_{\to m}(\overline{\boldsymbol{x}}_m, \overline{r}_m) \mathrm{d} \overline{\boldsymbol{x}}_m \prod_{l' \in \overline{\mathcal{L}}_m \backslash l} \overline{m}^{\mathrm{MF}}_{\rightarrow ml'}(\overline{a}_{ml'})  \prod_{o=m}^M ( \overline{m}^{\mathrm{BP},[\iota]}_{ml' \leftarrow o  } (\overline{a}_{ml'})  )^{\rho_\varphi}. 
\end{equation}
Using $\overline{h}_m\left(\overline{\boldsymbol{x}}_m, \overline{r}_m, \overline{\boldsymbol{a}}_m \right)$ and $\overline{m}^{\mathrm{MF}}_{\to m}(\overline{\boldsymbol{x}}_m, \overline{r}_m) = \overline{f}_m(\overline{\boldsymbol{x}}_m, \overline{r}_m)$ from \cite{MJDL}, we obtain
\begin{equation}
	\begin{aligned}
		&\sum_{\overline{r}_m} \int \overline{h}_m\left(\overline{\boldsymbol{x}}_m, \overline{r}_m, \overline{\boldsymbol{a}}_m \right) \overline{m}^{\mathrm{MF}}_{\to m}(\overline{\boldsymbol{x}}_m, \overline{r}_m) \mathrm{d} \overline{\boldsymbol{x}}_m \\
		=&\begin{cases}
        	\frac{\lambda_{\mathrm{n}} \overline{c}^\gamma_m p_D(\overline{\ell}_m - \|\overline{\boldsymbol{a}}_m\|_0)!}{\overline{\ell}_m!} \int \frac{ \overline{\gamma}_m^{\|\overline{\boldsymbol{a}}_m\|_0} \exp(-\overline{\gamma}_m) \overline{p}_m(\overline{\gamma}_m) } {(1-p_D + p_D \exp(-\overline{\gamma}_m))}  \mathrm{d}\overline{\gamma}_m, &  \|\overline{\boldsymbol{a}}_m\|_0 > 0 \\
        	1, & \|\overline{\boldsymbol{a}}_m\|_0 = 0
        \end{cases} \\
        \approx& \begin{cases}
        	\frac{\lambda_{\mathrm{n}} \overline{c}^\gamma_m p_D(\overline{\ell}_m - \|\overline{\boldsymbol{a}}_m\|_0)!}{(1-p_D)\overline{\ell}_m!} \int  \overline{\gamma}_m^{\|\overline{\boldsymbol{a}}_m\|_0} \exp(-\overline{\gamma}_m) \overline{p}_m(\overline{\gamma}_m) \mathrm{d}\overline{\gamma}_m, &  \|\overline{\boldsymbol{a}}_m\|_0 > 0 \\
        	1, & \|\overline{\boldsymbol{a}}_m\|_0 = 0
        \end{cases} \\
        =& \begin{cases}
        	\frac{\lambda_{\mathrm{n}} \overline{c}^\gamma_m p_D(\overline{\ell}_m - \|\overline{\boldsymbol{a}}_m\|_0)!}{(1 - p_D) \overline{\ell}_m!} \frac{ (\overline{\beta}_m)^{\overline{\alpha}_m} \Gamma(\|\overline{\boldsymbol{a}}_m\|_0 + \overline{\alpha}_m) }{(\overline{\beta}_m+1)^{\|\overline{\boldsymbol{a}}_m\|_0 + \overline{\alpha}_m}\Gamma(\overline{\alpha}_m)} , &  \|\overline{\boldsymbol{a}}_m\|_0 > 0 \\
        	1, & \|\overline{\boldsymbol{a}}_m\|_0 = 0
        \end{cases} \\
        =& \overline{p}_m(\|\overline{\boldsymbol{a}}_m\|_0).
    \end{aligned}
\end{equation}
Note that $\overline{p}_m(\overline{\gamma}_m) = \mathrm{G}(\overline{\gamma}_m; \overline{\alpha}_m, \overline{\beta}_m)$, and the integrals in the above formula can be computed using \eqref{eqn_et_legacy_integral}.

Using $\overline{p}_m(\|\overline{\boldsymbol{a}}_m\|_0)$ and \eqref{eqn_message_normalization_b2a_new} in \eqref{eqn_npt_message_h2a_1}, we obtain 
\begin{equation} 
	\begin{aligned}
		\overline{m}^{\mathrm{BP},[\iota]}_{\rightarrow ml}(\overline{a}_{ml}) &= \sum_{\overline{\boldsymbol{a}}_{m l \sim}} \overline{p}_m(\|\overline{\boldsymbol{a}}_m\|_0) \prod_{l' \in \overline{\mathcal{L}}_m \backslash l} \overline{m}^{\mathrm{MF}}_{\rightarrow ml'}(\overline{a}_{ml'}) \prod_{o=m}^M \left( \overline{\mathrm{m}}^{\mathrm{BP},[\iota]}_{ml' \leftarrow o  } (\overline{a}_{ml'}) \overline{m}^{\mathrm{BP},[\iota],0}_{ml' \leftarrow o  } \right)^{\rho_\varphi} \\
		&= C_2 \sum_{\overline{\boldsymbol{a}}_{m l \sim}} \overline{p}_m(\|\overline{\boldsymbol{a}}_m\|_0) \prod_{l' \in \overline{\mathcal{L}}_m \backslash l} \overline{m}^{\mathrm{MF}}_{\rightarrow ml'}(\overline{a}_{ml'})  \left( \overline{\mathrm{m}}^{\mathrm{BP},[\iota]}_{ml' \leftarrow \overline{a}_{ml'}} \right)^{\rho_\varphi} 
    \end{aligned}
\end{equation}
where $C_2 = \prod_{l' \in \overline{\mathcal{L}}_m \backslash l} \prod_{o=m}^M \left( \overline{m}^{\mathrm{BP},[\iota],0}_{ml' \leftarrow o  } \right)^{\rho_\varphi}$ and we introduce $\overline{\mathrm{m}}^{\mathrm{BP},[\iota]}_{ml' \leftarrow 0} = 1$. 

We also introduce a binary vector $\overline{\boldsymbol{c}}_m = [\overline{c}_{m1},\dots,\overline{c}_{m\overline{\ell}_m}]^T$, with entries $\overline{c}_{ml}\in\{0,1\}$, $l\in\overline{\mathcal{L}}_m$. Using the indicator function $\Phi_1(\cdot)$ in \eqref{eqn_lpt_indicator1}, we obtain
\begin{equation} \label{eqn_npt_message_h2a_2}
	\begin{aligned}
		\overline{m}^{\mathrm{BP},[\iota]}_{\rightarrow ml}(\overline{a}_{ml}) &= C_2\sum_{\overline{\boldsymbol{c}}_m} \sum_{\overline{\boldsymbol{a}}_{m l \sim}} \overline{p}_m(\|\overline{\boldsymbol{a}}_m\|_0) \bigg( \prod_{l' \in \overline{\mathcal{L}}_m} \Phi_1(\overline{a}_{ml'}, \overline{c}_{ml'}) \bigg) \prod_{l' \in \overline{\mathcal{L}}_m \backslash l} \overline{m}^{\mathrm{MF}}_{\rightarrow ml'}(\overline{a}_{ml'}) \left( \overline{\mathrm{m}}^{\mathrm{BP},[\iota]}_{ml' \leftarrow \overline{a}_{ml'}} \right)^{\rho_\varphi}  \\
		&= C_2 \sum_{\overline{c}_{ml}} \Phi_1(\overline{a}_{ml}, \overline{c}_{ml}) \sum_{\overline{\boldsymbol{c}}_{ml\sim}} \overline{p}_m(\|\overline{\boldsymbol{a}}_m\|_0) \prod_{l' \in \overline{\mathcal{L}}_m \backslash l} \sum_{\overline{a}_{m l'}} \Phi_1(\overline{a}_{ml'}, \overline{c}_{ml'}) \overline{m}^{\mathrm{MF}}_{\rightarrow ml'}(\overline{a}_{ml'})  \left( \overline{\mathrm{m}}^{\mathrm{BP},[\iota]}_{ml' \leftarrow \overline{a}_{ml'}} \right)^{\rho_\varphi}  \\
		&= C_2 \sum_{\overline{c}_{ml}} \Phi_1(\overline{a}_{ml}, \overline{c}_{ml}) \sum_{\overline{\boldsymbol{c}}_{ml\sim}} \overline{p}_m(\|\overline{\boldsymbol{a}}_m\|_0) \prod_{l' \in \overline{\mathcal{L}}_m \backslash l} \left( \widetilde{\overline{m}}^{\mathrm{BP},[\iota]}_{\rightarrow ml'} \right)^{\overline{c}_{ml'}}
    \end{aligned}
\end{equation}
where $\widetilde{\overline{m}}^{\mathrm{BP},[\iota]}_{\rightarrow ml'} = \sum_{m' = m}^M \overline{m}^{\mathrm{MF}}_{\rightarrow ml'}(m')  \left( \overline{\mathrm{m}}^{\mathrm{BP},[\iota]}_{ml' \leftarrow m'} \right)^{\rho_\varphi}$, and note that
\begin{equation} \nonumber
	\sum_{\overline{a}_{m l'}} \Phi_1(\overline{a}_{ml'}, 0) \overline{m}^{\mathrm{MF}}_{\rightarrow ml'}(\overline{a}_{ml'})  \left( \overline{\mathrm{m}}^{\mathrm{BP},[\iota]}_{ml' \leftarrow \overline{a}_{ml'}} \right)^{\rho_\varphi} = 1.
\end{equation}
Substituting the indicator function $\Phi_2(\cdot)$ in \eqref{eqn_lpt_indicator2} into \eqref{eqn_npt_message_h2a_2}, we obtain
\begin{equation} 
    \overline{m}^{\mathrm{BP},[\iota]}_{\rightarrow ml}(\overline{a}_{ml}) = C_2 \sum_{\overline{c}_{ml}} \Phi_1(\overline{a}_{ml}, \overline{c}_{ml}) \sum^{\overline{\ell}_m - 1}_{\varepsilon = 0} \overline{p}_m(\varepsilon + \overline{c}_{ml}) \sum_{\overline{\boldsymbol{c}}_{ml\sim}}  \Phi_2(\overline{\boldsymbol{c}}_{ml\sim}, \varepsilon )  \prod_{l' \in \overline{\mathcal{L}}_m \backslash l} \left( \widetilde{\overline{m}}^{\mathrm{BP},[\iota]}_{\rightarrow ml'} \right)^{\overline{c}_{ml'}}.
\end{equation}

Considering that for each $\varepsilon$ there are $C^{\overline{\ell}_m - 1}_\varepsilon$ different vectors $\overline{\boldsymbol{c}}_{ml\sim}$ for which $\Phi_{2}(\overline{\boldsymbol{c}}_{ml\sim},\varepsilon)$ is one and that the $\widetilde{\overline{m}}^{\mathrm{BP},[\iota]}_{\rightarrow ml'}$ are identical for all $l'\in\overline{\mathcal{L}}_m$, we obtain
\begin{equation}
	\begin{aligned}
    \overline{m}^{\mathrm{BP},[\iota]}_{\rightarrow ml}(\overline{a}_{ml}) &= C_2\sum_{\overline{c}_{ml}} \Phi_1(\overline{a}_{ml}, \overline{c}_{ml}) \sum^{\overline{\ell}_m - 1}_{\varepsilon = 0}  \frac{(\overline{\ell}_m - 1)! \overline{p}_m(\varepsilon + \overline{c}_{ml})}{\varepsilon!(\overline{\ell}_m - \varepsilon - 1)!}  \prod_{l' \in \overline{\mathcal{L}}_m \backslash l} \left( \widetilde{\overline{m}}^{\mathrm{BP},[\iota]}_{\rightarrow ml'} \right)^{\overline{c}_{ml'}} \\
    &= \begin{cases}
    	C_2 \sum_{\varepsilon = 0}^{\overline{\ell}_m - 1} \frac{\overline{f}^1_m(\varepsilon + 1)}{\varepsilon!} \left( \widetilde{\overline{m}}^{\mathrm{BP},[\iota]}_{\rightarrow ml'} \right)^{\varepsilon},& \overline{a}_{ml} \in\{m,\dots,M\} \\
    	C_2 \sum_{\varepsilon = 0}^{\overline{\ell}_m - 1} \frac{\overline{f}^0_m(\varepsilon)}{\varepsilon!} \left( \widetilde{\overline{m}}^{\mathrm{BP},[\iota]}_{\rightarrow ml'} \right)^{\varepsilon},& \overline{a}_{ml} = 0
    \end{cases}
    \end{aligned}
\end{equation}
where
\begin{equation}
	\begin{aligned}
		\overline{f}^1_m(\varepsilon) &= \frac{\lambda_{\mathrm{n}} \overline{c}^\gamma_m p_D}{(1-p_D)\overline{\ell}_m}\frac{ (\overline{\beta}_m)^{\overline{\alpha}_m} \Gamma(\varepsilon) }{(\overline{\beta}_m+1)^{\varepsilon + \overline{\alpha}_m}\Gamma(\overline{\alpha}_m)} \\ 
		\overline{f}^0_m(\varepsilon) 
		&= \begin{cases}
			\frac{\lambda_{\mathrm{n}} \overline{c}^\gamma_m p_D (\overline{\ell}_m - \varepsilon)}{(1-p_D)\overline{\ell}_m}\frac{ (\overline{\beta}_m)^{\overline{\alpha}_m} \Gamma(\varepsilon) }{(\overline{\beta}_m+1)^{\varepsilon + \overline{\alpha}_m}\Gamma(\overline{\alpha}_m)},& \varepsilon>0 \\
			1,& \varepsilon = 0.
		\end{cases}
    \end{aligned}
\end{equation}
Finally, the two distinct values in \eqref{eqn_distinctValues} for $\overline{m}^{\mathrm{BP},[\iota]}_{\rightarrow ml}(\overline{a}_{ml})$ are given by
\begin{equation}
	\begin{aligned}
		\overline{m}^{\mathrm{BP},[\iota], 1}_{\rightarrow ml} &= C_2 \sum_{\varepsilon = 0}^{\overline{\iota}_m - 1} \frac{1}{\varepsilon!} \overline{f}^1_m(\varepsilon + 1) \left( \widetilde{\overline{m}}^{\mathrm{BP},[\iota]}_{\rightarrow ml'} \right)^{\varepsilon} \\
		\overline{m}^{\mathrm{BP},[\iota], 0}_{\rightarrow ml} &= C_2 \sum_{\varepsilon = 0}^{\overline{\iota}_m - 1} \frac{1}{\varepsilon!} \overline{f}^0_m(\varepsilon) \left( \widetilde{\overline{m}}^{\mathrm{BP},[\iota]}_{\rightarrow ml'} \right)^{\varepsilon}
	\end{aligned}
\end{equation}
and the rescaled message $\overline{\mathrm{m}}^{\mathrm{BP},[\iota]}_{\rightarrow ml}$ is given by 
\begin{equation}
	\overline{\mathrm{m}}^{\mathrm{BP},[\iota]}_{\rightarrow ml} = \frac{ \overline{m}^{\mathrm{BP},[\iota], 1}_{\rightarrow ml} } { \overline{m}^{\mathrm{BP},[\iota], 0}_{\rightarrow ml} } = \frac{ \sum_{\varepsilon = 0}^{\overline{\ell}_m - 1} \overline{f}^1_m(\varepsilon + 1) \left( \widetilde{\overline{m}}^{\mathrm{BP},[\iota]}_{\rightarrow ml'} \right)^{\varepsilon} / \varepsilon!} { \sum_{\varepsilon = 0}^{\overline{\ell}_m - 1} \overline{f}^0_m(\varepsilon) \left( \widetilde{\overline{m}}^{\mathrm{BP},[\iota]}_{\rightarrow ml'} \right)^{\varepsilon} / \varepsilon! }.
\end{equation}

\subsubsection{The rescaled message \texorpdfstring{$\overline{\mathrm{m}}^{\mathrm{BP},[\iota]}_{ml\rightarrow o}$}{}}
According to equation (38) in \cite{MJDL}, the two distinct values in \eqref{eqn_distinctValues} for message $\overline{m}^{\mathrm{BP},[\iota]}_{ml\to o}(\boldsymbol{b}_o)$ are given by 
\begin{equation} \label{eqn_unnormalized_message_a2b_new}
    \begin{aligned}
        \overline{m}^{\mathrm{BP},[\iota],1}_{ml\rightarrow o} &= \left( \overline{m}^{\mathrm{BP},[\iota],1}_{ml \leftarrow o } \right)^{\rho_\varphi-1} \overline{m}^{\mathrm{MF}}_{\rightarrow ml}(o) \overline{m}^{\mathrm{BP},[\iota]}_{\rightarrow ml}(o)  \prod_{o'=m, o'\neq o}^{M} \left( \overline{m}^{\mathrm{BP},[\iota],0}_{ml \leftarrow o' }  \right)^{\rho_\varphi} \\
        \overline{m}^{\mathrm{BP},[\iota],0}_{ml\rightarrow o} &= \left( \overline{m}^{\mathrm{BP},[\iota],0}_{ml \leftarrow o } \right)^{\rho_\varphi-1} \sum_{\overline{a}_{m l} \in \overline{\mathcal{A}}_m \backslash o} \overline{m}^{\mathrm{MF}}_{\rightarrow ml}(\overline{a}_{ml}) \overline{m}^{\mathrm{BP},[\iota]}_{\rightarrow ml}(\overline{a}_{ml}) \prod_{o'=m, o'\neq o}^{M} \left( \overline{m}^{\mathrm{BP},[\iota]}_{ml \leftarrow o' }(\overline{a}_{ml}) \right)^{\rho_\varphi} 
	\end{aligned}
\end{equation}
where $\overline{\mathcal{A}}_m = \{0, m, m+1,\dots,M\}$. Substituting \eqref{eqn_unnormalized_message_a2b_new} into \eqref{eqn_message_normalization_a2b_new}, and using \eqref{eqn_message_normalization_b2a_new} and \eqref{eqn_message_normalization_h2a_new}, we obtain 
\begin{equation}
	\begin{aligned}
		\overline{\mathrm{m}}^{\mathrm{BP},[\iota]}_{ml\rightarrow o} &= \frac{ \left( \overline{m}^{\mathrm{BP},[\iota],1}_{ml \leftarrow o } \right)^{\rho_\varphi-1} \overline{m}^{\mathrm{MF}}_{\rightarrow ml}(o) \overline{m}^{\mathrm{BP},[\iota]}_{\rightarrow ml}(o)  \prod_{o'=m, o'\neq o}^{M} \left( \overline{m}^{\mathrm{BP},[\iota],0}_{ml \leftarrow o' } \right)^{\rho_\varphi} }{ \left( \overline{m}^{\mathrm{BP},[\iota],0}_{ml \leftarrow o } \right)^{\rho_\varphi-1} \sum_{\overline{a}_{m l} \in \overline{\mathcal{A}}_m \backslash o} \overline{m}^{\mathrm{MF}}_{\rightarrow ml}(\overline{a}_{ml}) \overline{m}^{\mathrm{BP},[\iota]}_{\rightarrow ml}(\overline{a}_{ml}) \prod_{o'=m, o'\neq o}^{M} \left( \overline{m}^{\mathrm{BP},[\iota]}_{ml \leftarrow o' }(\overline{a}_{ml}) \right)^{\rho_\varphi} } \\
		&= \frac{ \left( \overline{\mathrm{m}}^{\mathrm{BP},[\iota]}_{ml \leftarrow o } \right)^{\rho_\varphi-1} \overline{m}^{\mathrm{MF}}_{\rightarrow ml}(o) \overline{\mathrm{m}}^{\mathrm{BP},[\iota]}_{\rightarrow ml} \overline{m}^{\mathrm{BP},[\iota],0}_{\rightarrow ml}  \prod_{o'=m, o'\neq o}^{M} \left( \overline{m}^{\mathrm{BP},[\iota],0}_{ml \leftarrow o'} \right)^{\rho_\varphi} } { \sum_{\overline{a}_{m l} \in \overline{\mathcal{A}}_m \backslash o} \overline{m}^{\mathrm{MF}}_{\rightarrow ml}(\overline{a}_{ml}) \overline{\mathrm{m}}^{\mathrm{BP},[\iota]}_{\rightarrow ml}(\overline{a}_{ml}) \overline{m}^{\mathrm{BP},[\iota],0}_{\rightarrow ml}  \prod_{o'=m, o'\neq o}^{M} \left( \overline{\mathrm{m}}^{\mathrm{BP},[\iota]}_{ml \leftarrow o' }(\overline{a}_{ml}) \overline{m}^{\mathrm{BP},[\iota],0}_{ml \leftarrow o' } \right)^{\rho_\varphi} } \\
		&= \frac{ \left( \overline{\mathrm{m}}^{\mathrm{BP},[\iota]}_{ml \leftarrow o } \right)^{\rho_\varphi-1} \overline{m}^{\mathrm{MF}}_{\rightarrow ml}(o) \overline{\mathrm{m}}^{\mathrm{BP},[\iota]}_{\rightarrow ml} } { \sum_{\overline{a}_{m l} = m, \overline{a}_{m l} \neq o}^M \overline{m}^{\mathrm{MF}}_{\rightarrow ml}(\overline{a}_{ml}) \overline{\mathrm{m}}^{\mathrm{BP},[\iota]}_{\rightarrow ml}(\overline{a}_{ml}) \prod_{o'=m, o'\neq o}^{M} \left( \overline{\mathrm{m}}^{\mathrm{BP},[\iota]}_{ml \leftarrow o' }(\overline{a}_{ml}) \right)^{\rho_\varphi} } \\
		&= \frac{ \left( \overline{\mathrm{m}}^{\mathrm{BP},[\iota]}_{ml \leftarrow o } \right)^{\rho_\varphi-1} \overline{m}^{\mathrm{MF}}_{\rightarrow ml}(o) \overline{\mathrm{m}}^{\mathrm{BP},[\iota]}_{\rightarrow ml} } { \sum_{o'=m, o'\neq o}^M \overline{m}^{\mathrm{MF}}_{\rightarrow ml}(o') \overline{\mathrm{m}}^{\mathrm{BP},[\iota]}_{\rightarrow ml}  \left( \overline{\mathrm{m}}^{\mathrm{BP},[\iota]}_{ml \leftarrow o' } \right)^{\rho_\varphi} }.
	\end{aligned}
\end{equation}
Note that $\overline{m}^{\mathrm{MF}}_{\rightarrow ml}(0)= 0$ and $\overline{\mathrm{m}}^{\mathrm{BP},[\iota]}_{ml\leftarrow o}(\overline{a}_{ml}\neq o) = 1$.  

\subsection{The messages \texorpdfstring{$\underline{m}^{\mathrm{BP}}_{\leftarrow n}(\underline{\boldsymbol{y}}_n)$}{} and \texorpdfstring{$\overline{m}^{\mathrm{BP}}_{\leftarrow m}(\overline{\boldsymbol{y}}_m)$}{}}
After the final iteration of the rescaled data association messages, we proceed to calculate the messages $\underline{m}^{\mathrm{BP}}_{\leftarrow n}(\underline{\boldsymbol{y}}_n)$ and $\overline{m}^{\mathrm{BP}}_{\leftarrow m}(\overline{\boldsymbol{y}}_m)$. The derivations of these messages are similar to that of the rescaled messages $\underline{\mathrm{m}}^{\mathrm{BP},[\iota]}_{\to nl}$ and $\overline{\mathrm{m}}^{\mathrm{BP},[\iota]}_{\to ml}$. 

\subsubsection{The message \texorpdfstring{$\underline{m}^{\mathrm{BP}}_{\leftarrow n}(\underline{\boldsymbol{y}}_n)$}{}}
According to (21) in \cite{MJDL}, we have
\begin{equation}
	\begin{aligned}
		\underline{m}^{\mathrm{BP}}_{\leftarrow n}(\underline{\boldsymbol{y}}_n) &= \sum_{\boldsymbol{\underline{a}}_{n}} (\underline{h}_n(\underline{\boldsymbol{y}}_n, \underline{\boldsymbol{a}}_n ))^{\frac{1}{\rho_h}} \prod_{l \in \underline{\mathcal{L}}_n} \left(\underline{m}^{\mathrm{BP},[I_{\mathrm{BP}}]}_{\rightarrow nl}(\underline{a}_{nl})\right)^{\rho_h - 1} \underline{m}^{\mathrm{MF}}_{\rightarrow nl}(\underline{a}_{nl}) \prod_{m=1}^M \left( \underline{m}^{\mathrm{BP},[I_{\mathrm{BP}}]}_{nl \leftarrow m } (\underline{a}_{nl})  \right)^{\rho_\varphi}  \\
		&= C_1 \sum_{\boldsymbol{\underline{a}}_{n}} \underline{h}_n(\underline{\boldsymbol{y}}_n, \underline{\boldsymbol{a}}_n ) \prod_{l \in \underline{\mathcal{L}}_n } \underline{m}^{\mathrm{MF}}_{\rightarrow nl}(\underline{a}_{nl}) \left( \underline{\mathrm{m}}^{\mathrm{BP},[I_{\mathrm{BP}}]}_{nl \leftarrow \underline{a}_{nl}}  \right)^{\rho_\varphi}.  
	\end{aligned}
\end{equation}
Note that $\rho_h=1$. For $\underline{r}_n=0$, we have $\underline{m}^{\mathrm{BP}}_{\leftarrow n}(\underline{\boldsymbol{x}}_n, 0) = C$ (obtained at $\boldsymbol{\underline{a}}_{n} = \boldsymbol{0}$). For $\underline{r}_n=1$, incorporating the indicator function $\Phi_1(\cdot)$ given by \eqref{eqn_lpt_indicator1} into thr above formula, we obtain
\begin{equation}
	\begin{aligned}
		\underline{m}^{\mathrm{BP}}_{\leftarrow n}(\underline{\boldsymbol{x}}_n, 1) &= C_1 \sum_{\underline{\boldsymbol{c}}_{n}}  \sum_{\boldsymbol{\underline{a}}_{n}} \bigg( \prod_{l \in \underline{\mathcal{L}}_n } \Phi_1(\underline{a}_{nl}, \underline{\boldsymbol{c}}_{nl}) \bigg) \underline{h}_n(\underline{\boldsymbol{x}}_n, 1, \underline{\boldsymbol{a}}_n ) \prod_{l \in \underline{\mathcal{L}}_n } \underline{m}^{\mathrm{MF}}_{\rightarrow nl}(\underline{a}_{nl}) \left( \underline{\mathrm{m}}^{\mathrm{BP},[I_{\mathrm{BP}}]}_{nl \leftarrow \underline{a}_{nl}}  \right)^{\rho_\varphi} \\ 
		&= C_1 \sum_{\underline{\boldsymbol{c}}_{n}} \underline{h}_n(\underline{\boldsymbol{x}}_n, 1, \underline{\boldsymbol{c}}_n ) \prod_{l \in \underline{\mathcal{L}}_n } \sum_{\underline{a}_{nl}=0}^M \Phi_1(\underline{a}_{nl}, \underline{c}_{nl}) \underline{m}^{\mathrm{MF}}_{\rightarrow nl}(\underline{a}_{nl}) \left( \underline{\mathrm{m}}^{\mathrm{BP},[I_{\mathrm{BP}}]}_{nl \leftarrow \underline{a}_{nl}}  \right)^{\rho_\varphi} \\
		&= C_1 \sum_{\underline{\boldsymbol{c}}_{n}} \underline{h}_n(\underline{\boldsymbol{x}}_n, 1, \underline{\boldsymbol{c}}_n ) \prod_{l \in \underline{\mathcal{L}}_n } \left( \widetilde{\underline{m}}^{\mathrm{BP},[I_{\mathrm{BP}}]}_{\rightarrow nl} \right)^{\underline{c}_{nl}}
	\end{aligned}
\end{equation}
where $\widetilde{\underline{m}}^{\mathrm{BP},[I_{\mathrm{BP}}]}_{\rightarrow nl'} = \sum_{m = 1}^M \underline{m}^{\mathrm{MF}}_{\rightarrow nl'}(m) \left( \underline{\mathrm{m}}^{\mathrm{BP},[I_{\mathrm{BP}}]}_{nl' \leftarrow m} \right)^{\rho_\varphi}$. Incorporating the indicator function $\Phi_2(\cdot)$ given by \eqref{eqn_lpt_indicator2} into the above formula, we obtain 
\begin{equation}
	\underline{m}^{\mathrm{BP}}_{\leftarrow n}(\underline{\boldsymbol{x}}_n, 1) = C_1 \sum_{\varepsilon=0}^{\underline{\ell}_n} \sum_{\underline{\boldsymbol{c}}_{n}} \Phi_2(\underline{\boldsymbol{c}}_{n},\varepsilon) \underline{h}_n(\underline{\boldsymbol{x}}_n, 1, \underline{\boldsymbol{c}}_n ) \prod_{l \in \underline{\mathcal{L}}_n } \left( \widetilde{\underline{m}}^{\mathrm{BP},[I_{\mathrm{BP}}]}_{\rightarrow nl} \right)^{\underline{c}_{nl}}.
\end{equation}
Since for each $\varepsilon$ there are $C^{\underline{\ell}_n}_\varepsilon$ different vectors $\underline{\boldsymbol{c}}_{n}$ for which $\Phi_{2}(\underline{\boldsymbol{c}}_{n},\varepsilon)$ is one and that the $\widetilde{\underline{m}}^{\mathrm{BP},[I_{\mathrm{BP}}]}_{\rightarrow nl}$ are identical for all $l\in\underline{\mathcal{L}}_n$, we obtain
\begin{equation}
	\underline{m}^{\mathrm{BP}}_{\leftarrow n}(\underline{\boldsymbol{x}}_n, 1) = C_1 \sum_{\varepsilon=0}^{\underline{\ell}_n} \underline{f}_n(\varepsilon | \underline{\gamma}_n) \left( \widetilde{\underline{m}}^{\mathrm{BP},[I_{\mathrm{BP}}]}_{\rightarrow nl} \right)^\varepsilon
\end{equation}
where 
\begin{equation}
	\underline{f}_n(\varepsilon | \underline{\gamma}_n) = 
	\begin{cases}
        \frac{\underline{c}^\gamma_n p_D \underline{\gamma}_n^{\varepsilon} \exp(-\underline{\gamma}_n)}{\varepsilon!} , &  \varepsilon > 0 \\
        \underline{c}^\gamma_n (1-p_D + p_D \exp(-\underline{\gamma}_n)), & \varepsilon = 0.
    \end{cases}
\end{equation}
Finally, we have
\begin{equation}
	\begin{aligned}
		\underline{m}^{\mathrm{BP}}_{\leftarrow n}(\underline{\boldsymbol{y}}_n) &= 
		\begin{cases}
			C_1,& \underline{r}_n = 0, \\
			C_1 \sum_{\varepsilon=0}^{\underline{\ell}_n} \underline{f}_n(\varepsilon | \underline{\gamma}_n) \left( \widetilde{\underline{m}}^{\mathrm{BP},[I_{\mathrm{BP}}]}_{\rightarrow nl} \right)^\varepsilon, & \underline{r}_n = 1
		\end{cases} \\
		&\propto \begin{cases}
			1,& \underline{r}_n = 0, \\
			\sum_{\varepsilon=0}^{\underline{\ell}_n} \underline{f}_n(\varepsilon | \underline{\gamma}_n) \left( \widetilde{\underline{m}}^{\mathrm{BP},[I_{\mathrm{BP}}]}_{\rightarrow nl} \right)^\varepsilon, & \underline{r}_n = 1
		\end{cases} 
	\end{aligned}
\end{equation}

\subsubsection{The message \texorpdfstring{$\overline{m}^{\mathrm{BP}}_{\leftarrow m}(\overline{\boldsymbol{y}}_m)$}{}}
According to (21) in \cite{MJDL}, we have
\begin{equation}
	\begin{aligned} 
        \overline{m}^{\mathrm{BP}}_{\leftarrow m}(\overline{\boldsymbol{y}}_m) &= \sum_{\overline{\boldsymbol{a}}_{m }} (\overline{h}_n\left(\overline{\boldsymbol{y}}_m, \overline{\boldsymbol{a}}_m \right))^{\frac{1}{\rho_h}}  \prod_{l \in \overline{\mathcal{L}}_m }  \left(\overline{m}^{\mathrm{BP},[I_{\mathrm{BP}}]}_{\rightarrow ml}(\overline{a}_{ml})\right)^{\rho_h-1} \overline{m}^{\mathrm{MF}}_{\rightarrow ml}(\overline{a}_{ml}) \prod_{o=m}^M \left( \overline{m}^{\mathrm{BP},[I_{\mathrm{BP}}]}_{ml \leftarrow o  } (\overline{a}_{ml})  \right)^{\rho_\varphi}  \\
        &= C_2 \sum_{\overline{\boldsymbol{a}}_{m}} \overline{h}_m \left(\overline{\boldsymbol{y}}_m, \overline{\boldsymbol{a}}_m \right) \prod_{l \in \overline{\mathcal{L}}_m } \overline{m}^{\mathrm{MF}}_{\rightarrow ml}(\overline{a}_{ml}) \left( \overline{\mathrm{m}}^{\mathrm{BP},[I_{\mathrm{BP}}]}_{ml \leftarrow \overline{a}_{ml} }  \right)^{\rho_\varphi}
	\end{aligned}
\end{equation}
where the constant $C_2 = \prod_{l \in \overline{\mathcal{L}}_m } \prod_{o=m}^M \left(\overline{m}^{\mathrm{BP},[I_{\mathrm{BP}}], 0}_{ml \leftarrow o}  \right)^{\rho_\varphi} $ and note that $\rho_h=1$. For $\overline{r}_m=0$, we have $\underline{m}^{\mathrm{BP}}_{\leftarrow n}(\underline{\boldsymbol{x}}_n, 0) = 0$ (obtained at $\overline{\boldsymbol{a}}_{m } = \boldsymbol{0}$). For $\overline{r}_m=1$, incorporating the indicator function $\Phi_1(\cdot)$ given by \eqref{eqn_lpt_indicator1} into the above formula, we obtain
\begin{equation}
	\begin{aligned} 
        \overline{m}^{\mathrm{BP}}_{\leftarrow m}(\overline{\boldsymbol{x}}_m, 1) &= C_2 \sum_{\overline{\boldsymbol{c}}_{m }} \sum_{\overline{\boldsymbol{a}}_{m }} \left( \prod_{l \in \overline{\mathcal{L}}_m } \Phi_1(\overline{a}_{ml}, \overline{c}_{ml}) \right) \overline{h}_m \left(\overline{\boldsymbol{x}}_m, 1, \overline{\boldsymbol{a}}_m \right)  \prod_{l \in \overline{\mathcal{L}}_m } \overline{m}^{\mathrm{MF}}_{\rightarrow ml}(\overline{a}_{ml}) \left( \overline{\mathrm{m}}^{\mathrm{BP},[I_{\mathrm{BP}}]}_{ml \leftarrow \overline{a}_{ml} }  \right)^{\rho_\varphi} \\
        &=  C_2 \sum_{\overline{\boldsymbol{c}}_{m }} \overline{h}_n\left(\overline{\boldsymbol{x}}_m, 1, \overline{\boldsymbol{a}}_m \right) \prod_{l \in \overline{\mathcal{L}}_m } \sum_{\overline{a}_{ml}\in\overline{\mathcal{A}}_m} \Phi_1(\overline{a}_{ml}, \overline{c}_{ml}) \overline{m}^{\mathrm{MF}}_{\rightarrow ml}(\overline{a}_{ml}) \left( \overline{\mathrm{m}}^{\mathrm{BP},[I_{\mathrm{BP}}]}_{ml \leftarrow \overline{a}_{ml} }  \right)^{\rho_\varphi} \\
        &= C \sum_{\overline{\boldsymbol{c}}_{m }} \overline{h}_m \left(\overline{\boldsymbol{x}}_m, 1, \overline{\boldsymbol{a}}_m \right) \prod_{l \in \overline{\mathcal{L}}_m } \left( \widetilde{\overline{m}}^{\mathrm{BP},[I_{\mathrm{BP}}]}_{\rightarrow ml}  \right)^{\overline{c}_{ml}}
	\end{aligned}
\end{equation}
where $\widetilde{\overline{m}}^{\mathrm{BP}}_{\rightarrow ml} = \sum_{m' = m}^M \overline{m}^{\mathrm{MF}}_{\rightarrow ml}(m') \left( \overline{\mathrm{m}}^{\mathrm{BP},[I_{\mathrm{BP}}]}_{ml \leftarrow m'} \right)^{\rho_\varphi}$. Substituting the indicator function $\Phi_2(\cdot)$ given by \eqref{eqn_lpt_indicator2} into the above formula, we obtain
\begin{equation}
    \overline{m}^{\mathrm{BP}}_{\leftarrow m}(\overline{\boldsymbol{x}}_m, 1) = C_2 \sum_{\varepsilon=0}^{\overline{\ell}_m}\sum_{\overline{\boldsymbol{c}}_{m }} \Phi_2(\overline{\boldsymbol{c}}_{m},\varepsilon)\overline{h}_m\left(\overline{\boldsymbol{x}}_m, 1, \overline{\boldsymbol{a}}_m \right) \prod_{l \in \overline{\mathcal{L}}_m } \left( \widetilde{\overline{m}}^{\mathrm{BP},[I_{\mathrm{BP}}]}_{\rightarrow ml}  \right)^{\overline{c}_{ml}}.
\end{equation}
Since for each $\varepsilon$ there are $C^{\overline{\ell}_m}_\varepsilon$ different vectors $\overline{\boldsymbol{c}}_{m}$ for which $\Phi_{2}(\overline{\boldsymbol{c}}_{m},\varepsilon)$ is one and that the $\widetilde{\overline{m}}^{\mathrm{BP},[I_{\mathrm{BP}}]}_{\rightarrow ml}$ are identical for all $l\in\overline{\mathcal{L}}_m$, we obtain
\begin{equation}
    \overline{m}^{\mathrm{BP},[I_{\mathrm{BP}}]}_{\leftarrow m}(\overline{\boldsymbol{x}}_m, 1) = C_2 \sum_{\varepsilon=0}^{\overline{\ell}_m} \overline{f}_m(\varepsilon | \overline{\gamma}_m) \left( \widetilde{\overline{m}}^{\mathrm{BP},[I_{\mathrm{BP}}]}_{\rightarrow ml} \right)^\varepsilon
\end{equation}
where
\begin{equation}
	\overline{f}_m(\varepsilon | \overline{\gamma}_m) = 
	\begin{cases}
		\frac{ \overline{c}^\gamma_m p_D \overline{\gamma}_m^{\varepsilon} \exp(-\overline{\gamma}_m) } {(1-p_D + p_D \exp(-\overline{\gamma}_m))\varepsilon!} , &  \varepsilon>0 \\
		0, &  \varepsilon=0.
	\end{cases}
\end{equation}
Finally, we have 
\begin{equation}
	\begin{aligned}
		\overline{m}^{\mathrm{BP}}_{\leftarrow m}(\overline{\boldsymbol{y}}_m) &= \begin{cases}
			0,& \overline{r}_m = 0 \\
			C_2 \sum_{\varepsilon=0}^{\overline{\ell}_m} \overline{f}_m(\varepsilon | \overline{\gamma}_m) \left( \widetilde{\overline{m}}^{\mathrm{BP},[I_{\mathrm{BP}}]}_{\rightarrow ml}  \right)^\varepsilon, & \overline{r}_m = 1
		\end{cases} \\
		&\propto \begin{cases}
			0,& \overline{r}_m = 0 \\
			\sum_{\varepsilon=0}^{\overline{\ell}_m} \overline{f}_m(\varepsilon | \overline{\gamma}_m) \left( \widetilde{\overline{m}}^{\mathrm{BP},[I_{\mathrm{BP}}]}_{\rightarrow ml}  \right)^\varepsilon, & \overline{r}_m = 1.
		\end{cases}
	\end{aligned}
\end{equation}

\subsection{The beliefs \texorpdfstring{$\underline{q}_n(\underline{\boldsymbol{y}}_n)$}{} and \texorpdfstring{$\overline{q}_m(\overline{\boldsymbol{y}}_m)$}{}}

\subsubsection{The belief \texorpdfstring{$\underline{q}_n(\underline{\boldsymbol{y}}_n)$}{}}
According to equation (20) in \cite{MJDL}, the belief for $\underline{\boldsymbol{y}}_n$ is given by 
\begin{equation} \label{eqn_lpt_belief}
	\begin{aligned}
		\underline{p}_n(\underline{\boldsymbol{y}}_n) &\propto \underline{m}^{\mathrm{MF}}_{\rightarrow n}(\underline{\boldsymbol{y}}_n) \left( \underline{m}^{\mathrm{BP}}_{\leftarrow n}(\underline{\boldsymbol{y}}_n) \right)^{\rho_h} \prod_{l=1}^{\underline{\ell}_n} \underline{m}^{\mathrm{MF}}_{\leftarrow nl}(\underline{\boldsymbol{y}}_n) \\
		&= \underline{p}^+_n(\underline{\boldsymbol{y}}_n) \underline{m}^{\mathrm{BP}}_{\leftarrow n}(\underline{\boldsymbol{y}}_n) \left( \underline{m}^{\mathrm{MF}}_{\leftarrow nl}(\underline{\boldsymbol{y}}_n) \right)^{\underline{\ell}_n}  \\
		&= \underline{p}^+_n(\underline{\boldsymbol{y}}_n) \underline{m}^{\mathrm{BP}}_{\leftarrow n}(\underline{\boldsymbol{y}}_n) \prod_{\underline{a}_{nl} = 0}^M \left(\underline{g}_{nl}(\underline{\boldsymbol{y}}_n, \underline{a}_{nl} ; \boldsymbol{z}) \right)^{\underline{\ell}_n \underline{q}_{nl}(\underline{a}_{nl})}  \\
		&\propto \begin{cases}
			\underline{p}^{+,0}_n p_{\mathrm{d}}(\underline{\boldsymbol{x}}_n),& \underline{r}_n = 0\\
            \underline{p}^{+,1}_n \underline{p}^+_n(\underline{\boldsymbol{x}}_n) \left( \sum_{\varepsilon=0}^{\underline{\ell}_n} \underline{f}_n(\varepsilon | \underline{\gamma}_n) \left( \widetilde{\underline{m}}^{\mathrm{BP},[I_{\mathrm{BP}}]}_{\rightarrow nl} \right)^\varepsilon \right) \left( \prod_{m=1}^M \left( \frac{p(\boldsymbol{z}_m | \underline{\boldsymbol{\xi}}_n, \underline{\boldsymbol{e}}_n)} {\kappa_{\mathrm{c}}} \right)^{\underline{\ell}_n \underline{q}_{nl}(m)} \right), & \underline{r}_n = 1
		\end{cases} 
	\end{aligned}
\end{equation}
where the formulas of the terms in the above equation refer to \cite{MJDL}, and note that $\underline{m}^{\mathrm{MF}}_{\leftarrow nl}(\underline{\boldsymbol{y}}_n)$ are identical for all $l\in\underline{\mathcal{L}}_n$ and $\underline{g}_n(\underline{\boldsymbol{x}}_n, \underline{r}_n=1,\underline{a}_{nl}= 0 ; \boldsymbol{z})=1$. We reformulate the terms for  $\underline{r}_n = 1$ as follows
\begin{equation} \label{eqn_lpt_jpdf}
	\begin{aligned}
		&\underline{p}^+_n(\underline{\boldsymbol{x}}_n) \left( \sum_{\varepsilon=0}^{\underline{\ell}_n}  \underline{f}_n(\varepsilon | \underline{\gamma}_n) \left( \widetilde{\underline{m}}^{\mathrm{BP},[I_{\mathrm{BP}}]}_{\rightarrow nl} \right)^\varepsilon \right) \left( \prod_{m=1}^M \left( \frac{p(\boldsymbol{z}_m | \underline{\boldsymbol{\xi}}_n, \underline{\boldsymbol{e}}_n)} {\kappa_{\mathrm{c}}} \right)^{\underline{\ell}_n \underline{q}_{nl}(m)} \right) \\ 
		=& \left( \sum_{\varepsilon=0}^{\underline{\ell}_n}  \underline{p}^+_n(\underline{\gamma}_n) \underline{f}_n(\varepsilon | \underline{\gamma}_n) \left( \widetilde{\underline{m}}^{\mathrm{BP},[I_{\mathrm{BP}}]}_{\rightarrow nl} \right)^\varepsilon \right) \left( \underline{p}^+_n(\underline{\boldsymbol{\xi}}_n) \underline{p}^+_n(\underline{\boldsymbol{e}}_n) \prod_{m=1}^M \left( \frac{p(\boldsymbol{z}_m | \underline{\boldsymbol{\xi}}_n, \underline{\boldsymbol{e}}_n)} {\kappa_{\mathrm{c}}} \right)^{\underline{\ell}_n \underline{q}_{nl}(m)} \right). 
	\end{aligned}
\end{equation}

Let $\underline{p}_n(\underline{\gamma}_n)$ be the belief for measurement rate given by 
\begin{equation} \label{eqn_lpt_belief_measRate}
	\underline{p}_n(\underline{\gamma}_n) = \sum_{\varepsilon=0}^{\underline{\ell}_n} \underline{p}^+_n(\underline{\gamma}_n) \underline{f}_n(\varepsilon | \underline{\gamma}_n) \left( \widetilde{\underline{m}}^{\mathrm{BP},[I_{\mathrm{BP}}]}_{\rightarrow nl} \right)^\varepsilon
\end{equation}
i.e., the first part of \eqref{eqn_lpt_jpdf}. From equation \eqref{eqn_et_legacy_integral} we know that
\begin{equation}
	\underline{\gamma}_n^{\varepsilon} \exp(-\underline{\gamma}_n) \mathrm{G}(\underline{\gamma}_n; \underline{\alpha}^+_n, \underline{\beta}^+_n) = \frac{ (\underline{\beta}^+_n)^{\underline{\alpha}^+_n} \Gamma(\varepsilon + \underline{\alpha}^+_n) }{(\underline{\beta}^+_n+1)^{\varepsilon + \underline{\alpha}^+_n}\Gamma(\underline{\alpha}^+_n)} \mathrm{G}(\underline{\gamma}_n; \varepsilon + \underline{\alpha}^+_n, \underline{\beta}^+_n + 1).
\end{equation}
Inserting the above equation into \eqref{eqn_lpt_belief_measRate}, we obtain
\begin{equation}
	\underline{p}_n(\underline{\gamma}_n) \propto \underline{c}^\gamma_n(1- p_D)\mathrm{G}(\underline{\gamma}_n; \underline{\alpha}^+_n, \underline{\beta}^+_n) + \underline{c}^\gamma_n\sum_{\varepsilon=0}^{\underline{\ell}_n} \underline{w}_n^\varepsilon \mathrm{G}(\underline{\gamma}_n; \underline{\alpha}^+_n + \varepsilon, \underline{\beta}^+_n + 1)  
\end{equation}
where
\begin{equation}
	\underline{w}_n^\varepsilon = \frac{ p_D \Gamma(\underline{\alpha}^+_n+\varepsilon)(\underline{\beta}^+_n)^{\underline{\alpha}^+_n} \left( \widetilde{\underline{m}}^{\mathrm{BP},[I_{\mathrm{BP}}]}_{\rightarrow nl} \right)^\varepsilon}{ \Gamma(\underline{\alpha}^+_n)(\underline{\beta}^+_n+1)^{\underline{\alpha}^+_n+\varepsilon} \varepsilon! }  
\end{equation}
and the normalizing constant of $\underline{p}_n(\underline{\gamma}_n)$ is given by
\begin{equation}
	\int \sum_{\varepsilon=0}^{\underline{\ell}_n} \underline{p}^+_n(\underline{\gamma}_n) \underline{f}_n(\varepsilon | \underline{\gamma}_n) \left( \widetilde{\underline{m}}^{\mathrm{BP},[I_{\mathrm{BP}}]}_{\rightarrow nl} \right)^\varepsilon \mathrm{d} \underline{\gamma}_n = \underline{c}^\gamma_n \left( 1 - p_D + \sum_{\varepsilon = 0}^{\underline{\ell}_n} \underline{w}_n^\varepsilon \right)
\end{equation}

The second part of \eqref{eqn_lpt_jpdf} can be reformulated as 
\begin{equation}
	\begin{aligned}
		&\underline{p}^+_n(\underline{\boldsymbol{\xi}}_n) \underline{p}^+_n(\underline{\boldsymbol{e}}_n) \prod_{m=1}^M \left( \frac{p(\boldsymbol{z}_m | \underline{\boldsymbol{\xi}}_n, \underline{\boldsymbol{e}}_n)} {\kappa_{\mathrm{c}}} \right)^{\underline{\ell}_n \underline{q}_{nl}(m)}  \\
		=& \kappa_{\mathrm{c}}^{-\sum_{m=1}^M \underline{\ell}_n \underline{q}_{nl}(m)}  \underline{p}^+_n(\underline{\boldsymbol{\xi}}_n) \underline{p}^+_n(\underline{\boldsymbol{e}}_n) \prod_{m=1}^M \left( \mathrm{N}(\boldsymbol{z}_m; \boldsymbol{H}\underline{\boldsymbol{\xi}}_n, s\underline{\boldsymbol{E}}_n + \boldsymbol{R}) \right)^{\underline{\ell}_n \underline{q}_{nl}(m)} \\
		=& \kappa_{\mathrm{c}}^{-\sum_{m=1}^M \underline{\ell}_n \underline{q}_{nl}(m)}  \underline{p}^+_n(\underline{\boldsymbol{\xi}}_n) \underline{p}^+_n(\underline{\boldsymbol{e}}_n) \prod_{m=1}^M p(\boldsymbol{z}_m | \underline{\boldsymbol{\xi}}_n, \underline{\boldsymbol{e}}_n) \\
		=& \kappa_{\mathrm{c}}^{-\sum_{m=1}^M \underline{\ell}_n \underline{q}_{nl}(m)} \underline{p}_n(\underline{\boldsymbol{\xi}}_n, \underline{\boldsymbol{e}}_n, \boldsymbol{z}) 
    \end{aligned}
\end{equation}
where $\underline{p}_n(\underline{\boldsymbol{\xi}}_n, \underline{\boldsymbol{e}}_n, \boldsymbol{z})$ is the joint density of kinematic state, extent and measurements. Finally, \eqref{eqn_lpt_belief} can be rewritten as
\begin{equation} \label{eqn_belief_lpt_state}
	\underline{q}_n(\underline{\boldsymbol{y}}_n) 
	\propto \begin{cases}
		\underline{p}^{+,0}_n p_{\mathrm{d}}(\underline{\boldsymbol{x}}_n),& \underline{r}_n = 0\\
		\underline{p}^{+,1}_n \underline{c}^\gamma_n ( 1 - p_D + \sum_{\varepsilon = 0}^{\underline{\ell}_n} \underline{w}_n^\varepsilon ) \kappa_{\mathrm{c}}^{-\sum_{m=1}^M \underline{\ell}_n \underline{q}_{nl}(m)} \underline{p}_n(\underline{\gamma}_n) \underline{p}_n(\underline{\boldsymbol{\xi}}_n, \underline{\boldsymbol{e}}_n, \boldsymbol{z} ), & \underline{r}_n = 1
	\end{cases} 
\end{equation}

\subsubsection{The belief \texorpdfstring{$\overline{q}_m(\overline{\boldsymbol{y}}_m)$}{}}
According to equation (20) in \cite{MJDL}, the belief for $\overline{\boldsymbol{y}}_m$ is given by 
\begin{equation} 
	\begin{aligned}
		\overline{q}_m(\overline{\boldsymbol{y}}_m) & \propto  \overline{m}^{\mathrm{MF}}_{\rightarrow m}(\overline{\boldsymbol{y}}_m) \left( \overline{m}^{\mathrm{BP}}_{\leftarrow m}(\overline{\boldsymbol{y}}_m) \right)^{\rho_h} \prod_{l=1}^{\overline{\ell}_m} \overline{m}^{\mathrm{MF}}_{\leftarrow ml}(\overline{\boldsymbol{y}}_m) \\
		&= \overline{p}_m(\overline{\boldsymbol{x}}_m) \overline{m}^{\mathrm{BP}}_{\leftarrow m}(\overline{\boldsymbol{y}}_m)  (\overline{m}^{\mathrm{MF}}_{\leftarrow ml}(\overline{\boldsymbol{y}}_m))^{\overline{\ell}_m}  \\
		&= \overline{p}_m(\overline{\boldsymbol{x}}_m) \overline{m}^{\mathrm{BP}}_{\leftarrow m}(\overline{\boldsymbol{y}}_m) \prod_{o = m}^{M} \left( \frac{p(\boldsymbol{z}_o | \overline{\boldsymbol{\xi}}_m, \overline{\boldsymbol{e}}_m)} {\kappa_{\mathrm{c}} } \right)^{\overline{\ell}_m \overline{q}_{ml}(o)} \\
		&\propto \begin{cases}
			0, & \overline{r}_m = 0\\
            \lambda_{\mathrm{n}} \overline{p}_m(\overline{\boldsymbol{x}}_m) \left( \sum_{\varepsilon=1}^{\overline{\ell}_m} \overline{f}_m(\varepsilon | \overline{\gamma}_m) \left( \widetilde{\overline{m}}^{\mathrm{BP},[I_{\mathrm{BP}}]}_{\rightarrow ml}  \right)^\varepsilon \right)\left( \prod_{o=m}^M \left( \frac{p(\boldsymbol{z}_o | \overline{\boldsymbol{\xi}}_m, \overline{\boldsymbol{e}}_m)} {\kappa_{\mathrm{c}}} \right)^{\overline{\ell}_m \overline{q}_{ml}(o)} \right), & \overline{r}_m = 1
		\end{cases} 
	\end{aligned}
\end{equation}
where the formulas of the terms in the above equation refer to \cite{MJDL}. Following the derivation of $\underline{p}_n(\underline{\boldsymbol{y}}_n)$, it is straightforward to obtain that
\begin{equation} 
	\overline{q}_m(\overline{\boldsymbol{y}}_m) \propto \begin{cases}
		0, & \overline{r}_m = 0\\
		\overline{p}_m(\overline{\gamma}_m) \overline{p}_m(\overline{\boldsymbol{\xi}}_m, \overline{\boldsymbol{e}}_m, \boldsymbol{z} ), & \overline{r}_m = 1
	\end{cases} 
\end{equation}
where 
\begin{equation} 
	\begin{aligned}
		\overline{p}_m(\overline{\gamma}_m) &\propto \sum_{\varepsilon=1}^{\overline{\ell}_m} \overline{w}_m^\varepsilon \mathrm{G}(\overline{\gamma}_m; \overline{\alpha}_m + \varepsilon, \overline{\beta}_m + 1)  \\
		\overline{w}_m^\varepsilon &\propto \frac{ \Gamma(\overline{\alpha}_m+\varepsilon) ( \widetilde{\underline{m}}^{\mathrm{BP},[I_{\mathrm{BP}}]}_{\rightarrow ml} )^\varepsilon }{ (\overline{\beta}_m+1)^{\overline{\alpha}_m+\varepsilon} \varepsilon! } \\
		\overline{p}_m(\overline{\boldsymbol{\xi}}_m, \overline{\boldsymbol{e}}_m, \boldsymbol{z} ) &= \overline{p}_m(\overline{\boldsymbol{\xi}}_m) \overline{p}_m(\overline{\boldsymbol{e}}_m) \prod_{o=m}^M ( \mathrm{N}(\boldsymbol{z}_o; \boldsymbol{H} \overline{\boldsymbol{\xi}}_m, s\overline{\boldsymbol{E}}_m + \boldsymbol{R}) )^{\overline{\ell}_m\overline{q}_{ml}(o)}.
	\end{aligned}
\end{equation}

\subsection{MF approximation to the density \texorpdfstring{$\underline{p}_n(\underline{\boldsymbol{\xi}}_n, \underline{\boldsymbol{e}}_n, \boldsymbol{z})$}{} }
\subsubsection{The approximate densities}
The joint density $p(\underline{\boldsymbol{\xi}}_n, \underline{\boldsymbol{e}}_n, \boldsymbol{\zeta}, \boldsymbol{z})$ in equation (61) of \cite{MJDL} is given by
\begin{equation} \label{eqn_mf_jpdf}
	\begin{aligned}
		p(\underline{\boldsymbol{\xi}}_n, \underline{\boldsymbol{e}}_n, \boldsymbol{z}, \boldsymbol{y}) =& \underline{p}^+_n(\underline{\boldsymbol{\xi}}_n) \underline{p}^+_n(\underline{\boldsymbol{e}}_n) \prod_{m=1}^M \left( \mathrm{N}(\boldsymbol{z}_m; \boldsymbol{H}\underline{\boldsymbol{\xi}}_n, s\underline{\boldsymbol{E}}_n + \boldsymbol{R}) \right)^{\omega_m}  \\
		=& \mathrm{N}(\underline{\boldsymbol{\xi}}_n; \underline{\boldsymbol{\mu}}^+_n, \underline{\boldsymbol{\Sigma}}^+_n) \mathrm{IW}(\underline{\boldsymbol{E}}_n; \underline{\nu}^+_n, \underline{\boldsymbol{V}}^+_n) \prod_{m=1}^M C^{\omega_m}_{s\underline{\boldsymbol{E}}_n + \boldsymbol{R}}  \mathrm{N}\left(\boldsymbol{z}_m; \boldsymbol{\zeta}_m, \frac{\boldsymbol{R}}{\omega_m}\right) \mathrm{N}\left(\boldsymbol{\zeta}_m; \boldsymbol{H} \underline{\boldsymbol{\xi}}_n, \frac{s\underline{\boldsymbol{E}}_n}{\omega_m} \right) 
    \end{aligned}
\end{equation}
where 
\begin{equation} \nonumber
	C^{\omega_m}_{s\underline{\boldsymbol{E}}_n + \boldsymbol{R}} = \frac{|2\pi(s\underline{\boldsymbol{E}}_n + \boldsymbol{R}) \omega_m^{-1}|^{\frac{1}{2}}}{|2\pi(s\underline{\boldsymbol{E}}_n + \boldsymbol{R})|^{\frac{\omega_m}{2}}}. 
\end{equation}

We approximate the joint density $p(\underline{\boldsymbol{\xi}}_n, \underline{\boldsymbol{e}}_n, \boldsymbol{\zeta})$ of the latent variables as $q_{\boldsymbol{\xi}}(\underline{\boldsymbol{\xi}}_n) q_{\boldsymbol{e}}(\underline{\boldsymbol{e}}_n) q_{\boldsymbol{\zeta}}(\boldsymbol{\zeta})$. According to the variational equation (7) in \cite{MJDL}, the optimal estimates of $q_{\boldsymbol{\xi}}(\cdot)$, $q_{\boldsymbol{e}}(\cdot)$ and $q_{\boldsymbol{\zeta}}(\cdot)$ are obtained as  
\begin{equation} \label{eqn_vb_approximation}
	\begin{aligned}
		q_{\boldsymbol{\xi}}(\underline{\boldsymbol{\xi}}_n) &= \mathbb{E}_{\underline{\boldsymbol{e}}_n, \boldsymbol{\zeta}} [ \ln p(\underline{\boldsymbol{\xi}}_n, \underline{\boldsymbol{e}}_n, \boldsymbol{\zeta}, \boldsymbol{z}) ] + c_{\underline{\boldsymbol{\xi}}_n} \\
		q_{\boldsymbol{e}}(\underline{\boldsymbol{e}}_n) &= \mathbb{E}_{\underline{\boldsymbol{\xi}}_n, \boldsymbol{\zeta}} [ \ln p(\underline{\boldsymbol{\xi}}_n, \underline{\boldsymbol{e}}_n, \boldsymbol{\zeta}, \boldsymbol{z}) ] + c_{\underline{\boldsymbol{e}}_n} \\
		q_{\boldsymbol{\zeta}}(\boldsymbol{\zeta}) &= \mathbb{E}_{\underline{\boldsymbol{\xi}}_n, \underline{\boldsymbol{e}}_n} [ \ln p(\underline{\boldsymbol{\xi}}_n, \underline{\boldsymbol{e}}_n, \boldsymbol{\zeta}, \boldsymbol{z}) ] + c_{\boldsymbol{\zeta}} \\
	\end{aligned}
\end{equation}
where $c_{\underline{\boldsymbol{\xi}}_n}$, $c_{\underline{\boldsymbol{e}}_n}$ and $c_{\boldsymbol{\zeta}}$ are constants with respect to $\underline{\boldsymbol{\xi}}_n$, $\underline{\boldsymbol{e}}_n$ and $\boldsymbol{\zeta}$. We employ fixed-point iterations to solve \eqref{eqn_vb_approximation}.

Using equation \eqref{eqn_vb_approximation}, the logarithm of the $\iota$th iterative density $q^{[\iota]}_{\boldsymbol{\xi}}(\underline{\boldsymbol{\xi}}_n)$ is given by
\begin{equation} \label{eqn_vb_posterior_x}
	\begin{aligned}
		&\ln q^{[\iota]}_{\boldsymbol{\xi}}(\underline{\boldsymbol{\xi}}_n) \\
		=& \mathbb{E}_{q^{[\iota-1]}_{\boldsymbol{e}}, q^{[\iota-1]}_{\boldsymbol{\zeta}}} \left[ \ln \prod_{m=1}^M \mathrm{N}\left(\boldsymbol{\zeta}_m; \boldsymbol{H} \underline{\boldsymbol{\xi}}_n, \frac{s\underline{\boldsymbol{E}}_n}{\omega_m} \right) \right] + \ln \mathrm{N}(\underline{\boldsymbol{\xi}}_n; \underline{\boldsymbol{\mu}}^+_n, \underline{\boldsymbol{\Sigma}}^+_n) + c_{\underline{\boldsymbol{\xi}}_n} \\
		=& \mathbb{E}_{q^{[\iota-1]}_{\boldsymbol{e}}, q^{[\iota-1]}_{\boldsymbol{\zeta}}} \left[ \ln \left( \mathrm{N} \left(\frac{\sum_{m=1}^{M} \omega_m \boldsymbol{\zeta}_m }{\sum_{m=1}^{M} \omega_m} ; \boldsymbol{H} \underline{\boldsymbol{\xi}}_n, \frac{s \underline{\boldsymbol{E}}_n}{\sum_{m=1}^{M} \omega_m} \right) \right) \right] + \ln \mathrm{N}(\underline{\boldsymbol{\xi}}_n; \underline{\boldsymbol{\mu}}^+_n, \underline{\boldsymbol{\Sigma}}^+_n) + c_{\underline{\boldsymbol{\xi}}_n}  \\
		=& -\frac{1}{2}  \operatorname{tr} \left[ \left( \frac{\sum_{m=1}^{M} \omega_m \mathbb{E}_{q^{[\iota-1]}_{\boldsymbol{\zeta}}} \left[ \boldsymbol{\zeta}_m \right] }{\sum_{m=1}^{M} \omega_m} - \boldsymbol{H} \underline{\boldsymbol{\xi}}_n \right)\left( \frac{\sum_{m=1}^{M} \omega_m \mathbb{E}_{q^{[\iota-1]}_{\boldsymbol{\zeta}}} \left[ \boldsymbol{\zeta}_m \right] }{\sum_{m=1}^{M} \omega_m} - \boldsymbol{H} \underline{\boldsymbol{\xi}}_n \right)^T \sum_{m=1}^{M} \omega_m \mathbb{E}_{q^{[\iota-1]}_{\boldsymbol{e}}} \left[ (s\underline{\boldsymbol{E}}_n)^{-1} \right]  \right] \\
		& + \ln \mathrm{N}(\underline{\boldsymbol{\xi}}_n; \underline{\boldsymbol{\mu}}^+_n, \underline{\boldsymbol{\Sigma}}^+_n) + c_{\underline{\boldsymbol{\xi}}_n} \\
		=& \ln \mathrm{N}\left( \frac{\sum_{m=1}^{M} \omega_m \mathbb{E}_{q^{[\iota-1]}_{\boldsymbol{\zeta}}} \left[ \boldsymbol{\zeta}_m \right] }{\sum_{m=1}^{M} \omega_m}; \boldsymbol{H} \underline{\boldsymbol{\xi}}_n, \frac{\mathbb{E}_{q^{[\iota-1]}_{\boldsymbol{e}}} \left[ (s\underline{\boldsymbol{E}}_n)^{-1} \right]^{-1}}{\sum_{m=1}^{M} \omega_m} \right) + \ln \mathrm{N}(\underline{\boldsymbol{\xi}}_n; \underline{\boldsymbol{\mu}}^+_n, \underline{\boldsymbol{\Sigma}}^+_n) + c_{\underline{\boldsymbol{\xi}}_n}
	\end{aligned}
\end{equation}
where we have used the following formula \cite{GLG2024}
\begin{equation}
	\prod_{m=1}^M \mathrm{N}\left(\boldsymbol{\zeta}_m; \boldsymbol{H} \underline{\boldsymbol{\xi}}_n, \frac{s\underline{\boldsymbol{E}}_n}{\omega_m} \right) \propto \mathrm{N} \left(\frac{\sum_{m=1}^{M} \omega_m \boldsymbol{\zeta}_m }{\sum_{m=1}^{M} \omega_m} ; \boldsymbol{H} \underline{\boldsymbol{\xi}}_n, \frac{s \underline{\boldsymbol{E}}_n}{\sum_{m=1}^{M} \omega_m} \right).
\end{equation}

Taking the exponential of both sides of \eqref{eqn_vb_posterior_x}, normalizing, and using the Kalman filter measurement update equation, we obtain
\begin{equation} \label{eqn_vb_iteration_xi}
	\begin{aligned}
		q^{[\iota]}_{\boldsymbol{\xi}}(\underline{\boldsymbol{\xi}}_n) &= \mathrm{N}(\underline{\boldsymbol{\xi}}_n; \underline{\boldsymbol{\mu}}_n^{[\iota]}, \underline{\boldsymbol{\Sigma}}_n^{[\iota]}) \\
		\underline{\boldsymbol{\mu}}_n^{[\iota]} &= \underline{\boldsymbol{\Sigma}}_n^{[\iota]} \bigg( (\underline{\boldsymbol{\Sigma}}_n^+)^{-1}\underline{\boldsymbol{\mu}}^+_n + \frac{1}{s} \boldsymbol{H}^T \mathbb{E}_{q^{[\iota-1]}_{\boldsymbol{e}}} \left[ \underline{\boldsymbol{E}}_n^{-1} \right] \sum_{m=1}^M \omega_m \mathbb{E}_{q^{[\iota-1]}_{\boldsymbol{\zeta}}} \left[ \boldsymbol{\zeta}_m \right] \bigg)\\
		\underline{\boldsymbol{\Sigma}}_n^{[\iota]} &= \Big( (\underline{\boldsymbol{\Sigma}}^+_n)^{-1} + \frac{1}{s} \sum_{m=1}^{M} \omega_m \boldsymbol{H}^T \mathbb{E}_{q^{[\iota-1]}_{\boldsymbol{e}}} \left[ \underline{\boldsymbol{E}}_n^{-1} \right]\boldsymbol{H} \Big)^{-1} .
	\end{aligned}
\end{equation}

As we assume that $\boldsymbol{R}$ is relatively small compared to $s\underline{\boldsymbol{E}}_n$, the product $\prod_{m=1}^M C^{\omega_m}_{s\underline{\boldsymbol{E}}_n + \boldsymbol{R}} \mathrm{N}(\boldsymbol{\zeta}_m; \boldsymbol{H} \underline{\boldsymbol{\xi}}_n, \frac{s\underline{\boldsymbol{E}}_n}{\omega_m})  $ in \eqref{eqn_mf_jpdf} can be reformulated as follows
\begin{equation}
	\begin{aligned}
		&\prod_{m=1}^M C^{\omega_m}_{s\underline{\boldsymbol{E}}_n + \boldsymbol{R}} \mathrm{N}\left(\boldsymbol{\zeta}_m; \boldsymbol{H} \underline{\boldsymbol{\xi}}_n, \frac{s\underline{\boldsymbol{E}}_n}{\omega_m}\right) \\
		=& \prod_{m=1}^M \frac{|2\pi(s\underline{\boldsymbol{E}}_n + \boldsymbol{R}) \omega_m^{-1}|^{\frac{1}{2}}}{|2\pi(s\underline{\boldsymbol{E}}_n + \boldsymbol{R})|^{\frac{\omega_m}{2}}} |2\pi(s\underline{\boldsymbol{E}}_n) \omega_m^{-1}|^{-\frac{1}{2}} \exp\left( -\frac{\omega_m}{2} (\boldsymbol{\zeta}_m - \boldsymbol{H} \underline{\boldsymbol{\xi}}_n)^T (s\underline{\boldsymbol{E}}_n)^{-1} (\boldsymbol{\zeta}_m - \boldsymbol{H} \underline{\boldsymbol{\xi}}_n) \right) \\
		\approx& \prod_{m=1}^M \frac{|2\pi(s\underline{\boldsymbol{E}}_n ) \omega_m^{-1}|^{\frac{1}{2}}}{|2\pi(s\underline{\boldsymbol{E}}_n )|^{\frac{\omega_m}{2}}} |2\pi(s\underline{\boldsymbol{E}}_n) \omega_m^{-1}|^{-\frac{1}{2}} \exp\left( -\frac{\omega_m}{2} (\boldsymbol{\zeta}_m - \boldsymbol{H} \underline{\boldsymbol{\xi}}_n)^T (s\underline{\boldsymbol{E}}_n)^{-1} (\boldsymbol{\zeta}_m - \boldsymbol{H} \underline{\boldsymbol{\xi}}_n) \right) \\
		=& \prod_{m=1}^M |2\pi(s\underline{\boldsymbol{E}}_n )|^{-\frac{\omega_m}{2}}  \exp\left( -\frac{\omega_m}{2} (\boldsymbol{\zeta}_m - \boldsymbol{H} \underline{\boldsymbol{\xi}}_n)^T (s\underline{\boldsymbol{E}}_n)^{-1} (\boldsymbol{\zeta}_m - \boldsymbol{H} \underline{\boldsymbol{\xi}}_n) \right) .
	\end{aligned}
\end{equation}
Using equation \eqref{eqn_vb_approximation} and the above formula, the logarithm of the $\iota $th iterative density $q^{[\iota]}_{\boldsymbol{e}}(\underline{\boldsymbol{e}}_n)$ is given by
\begin{equation} \label{eqn_vb_posterior_e}
	\begin{aligned}
		\ln q^{[\iota]}_{\boldsymbol{e}}(\underline{\boldsymbol{e}}_n) =& \mathbb{E}_{q^{[\iota-1]}_{\boldsymbol{\xi}}, q^{[\iota-1]}_{\boldsymbol{\zeta}}} \left[ \ln \prod_{m=1}^M C^{\omega_m}_{s\underline{\boldsymbol{E}}_n + \boldsymbol{R}} \mathrm{N}\left(\boldsymbol{\zeta}_m; \boldsymbol{H} \underline{\boldsymbol{\xi}}_n, \frac{s\underline{\boldsymbol{E}}_n}{\omega_m}\right) \right] + \ln \mathrm{IW}(\underline{\boldsymbol{E}}_n; \underline{\nu}^+_n, \underline{\boldsymbol{V}}^+_n) + c_{\underline{\boldsymbol{e}}_n} \\ 
		=& -\frac{\sum_{m=1}^{M}\omega_m}{2} \ln( |\underline{\boldsymbol{E}}_n| ) -\frac{1}{2} \operatorname{tr} \left[ \sum_{m=1}^{M} \frac{\omega_m}{s} \mathbb{E}_{q^{[\iota-1]}_{\boldsymbol{\xi}}, q^{[\iota-1]}_{\boldsymbol{\zeta}}} \left[ (\boldsymbol{\zeta}_m - \boldsymbol{H} \underline{\boldsymbol{\xi}}_n)(\boldsymbol{\zeta}_m - \boldsymbol{H} \underline{\boldsymbol{\xi}}_n)^T \right] (\underline{\boldsymbol{E}}_n)^{-1} \right] \\
		&+ \ln \mathrm{IW}(\underline{\boldsymbol{E}}_n; \underline{\nu}^+_n, \underline{\boldsymbol{V}}^+_n) + c_{\underline{\boldsymbol{e}}_n} \\
	\end{aligned}
\end{equation}
Observing that the right hand side of the equation \eqref{eqn_vb_posterior_e} is in the form of the logarithm of an IW density for $\underline{\boldsymbol{E}}_n$, we obtain
\begin{equation} \label{eqn_vb_iteration_e}
	\begin{aligned}
		q^{[\iota]}_{\boldsymbol{e}}(\underline{\boldsymbol{e}}_n) =& \mathrm{IW}(\underline{\boldsymbol{E}}_n; \underline{\nu}^{[\iota]}_n, \underline{\boldsymbol{V}}^{[\iota]}_n)\\ 
		\underline{\nu}^{[\iota]}_n =& \underline{\nu}^+_n + \sum_{m=1}^M \omega_m \\
		\underline{\boldsymbol{V}}^{[\iota]}_n =& \underline{\boldsymbol{V}}^+_n + \frac{1}{s} \sum_{m=1}^M \omega_m \mathbb{E}_{q^{[\iota-1]}_{\boldsymbol{\xi}}, q^{[\iota-1]}_{\boldsymbol{\zeta}}} \left[ (\boldsymbol{\zeta}_m - \boldsymbol{H} \underline{\boldsymbol{\xi}}_n)(\boldsymbol{\zeta}_m - \boldsymbol{H} \underline{\boldsymbol{\xi}}_n)^T \right] .  \\
	\end{aligned}
\end{equation}

Using equation \eqref{eqn_vb_approximation}, the logarithm of the $\iota$th iterative density $q^{[\iota]}_{\boldsymbol{\zeta}}(\boldsymbol{\zeta})$ is given by
\begin{equation}
	\begin{aligned}
		&\ln q^{[\iota]}_{\boldsymbol{\zeta}}(\boldsymbol{\zeta}) \\
		=& \ln \prod_{m=1}^M  \mathrm{N}\left(\boldsymbol{z}_m; \boldsymbol{\zeta}_m, \frac{\boldsymbol{R}}{\omega_m}\right) + \mathbb{E}_{q^{[\iota-1]}_{\boldsymbol{\xi}}, q^{[\iota-1]}_{\boldsymbol{e}}} \left[ \ln \prod_{m=1}^M \mathrm{N}\left(\boldsymbol{\zeta}_m; \boldsymbol{H} \underline{\boldsymbol{\xi}}_n, \frac{s\underline{\boldsymbol{E}}_n}{\omega_m} \right)  \right] + c_{\boldsymbol{\zeta}} \\
		=& \ln \prod_{m=1}^M  \mathrm{N}\left(\boldsymbol{z}_m; \boldsymbol{\zeta}_m, \frac{\boldsymbol{R}}{\omega_m}\right) + \sum_{m=1}^{M}-\frac{1}{2} \operatorname{tr} \left[ \omega_m (\boldsymbol{\zeta}_m - \boldsymbol{H} \mathbb{E}_{q^{[\iota-1]}_{\boldsymbol{\xi}}}(\underline{\boldsymbol{\xi}}_n))(\boldsymbol{\zeta}_m - \boldsymbol{H} \mathbb{E}_{q^{[\iota-1]}_{\boldsymbol{\xi}}}(\underline{\boldsymbol{\xi}}_n))^T \mathbb{E}_{q^{[\iota-1]}_{\boldsymbol{e}}} \left[ (s\underline{\boldsymbol{E}}_n)^{-1} \right] \right] + c_{\boldsymbol{\zeta}} \\
		=& \sum_{m=1}^{M} \left[ \ln \mathrm{N}\left(\boldsymbol{z}_m; \boldsymbol{\zeta}_m, \frac{\boldsymbol{R}}{\omega_m}\right) + \ln\mathrm{N}\left( \boldsymbol{\zeta}_m; \boldsymbol{H} \mathbb{E}_{q^{[\iota-1]}_{\boldsymbol{\xi}}}(\underline{\boldsymbol{\xi}}_n), \omega_m^{-1} \mathbb{E}_{q^{[\iota-1]}_{\boldsymbol{e}}} \left[ (s\underline{\boldsymbol{E}}_n)^{-1} \right]^{-1} \right) \right] + c_{\boldsymbol{\zeta}}
	\end{aligned}
\end{equation}
Taking the exponential of both sides of the above equation, normalizing and using the Kalman filter measurement update equation, we obtain the density of $\boldsymbol{\zeta}$ as 
\begin{equation} \label{eqn_vb_iteration_zeta}
	\begin{aligned}
		q^{[\iota]}_{\boldsymbol{\zeta}}(\boldsymbol{\zeta}) &= \prod_{m=1}^M \mathrm{N}(\boldsymbol{\zeta}_m; \widetilde{\boldsymbol{\mu}}_m^{[\iota]}, \widetilde{\boldsymbol{\Sigma}}_m^{[\iota]}) \\
		\widetilde{\boldsymbol{\mu}}_m^{[\iota]} &= \widetilde{\boldsymbol{\Sigma}}_m^{[\iota]} \Big( \frac{\omega_m}{s} \mathbb{E}_{q^{[\iota-1]}_{\boldsymbol{e}}} \left[ \underline{\boldsymbol{E}}_n^{-1} \right] \boldsymbol{H}\mathbb{E}_{q^{[\iota-1]}_{\boldsymbol{\xi}}}(\underline{\boldsymbol{\xi}}_n) + \omega_m \boldsymbol{R}^{-1}\boldsymbol{z}_m \Big) \\
		\widetilde{\boldsymbol{\Sigma}}_m^{[\iota]} &= \Big( \frac{\omega_m}{s} \mathbb{E}_{q^{[\iota-1]}_{\boldsymbol{e}}} \left[ \underline{\boldsymbol{E}}_n^{-1} \right] + \omega_m \boldsymbol{R}^{-1} \Big)^{-1} .
	\end{aligned}
\end{equation}
For the fixed-point iterations \eqref{eqn_vb_iteration_xi}, \eqref{eqn_vb_iteration_e} and \eqref{eqn_vb_iteration_zeta}, the involved expectations are computed as
\begin{equation} \nonumber
	\begin{aligned}
		&\mathbb{E}_{q^{[\iota-1]}_{\boldsymbol{\xi}}}(\underline{\boldsymbol{\xi}}_n) = \underline{\boldsymbol{\mu}}_n^{[\iota-1]}, \quad \mathbb{E}_{q^{[\iota-1]}_{\boldsymbol{\zeta}}} \left[ \boldsymbol{\zeta}_m \right] = \widetilde{\boldsymbol{\mu}}_m^{[\iota-1]}, \quad \mathbb{E}_{q^{[\iota-1]}_{\boldsymbol{e}}} \left[ \underline{\boldsymbol{E}}_n^{-1} \right] = (\underline{\nu}^{[\iota-1]}_n -d_{\boldsymbol{z}} -1)\left( \underline{\boldsymbol{V}}^{[\iota-1]}_n \right)^{-1}, \\
		&\mathbb{E}_{q^{[\iota-1]}_{\boldsymbol{\xi}}, q^{[\iota-1]}_{\boldsymbol{\zeta}}} \left[ (\boldsymbol{\zeta}_m - \boldsymbol{H} \underline{\boldsymbol{\xi}}_n)(\boldsymbol{\zeta}_m - \boldsymbol{H} \underline{\boldsymbol{\xi}}_n)^T \right] = \boldsymbol{H}\underline{\boldsymbol{\Sigma}}_n^{[\iota-1]}\boldsymbol{H}^T + \widetilde{\boldsymbol{\Sigma}}_m^{[\iota-1]} + (\widetilde{\boldsymbol{\mu}}_m^{[\iota-1]} - \boldsymbol{H}\underline{\boldsymbol{\mu}}_n^{[\iota-1]})(\widetilde{\boldsymbol{\mu}}_m^{[\iota-1]} - \boldsymbol{H}\underline{\boldsymbol{\mu}}_n^{[\iota-1]})^T .
	\end{aligned}
\end{equation}

\subsubsection{The predictive likelihood}
The predictive likelihood $p(\boldsymbol{z}) = \int \underline{p}_n(\underline{\boldsymbol{\xi}}_n, \underline{\boldsymbol{e}}_n, \boldsymbol{z}) \mathrm{d}\underline{\boldsymbol{\xi}}_n \mathrm{d}\underline{\boldsymbol{e}}_n$ can not be computed analytically. Nevertheless, the variational approach provides an approximate solution to it. The predictive log-likelihood $\ln p(\boldsymbol{z}) $ can be written using the approximate density $q(\cdot)$ as \cite{BMN2006}
\begin{equation} \label{eqn_logPredictiveLikelihood}
	\ln p(\boldsymbol{z}) = \mathrm{PL}(q) + \mathrm{KL}(q(\underline{\boldsymbol{\xi}}_n, \underline{\boldsymbol{e}}_n, \boldsymbol{\zeta}) \| p(\underline{\boldsymbol{\xi}}_n, \underline{\boldsymbol{e}}_n, \boldsymbol{\zeta}, \boldsymbol{z}))
\end{equation}
where $\mathrm{KL}(q(\cdot)||p(\cdot))$ is the Kullback-Leibler (KL) divergence between $q(\cdot)$ and $p(\cdot)$, and
\begin{equation}
	\begin{aligned}
        \mathrm{PL}(q) &= \int q_{\boldsymbol{\xi}}(\underline{\boldsymbol{\xi}}_n) q_{\boldsymbol{e}}(\underline{\boldsymbol{e}}_n) q_{\boldsymbol{\zeta}}(\boldsymbol{\zeta}) \ln \frac{p(\underline{\boldsymbol{\xi}}_n, \underline{\boldsymbol{e}}_n, \boldsymbol{\zeta}, \boldsymbol{z})}{q_{\boldsymbol{\xi}}(\underline{\boldsymbol{\xi}}_n) q_{\boldsymbol{e}}(\underline{\boldsymbol{e}}_n) q_{\boldsymbol{\zeta}}(\boldsymbol{\zeta})} \mathrm{d} \underline{\boldsymbol{\xi}}_n \mathrm{d}\underline{\boldsymbol{e}}_n \mathrm{d}\boldsymbol{\zeta} \\
        &=\mathbb{E}_{q_{\boldsymbol{\xi}}, q_{\boldsymbol{e}}, q_{\boldsymbol{\zeta}}}\left[ \ln p(\underline{\boldsymbol{\xi}}_n, \underline{\boldsymbol{e}}_n, \boldsymbol{\zeta}, \boldsymbol{z}) \right] - \mathbb{E}_{q_{\boldsymbol{\xi}}, q_{\boldsymbol{e}}, q_{\boldsymbol{\zeta}}} \left[ \ln q_{\boldsymbol{\xi}}(\underline{\boldsymbol{\xi}}_n) q_{\boldsymbol{e}}(\underline{\boldsymbol{e}}_n) q_{\boldsymbol{\zeta}}(\boldsymbol{\zeta}) \right]
    \end{aligned}
\end{equation}
where $q_{\boldsymbol{\xi}}(\underline{\boldsymbol{\xi}}_n) = \mathrm{N}(\underline{\boldsymbol{\xi}}_n; \underline{\boldsymbol{\mu}}_n, \underline{\boldsymbol{\Sigma}}_n)$, $q_{\boldsymbol{e}}(\underline{\boldsymbol{e}}_n) = \mathrm{IW}(\underline{\boldsymbol{E}}_n; \underline{\nu}_n, \underline{\boldsymbol{V}}_n)$ and $q_{\boldsymbol{\zeta}}(\boldsymbol{\zeta}) = \prod_{m=1}^M \mathrm{N}(\boldsymbol{\zeta}_m; \widetilde{\boldsymbol{\mu}}_m, \widetilde{\boldsymbol{\Sigma}}_m) $ are given in \cite{MJDL}. 

Since the KL divergence in \eqref{eqn_logPredictiveLikelihood} is minimized, the predictive likelihood can be approximated as $p(\boldsymbol{z})\approx \exp(\mathrm{PL}(q))$. The two expectations in the above formula are calculated as follows
\begin{equation} \label{eqn_pl_2}
	\begin{aligned}
        &2\mathbb{E}_{q_{\boldsymbol{\xi}}, q_{\boldsymbol{e}}, q_{\boldsymbol{\zeta}}} \left[ \ln q_{\boldsymbol{\xi}}(\underline{\boldsymbol{\xi}}_n) q_{\boldsymbol{e}}(\underline{\boldsymbol{e}}_n) q_{\boldsymbol{\zeta}}(\boldsymbol{\zeta}) \right] \\
        = &- d_{\boldsymbol{\xi}}\ln2\pi - \ln|\underline{\boldsymbol{\Sigma}}_n| -\operatorname{tr}\left( (\underline{\boldsymbol{\Sigma}}_n)^{-1} \mathbb{E}_{q_{\boldsymbol{\xi}}} \left[ (\underline{\boldsymbol{\xi}}_n - \underline{\boldsymbol{\mu}}_n)(\underline{\boldsymbol{\xi}}_n - \underline{\boldsymbol{\mu}}_n)^T\right] \right) -(\underline{\nu}_n - d_{\boldsymbol{z}} - 1)d_{\boldsymbol{z}} \ln2\\
        &  + (\underline{\nu}_n - d_{\boldsymbol{z}} - 1)\ln|\underline{\boldsymbol{V}}_n| - 2\ln\Gamma_{d_{\boldsymbol{z}}}\left(\frac{(\underline{\nu}_n - d_{\boldsymbol{z}} - 1)}{2}\right) - \underline{\nu}_n \mathbb{E}_{q_{\boldsymbol{e}}} \left[ \ln|\underline{\boldsymbol{E}}_n| \right] - \operatorname{tr}\left( \underline{\boldsymbol{V}}_n \mathbb{E}_{q_{\boldsymbol{e}}} \left[ \underline{\boldsymbol{E}}_n^{-1}\right] \right) \\
        & + \sum_{m=1}^M -d_{\boldsymbol{z}}\ln2\pi - \ln|\widetilde{\boldsymbol{\Sigma}}_m| - \operatorname{tr}\left( (\widetilde{\boldsymbol{\Sigma}}_m)^{-1} \mathbb{E}_{q_{\boldsymbol{\zeta}}} \left[ (\boldsymbol{\zeta}_m - \widetilde{\boldsymbol{\mu}}_m)(\boldsymbol{\zeta}_m - \widetilde{\boldsymbol{\mu}}_m)^T \right] \right)
    \end{aligned}
\end{equation}
\begin{equation} \label{eqn_pl_1}
	\begin{aligned}
        &2\mathbb{E}_{q_{\boldsymbol{\xi}}, q_{\boldsymbol{e}}, q_{\boldsymbol{\zeta}}}\left[ \ln p(\underline{\boldsymbol{\xi}}_n, \underline{\boldsymbol{e}}_n, \boldsymbol{\zeta}, \boldsymbol{z}) \right]  \\
        =& - d_{\boldsymbol{\xi}} \ln2\pi - \ln|\underline{\boldsymbol{\Sigma}}^+_n| -\operatorname{tr}\left( \mathbb{E}_{q_{\boldsymbol{\xi}}} \left[(\underline{\boldsymbol{\Sigma}}^+_n)^{-1} (\underline{\boldsymbol{\xi}}_n - \underline{\boldsymbol{\mu}}^+_n)(\underline{\boldsymbol{\xi}}_n - \underline{\boldsymbol{\mu}}^+_n)^T \right] \right) -(\underline{\nu}^+_n - d_{\boldsymbol{z}} - 1)d_{\boldsymbol{z}} \ln2 \\
		& + (\underline{\nu}^+_n - d_{\boldsymbol{z}} - 1)\ln|\underline{\boldsymbol{V}}^+_n| -  2\ln\Gamma_{d_{\boldsymbol{z}}}\left(\frac{(\underline{\nu}^+_n - d_{\boldsymbol{z}} - 1)}{2}\right) - \underline{\nu}^+_n  \mathbb{E}_{q_{\boldsymbol{e}}} \left[ \ln|\underline{\boldsymbol{E}}_n| \right] - \operatorname{tr}\left( \underline{\boldsymbol{V}}^+_n \mathbb{E}_{q_{\boldsymbol{e}}} \left[ \underline{\boldsymbol{E}}_n^{-1} \right] \right) \\
		& + \sum_{m=1}^M \bigg[ -d_{\boldsymbol{z}}\ln2\pi - \ln|\boldsymbol{R}/\omega_m| - \omega_m\operatorname{tr}\left( \boldsymbol{R}^{-1} \mathbb{E}_{q_{\boldsymbol{\zeta}}} \left[ (\boldsymbol{z}_m - \boldsymbol{\zeta}_m)(\boldsymbol{z}_m - \boldsymbol{\zeta}_m)^T \right]\right) \bigg] \\
		& + \sum_{m=1}^M \bigg[ (1-\omega_m)(d_{\boldsymbol{z}}\ln2\pi + \mathbb{E}_{q_{\boldsymbol{e}}} \left[ \ln|s \underline{\boldsymbol{E}}_n + \boldsymbol{R}| \right] ) -d_{\boldsymbol{z}}\ln2\pi - d_{\boldsymbol{z}}\ln\omega_m - \mathbb{E}_{q_{\boldsymbol{e}}} \left[ \ln|s\underline{\boldsymbol{E}}_n/\omega_m| \right] \bigg. \\
		&-\bigg.  \omega_m\operatorname{tr}\left( \mathbb{E}_{q^{[\iota]}_{\boldsymbol{e}}} \left[ (s\underline{\boldsymbol{E}}_n)^{-1} \right] ~\mathbb{E}_{q_{\boldsymbol{\zeta}}, q_{\boldsymbol{\xi}}} \left[ (\boldsymbol{\zeta}_m - \boldsymbol{H}\underline{\boldsymbol{\xi}}_n)(\boldsymbol{\zeta}_m - \boldsymbol{H}\underline{\boldsymbol{\xi}}_n)^T  \right] \right) \bigg]
    \end{aligned}
\end{equation}

Subtracting \eqref{eqn_pl_2} from \eqref{eqn_pl_1}, and using the identities 
\begin{equation} \nonumber
	\underline{\nu}_n = \underline{\nu}^+ + \sum_{m=1}^M \omega_m, \quad \underline{\boldsymbol{V}}_n = \underline{\boldsymbol{V}}_n^+ + \frac{1}{s} \sum_{m=1}^M \omega_m \mathbb{E}_{q_{\boldsymbol{\zeta}}, q_{\boldsymbol{\xi}}} \left[ (\boldsymbol{\zeta}_m - \boldsymbol{H}\underline{\boldsymbol{\xi}}_n)(\boldsymbol{\zeta}_m - \boldsymbol{H}\underline{\boldsymbol{\xi}}_n)^T  \right]
\end{equation}
from equation \eqref{eqn_vb_iteration_e}, we obtain
\begin{equation} \label{eqn_pl_subtraction}
	\begin{aligned}
        &2\mathbb{E}_{q_{\boldsymbol{\xi}}, q_{\boldsymbol{e}}, q_{\boldsymbol{\zeta}}}\left[ \ln p(\underline{\boldsymbol{\xi}}_n, \underline{\boldsymbol{e}}_n, \boldsymbol{\zeta}, \boldsymbol{z}) \right] - 2\mathbb{E}_{q_{\boldsymbol{\xi}}, q_{\boldsymbol{e}}, q_{\boldsymbol{\zeta}}} \left[ \ln q_{\boldsymbol{\xi}}(\underline{\boldsymbol{\xi}}_n) q_{\boldsymbol{e}}(\underline{\boldsymbol{e}}_n) q_{\boldsymbol{\zeta}}(\boldsymbol{\zeta}) \right] \\
        =& - \ln|\underline{\boldsymbol{\Sigma}}^+_n| -\operatorname{tr}\left( (\underline{\boldsymbol{\Sigma}}^+_n)^{-1} \mathbb{E}_{ q_{\boldsymbol{\xi}}} \left[ (\underline{\boldsymbol{\xi}}_n - \underline{\boldsymbol{\mu}}^+_n)(\underline{\boldsymbol{\xi}}_n - \underline{\boldsymbol{\mu}}^+_n)^T \right] \right) + \ln|\underline{\boldsymbol{\Sigma}}_n| \\
		& +\operatorname{tr}\left( (\underline{\boldsymbol{\Sigma}}_n)^{-1} \mathbb{E}_{ q_{\boldsymbol{\xi}}} \left[ (\underline{\boldsymbol{\xi}}_n - \underline{\boldsymbol{\mu}}_n)(\underline{\boldsymbol{\xi}}_n - \underline{\boldsymbol{\mu}}_n)^T \right]  \right) + \left(\sum_{m=1}^M \omega_m \right)d_{\boldsymbol{z}}\ln2 + (\underline{\nu}^+_n - d_{\boldsymbol{z}} - 1)\ln|\underline{\boldsymbol{V}}^+_n| \\
		&- (\underline{\nu}_n - d_{\boldsymbol{z}} - 1)\ln|\underline{\boldsymbol{V}}_n| - 2\ln\Gamma_{d_{\boldsymbol{z}}}\left(\frac{\underline{\nu}^+_n - d_{\boldsymbol{z}} - 1}{2}\right) + 2\ln\Gamma_{d_{\boldsymbol{z}}}\left(\frac{\underline{\nu}_n - d_{\boldsymbol{z}} - 1}{2}\right) \\
		& + \sum_{m=1}^M \left[ -d_{\boldsymbol{z}}\ln2\pi - \ln|\boldsymbol{R}/\omega_m| - \omega_m\operatorname{tr}\left( \boldsymbol{R}^{-1} \mathbb{E}_{ q_{\boldsymbol{\zeta}}} \left[(\boldsymbol{z}_m - \boldsymbol{\zeta}_m)(\boldsymbol{z}_m - \boldsymbol{\zeta}_m)^T\right] \right) \right] \\
        &+ \sum_{m=1}^M \left[ (1-\omega_m)( \mathbb{E}_{ q_{\boldsymbol{e}}} \left[ \ln|s\underline{\boldsymbol{E}}_n + \boldsymbol{R}| \right]) -\omega_m d_{\boldsymbol{z}}\ln2\pi -2d_{\boldsymbol{z}}\ln\omega_m - d_{\boldsymbol{z}}\ln s \right] \\
        &+ \sum_{m=1}^M d_{\boldsymbol{z}}\ln2\pi + \ln|\widetilde{\boldsymbol{\Sigma}}_m| + \operatorname{tr}\left( (\widetilde{\boldsymbol{\Sigma}}_m)^{-1} \mathbb{E}_{ q_{\boldsymbol{\zeta}}} \left[(\boldsymbol{\zeta}_m - \widetilde{\boldsymbol{\mu}}_m)(\boldsymbol{\zeta}_m - \widetilde{\boldsymbol{\mu}}_m)^T \right] \right)
    \end{aligned}
\end{equation}
The expected values in the above equations are computed as follows
\begin{equation} \nonumber
	\begin{aligned}
		\mathbb{E}_{ q_{\boldsymbol{\xi}}} \left[ (\underline{\boldsymbol{\xi}}_n - \underline{\boldsymbol{\mu}}^+_n)(\underline{\boldsymbol{\xi}}_n - \underline{\boldsymbol{\mu}}^+_n)^T \right] &= (\underline{\boldsymbol{\mu}}_n - \underline{\boldsymbol{\mu}}^+_n)(\underline{\boldsymbol{\mu}}_n - \underline{\boldsymbol{\mu}}^+_n)^T + \underline{\boldsymbol{\Sigma}}_n \\
		\mathbb{E}_{ q_{\boldsymbol{\xi}}} \left[ (\underline{\boldsymbol{\xi}}_n - \underline{\boldsymbol{\mu}}_n)(\underline{\boldsymbol{\xi}}_n - \underline{\boldsymbol{\mu}}_n)^T \right] &= \underline{\boldsymbol{\Sigma}}_n \\
		\mathbb{E}_{ q_{\boldsymbol{\zeta}}} \left[ (\boldsymbol{z}_m - \boldsymbol{\zeta}_m)(\boldsymbol{z}_m - \boldsymbol{\zeta}_m)^T \right] &= (\boldsymbol{z}_m - \widetilde{\boldsymbol{\mu}}_m)(\boldsymbol{z}_m -\widetilde{\boldsymbol{\mu}}_m)^T + \widetilde{\boldsymbol{\Sigma}}_m \\
		\mathbb{E}_{ q_{\boldsymbol{\zeta}}} \left[  (\boldsymbol{\zeta}_m - \widetilde{\boldsymbol{\mu}}_m)(\boldsymbol{\zeta}_m - \widetilde{\boldsymbol{\mu}}_m)^T \right] &= \widetilde{\boldsymbol{\Sigma}}_m.
	\end{aligned}
\end{equation}

Substituting the above expected values into \eqref{eqn_pl_subtraction} we get
\begin{equation}
	\begin{aligned}
        &2\mathbb{E}_{q_{\boldsymbol{\xi}}, q_{\boldsymbol{e}}, q_{\boldsymbol{\zeta}}}\left[ \ln p(\underline{\boldsymbol{\xi}}_n, \underline{\boldsymbol{e}}_n, \boldsymbol{\zeta}, \boldsymbol{z}) \right] - 2\mathbb{E}_{q_{\boldsymbol{\xi}}, q_{\boldsymbol{e}}, q_{\boldsymbol{\zeta}}} \left[ \ln q_{\boldsymbol{\xi}}(\underline{\boldsymbol{\xi}}_n) q_{\boldsymbol{e}}(\underline{\boldsymbol{e}}_n) q_{\boldsymbol{\zeta}}(\boldsymbol{\zeta}) \right] \\
        =& 2\ln\mathrm{N}(\underline{\boldsymbol{\mu}}_n; \underline{\boldsymbol{\mu}}^+_n, \underline{\boldsymbol{\Sigma}}^+_n) + d_{\boldsymbol{\xi}}\ln2\pi - \operatorname{tr}\left( (\underline{\boldsymbol{\Sigma}}^+_n)^{-1} \underline{\boldsymbol{\Sigma}}_n \right) + \ln|\underline{\boldsymbol{\Sigma}}_n| + d_{\boldsymbol{\xi}}  + \left(\sum_{m=1}^M \omega_m \right)d_{\boldsymbol{z}}\ln2 \\
		&+ (\underline{\nu}^+_n - d_{\boldsymbol{z}} - 1)\ln|\underline{\boldsymbol{V}}^+_n| - (\underline{\nu}_n - d_{\boldsymbol{z}} - 1)\ln|\underline{\boldsymbol{V}}_n| - 2\ln\Gamma_{d_{\boldsymbol{z}}}\left(\frac{\underline{\nu}^+_n - d_{\boldsymbol{z}} - 1}{2}\right) + 2\ln\Gamma_{d_{\boldsymbol{z}}}\left(\frac{\underline{\nu}_n - d_{\boldsymbol{z}} - 1}{2}\right) \\
        &+ \sum_{m=1}^M \left( 2\ln \mathrm{N}(\boldsymbol{z}_m; \widetilde{\boldsymbol{\mu}}_m, \boldsymbol{R}/\omega_m) - \omega_m\operatorname{tr}(\boldsymbol{R}^{-1}\widetilde{\boldsymbol{\Sigma}}_m )\right) + \sum_{m=1}^M\ln|\widetilde{\boldsymbol{\Sigma}}_m | + Md_{\boldsymbol{z}}\ln2\pi  + M d_{\boldsymbol{z}}\\
        &+ \sum_{m=1}^M \left[ (1-\omega_m)( \mathbb{E}_{ q_{\boldsymbol{e}}} \left[ \ln|s\underline{\boldsymbol{E}}_n + \boldsymbol{R}| \right]) -\omega_m d_{\boldsymbol{z}}\ln2\pi -2d_{\boldsymbol{z}}\ln\omega_m \right] - Md_{\boldsymbol{z}}\ln s \\
    \end{aligned}
\end{equation}
Dividing both sides by 2 and taking exponential of both sides of the above formula, we obtain  
\begin{equation}
	\begin{aligned}
		p(\boldsymbol{z}) \approx&  \mathrm{N}(\underline{\boldsymbol{\mu}}_n; \underline{\boldsymbol{\mu}}^+_n, \underline{\boldsymbol{\Sigma}}^+_n) |2\pi\underline{\boldsymbol{\Sigma}}_n|^{\frac{1}{2}} \exp\left( -\frac{1}{2} \left( \operatorname{tr}\left( (\underline{\boldsymbol{\Sigma}}^+_n)^{-1} \underline{\boldsymbol{\Sigma}}_n \right) - d_{\boldsymbol{\xi}} \right) \right) \\
		&\times 2^{\frac{d_{\boldsymbol{z}}\sum_{m=1}^M \omega_m}{2}} \frac{|\underline{\boldsymbol{V}}^+_n|^{\frac{\underline{\nu}^+_n - d_{\boldsymbol{z}} - 1}{2}}}{|\underline{\boldsymbol{V}}_n|^{\frac{\underline{\nu}_n - d_{\boldsymbol{z}} - 1}{2}}} \frac{\Gamma_{d_{\boldsymbol{z}}}\left(\frac{\underline{\nu}_n - d_{\boldsymbol{z}} - 1}{2}\right)}{\Gamma_{d_{\boldsymbol{z}}}\left(\frac{\underline{\nu}^+_n - d_{\boldsymbol{z}} - 1}{2}\right)}  \prod_{m=1}^M \mathrm{N}(\boldsymbol{z}_m; \widetilde{\boldsymbol{\mu}}_m, \boldsymbol{R}/\omega_m) \exp\left( -\frac{1}{2}\omega_m\operatorname{tr}(\boldsymbol{R}^{-1}\widetilde{\boldsymbol{\Sigma}}_m) \right) \\
		&\times \exp \left(\frac{M-\sum_{m=1}^M\omega_m}{2}\mathbb{E}_{ q_{\boldsymbol{e}}} \left[ \ln|s\underline{\boldsymbol{E}}_n + \boldsymbol{R}| \right] \right) (2\pi)^{\frac{-d_{\boldsymbol{z}}\sum_{m=1}^M\omega_m}{2}} \left( \prod_{m=1}^M \omega_m^{-d_{\boldsymbol{z}}} \right) s^{- \frac{Md_{\boldsymbol{z}}}{2}}\\
		&\times (2\pi)^{\frac{Md_{\boldsymbol{z}}}{2}} \left(\prod_{m=1}^M |\widetilde{\boldsymbol{\Sigma}}_m|^{\frac{1}{2}} \right)\exp\left(\frac{M d_{\boldsymbol{z}}}{2}\right) \\
		=& |2\pi \underline{\boldsymbol{\Sigma}}_n|^{\frac{1}{2}} s^{-\frac{Md_{\boldsymbol{z}}}{2}} (2\pi)^{\frac{d_{\boldsymbol{z}}(M-\sum_{m=1}^M \omega_m)}{2}} 2^{\frac{d_{\boldsymbol{z}}\sum_{m=1}^M \omega_m}{2}} \prod_{m=1}^M |\widetilde{\boldsymbol{\Sigma}}_m|^{\frac{1}{2}} \omega_m^{-d_{\boldsymbol{z}}} \frac{|\underline{\boldsymbol{V}}^+_n|^{\frac{\underline{\nu}^+_n - d_{\boldsymbol{z}} - 1}{2}}}{|\underline{\boldsymbol{V}}_n|^{\frac{\underline{\nu}_n - d_{\boldsymbol{z}} - 1}{2}}} \frac{\Gamma_{d_{\boldsymbol{z}}}\left(\frac{\underline{\nu}_n - d_{\boldsymbol{z}} - 1}{2}\right)}{\Gamma_{d_{\boldsymbol{z}}}\left(\frac{\underline{\nu}^+_n - d_{\boldsymbol{z}} - 1}{2}\right)} \\
		&\times \exp \left( -\frac{\Psi}{2}  \right)  \mathrm{N}(\underline{\boldsymbol{\mu}}_n; \underline{\boldsymbol{\mu}}^+_n, \underline{\boldsymbol{\Sigma}}^+_n) \prod_{m=1}^M \mathrm{N}(\boldsymbol{z}_m; \widetilde{\boldsymbol{\mu}}_m, \boldsymbol{R}/\omega_m)
    \end{aligned}
\end{equation}
where 
\begin{equation} \nonumber
	\Psi = \operatorname{tr}\left( (\underline{\boldsymbol{\Sigma}}^+_n)^{-1} \underline{\boldsymbol{\Sigma}}_n \right) + \sum_{m=1}^M\omega_m \operatorname{tr}(\boldsymbol{R}^{-1}\widetilde{\boldsymbol{\Sigma}}_m) - d_{\boldsymbol{\xi}} - Md_{\boldsymbol{z}} - \left(M-\sum_{m=1}^M\omega_m\right)\mathbb{E}_{ q_{\boldsymbol{e}}} \left[ \ln|s\underline{\boldsymbol{E}}_n + \boldsymbol{R}| \right]
\end{equation}
and $\mathbb{E}_{ q_{\boldsymbol{e}}} \left[ \ln|s\underline{\boldsymbol{E}}_n + \boldsymbol{R}| \right]$ can be computed according to (28) in \cite{MJDL}.

\end{document}